%% file: main.tex
\documentclass[12pt,prd, aps, showpacs, superscriptaddress,floatfix,nofootinbib]{revtex4-1}
\usepackage{graphicx}
\usepackage{bm}
\usepackage{multirow}
\usepackage{epsfig}
\usepackage{psfrag}
\usepackage{soul}
\usepackage{multirow}
\usepackage{mathrsfs}
\usepackage{amsbsy}
\usepackage{amsmath, amssymb}
\usepackage{dcolumn}
\usepackage{epstopdf}
\usepackage{color}
\usepackage{ulem}
\usepackage{hhline}
\usepackage{changebar}
\usepackage{placeins}
\usepackage{pstricks}
\usepackage{color}
\usepackage{tablefootnote}

\usepackage{slashed}

\usepackage{flexisym,breqn,gensymb}
\usepackage{subcaption}
\usepackage[colorlinks=true, linkcolor=blue, filecolor=blue, urlcolor=blue, citecolor=blue,plainpages=false]{hyperref}

\newcommand\riken{RIKEN-BNL Research Center, Brookhaven National
  Laboratory, Upton, NY 11973, USA}
\newcommand\bnl{Brookhaven National Laboratory, Upton, NY 11973, USA}

\newcommand\cu{Physics Department, Columbia University, New York,
  NY 10027, USA}

\newcommand\uconn{Physics Department, University of Connecticut,
  Storrs, CT 06269-3046, USA}
\newcommand\soton{School of Physics and Astronomy, University of
  Southampton,  Southampton SO17 1BJ, UK}

\newcommand\rb{Universit\"at Regensburg, Fakult\"at f\"ur Physik, 93040, Regensburg, Germany}

\newcommand\cern{Theoretical Physics Department, CERN, 1211 Geneve 23, Switzerland} 
\newcommand\ar{Applied Research and Emerging Technology, Qlik, New York, NY 10118, USA}
\newcommand\CAL{Columbia Astrophysics Laboratory, Columbia University, New York, NY 10027, USA}
\newcommand\msu{Department of Computational Mathematics, Science and Engineering; Michigan State University, East Lansing, MI, 48824}

\input{user_def.tex}

\date{\today}

\begin{document}
\bibliographystyle{apsrev}

\title{Lattice determination of \texorpdfstring{$I=0$}{I=0} and 2 \texorpdfstring{$\pi\pi$}{pp} scattering phase shifts with a physical pion mass}

\author{T.~Blum}\affiliation{\uconn}\affiliation{\riken}
\author{P.A.~Boyle}\affiliation{\bnl}
\author{M.~Bruno}\affiliation{\cern}
\author{N.H.~Christ}\affiliation{\cu}
\author{D.~Hoying}\affiliation{\msu}
\author{C.~Kelly}\affiliation{\cu}\affiliation{\bnl}
\author{C.~Lehner}\affiliation{\bnl}\affiliation{\rb}
\author{R.D.~Mawhinney}\affiliation{\cu}
\author{A.S.~Meyer}\altaffiliation{present address: UC Berkeley and Lawrence Berkeley National Laboratory, Berkeley, CA 94720, USA}\affiliation{\bnl}
\author{D.J.~Murphy}\affiliation{\ar}\affiliation{\CAL}
\author{C.T.~Sachrajda}\affiliation{\soton}
\author{A.~Soni}\affiliation{\bnl}
\author{ T.~Wang}\affiliation{\cu}

\collaboration{RBC and UKQCD Collaborations}
%
%
\mbox{}\hfill\noaffiliation{CERN-TH-2021-039}


\maketitle
\centerline{ABSTRACT}
Phase shifts for $s$-wave $\pi\pi$ scattering in both the $I=0$ and $I=2$ channels are determined from a lattice QCD calculation performed on 741 gauge configurations obeying G-parity boundary conditions with a physical pion mass and lattice size of $32^3\times 64$.  These results support our recent study of direct CP violation in $K\to\pi\pi$ decay~\cite{Abbott:2020hxn}, improving our earlier 2015 calculation~\cite{Bai:2015nea}. The phase shifts are determined for both stationary and moving $\pi\pi$ systems, at three ($I=0$) and four ($I=2$) different total momenta. We implement several $\pi\pi$ interpolating operators including a scalar bilinear ``$\sigma$'' operator and paired single-pion bilinear operators with the constituent pions carrying various relative momenta. Several techniques, including correlated fitting and a bootstrap determination of p-values have been used to refine the results and a comparison with the generalized eigenvalue problem (GEVP) method is given. A detailed systematic error analysis is performed which allows phase shift results to be presented at a fixed energy.


\newpage

\section{Introduction}
\label{sec:Introduction}
The scattering of two pions is one of the simplest hadronic processes in QCD. Since the only meson involved, the pion, is the lightest hadron in the standard model and originates from the vacuum breaking of almost exact $SU(2)_L \times SU(2)_R$ chiral symmetry, the behavior of this process at low energy can be well described by chiral perturbation theory (ChPT)~\cite{Ananthanarayan:1998me}. Although this scattering is not directly measurable by experiment, information can be inferred indirectly from $K \rightarrow \pi\pi e\bar\nu_e$ (Ke4) decay~\cite{Batley:2007zz} and $\pi N \rightarrow \pi\pi N$ scattering~\cite{Estabrooks:1974vu}. Unfortunately from these experiments alone, it is difficult to obtain accurate $\pi\pi$ scattering phase shifts for a broad range of $\pi\pi$ energies, because, for example, of the limited energy available in $Ke4$ decays. \par

In addition to its intrinsic interest, $\pi\pi$ scattering is also an important ingredient in our recent calculation of the two-pion decay of the kaon~\cite{Abbott:2020hxn} and the parameter $\varepsilon'$, a highly sensitive measure of direct CP violation, which is a key component in the search for an explanation of the dominance of matter over antimatter in the present Universe. By comparing this lattice QCD result with the experimentally measured value we can gain a better understanding of CP violation in the standard model, with possible insights into the physics beyond it.  While this result for $\varepsilon'$ is reported in Reference~\cite{Abbott:2020hxn}, essential components of this calculation are presented in two companion papers, an extensive study of G-parity boundary conditions~\cite{Christ:2019sah} and the discussion of $\pi\pi$ scattering presented in this paper. \par

Because of the non-perturbative nature of QCD interactions, lattice QCD provides a unique, first-principles method with controlled systematic errors to determine the properties of low-energy QCD.  With lattice QCD and the finite-volume L\"uscher technique~\cite{Luscher:1990ux}, we can calculate the $\pi\pi$ scattering phase shifts within the energy region from $2m_{\pi}$ to approximately $4m_{\pi}$\footnote{For a recent example of an alternative method to extract scattering amplitudes from Euclidean correlators at energies above $4m_\pi$ see Refs.~\cite{Bulava:2019kbi} and \cite{Bruno:2020kyl}.}while the interaction energies near the $2m_\pi$ threshold can be used to determine the scattering lengths~\cite{Luscher:1986pf}.  Such calculations complement the existing determinations of the scattering lengths obtained using chiral perturbation theory~\cite{Colangelo_2000} and the dispersive calculations~\cite{Ananthanarayan:2000ht, Colangelo:2001df, GarciaMartin:2011cn, GarciaMartin:2011jx} of the energy dependence of the phase shift based on the Roy equations~\cite{Roy:1971tc} and experimental input. For a recent review discussing experimental and theoretical results see also Ref.~\cite{Pelaez:2015qba}.  The dispersive technique might also be applied to extrapolate the lattice determination of the energy dependence of the phase shifts above the $4m_\pi$ threshold. \par

Lattice QCD calculations of the $\pi\pi$ scattering phase shifts have been performed by a number of groups, but with pion masses heavier than the physical one so that a chiral extrapolation is required to obtain physical results, see {\it e.g.} Refs.~\cite{Feng:2009ij, Beane:2011sc, Dudek:2012gj, Kurth:2014fra, Bulava:2016mks, Briceno:2016mjc, Briceno:2017max, Fu:2017apw, Culver:2019qtx}. (The exceptions to this statement are our previous physical pion mass calculations of the $I = 2$ and $I = 0$ $s$-wave phase shifts at a single energy close to the kaon mass, which enter the calculation of $\Delta I = 3/2$~\cite{Blum:2011ng, Blum:2015ywa} and $\Delta I = 1/2$  $K\to\pi\pi$ decays~\cite{Bai:2015nea,Abbott:2020hxn} and a calculation of the $I=2$ scattering length~\cite{Fischer:2020jzp} at the physical pion mass.) This extrapolation is most likely valid for a scattering length calculation, but may become less trustworthy at higher energies where the accuracy of chiral perturbation theory becomes less certain. In this paper, we report the first lattice QCD calculation of both $I=0$ and $I=2$ $s$-wave $\pi\pi$ phase shifts performed over a range of two-pion energies with a physical pion mass so that a chiral extrapolation is no longer necessary. \par

As explained above, a central motivation for this study of $\pi\pi$ scattering is its importance for the calculation of the two-pion decay of the kaon~\cite{Abbott:2020hxn} where it enters in three different ways. 1) The lattice calculation of the $K\to\pi\pi$ decay matrix elements is performed in a finite volume while the matrix elements of interest are defined in infinite volume.  The Lellouch-L\"uscher (LL) factor which corrects for this difference is determined by the $\pi\pi$ interaction or, to be more specific, the derivative of the $\pi\pi$ scattering phase shift with respect to energy.  2) In order to determine the $K\to\pi\pi$ decay matrix elements we need to know the amplitude with which the two-pion interpolating operators create the normalized finite-volume $\pi\pi$ states.  The determination of these amplitudes is made difficult by excited state contamination as discussed in Section~\ref{sec:systematic_error}.  3). We need to know the finite-volume $\pi\pi$ state energies and the ground state energy should be close to but will not be exactly the same as the kaon mass. As is the case for the $K \rightarrow \pi\pi$ calculation, this work is performed on a lattice with G-parity boundary conditions (GPBC)~\cite{Christ:2019sah}. This choice is different from the periodic boundary conditions used in most lattice QCD calculations and we will discuss the advantages and drawbacks of this choice in Section~\ref{sec:lattice_detail}. \par

Our first calculation of $I=0$ $\pi\pi$ scattering and $K\to\pi\pi$ decay with physical kinematics~\cite{Bai:2015nea}  was published in 2015 and used the same $32^3\times 64$ lattice volume, M\"obius domain wall fermions and G-parity boundary conditions as the current calculation.  In the earlier calculation we used a single $I=0$ $\pi\pi$ interpolating operator to compute the scattering phase shift at an energy near the kaon mass using 216 configurations. The resulting $I=0$ $s$-wave phase shift, $\delta_0 = 23.8(4.9)(1.2)^\circ$ at center of mass energy $E_{\pi\pi}=498(11)$ MeV, was significantly lower than the dispersive prediction of approximately $39^\circ$ at the kaon mass. (Here and later in this paper when two errors are given, the first is statistical and the second systematic.)\par

Following our 2015 calculation and in light of the discrepancy between our results and the dispersive prediction, we devoted considerable effort to increasing our statistical precision. We found that with 1400 configurations and the same four-quark $\pi\pi$ interpolating operator, a single-state fit to our data continued to be accurate but gave an even lower phase shift of $19.1(2.5)^\circ$, increasing the disagreement with the dispersive result~\cite{Wang:2019nes}. As was the case with the original 216 configurations, performing a two-state fit to the two-point Green's function obtained from this single operator gave a ground state $\pi\pi$ energy and resulting $\pi\pi$ phase shift consistent with what was found from the single state fit. In addition to increasing the statistics we also experimented with adding a second, scalar bilinear $\pi\pi$ interpolating operator that we refer to as the $\sigma$ operator and describe in more detail in Section~\ref{sec:overview:ops}. Performing a two-state fit to the $2\times2$ matrix of two-point Green's functions obtained by including this operator revealed the presence of a previously unrecognized, nearby excited state, leading to a substantially smaller ground state energy and larger $I=0$ phase shift~\cite{Wang:2019nes}.  \par

Our present calculation builds upon this initial effort with increased statistics, additional $\pi\pi$ interpolating operators and a more accurate measure of the quality of the agreement between our data and our theoretical fitting formula. We also extend our calculation beyond a single $\pi\pi$ energy by computing $\pi\pi$ two-point correlation functions with the two pions carrying several values of the total momentum, allowing for an exploration of the scattering phase shifts at center-of-mass energies between approximately $2m_\pi$ up to the kaon mass. Recognizing the importance of multiple operators we further increase the number of independent interpolating operators for both the stationary and moving frame calculations of the $I=0$ and 2 phase shifts.  Here we present results from 741 configurations using three operators in the moving frame calculation and three (for $I=0$) and two (for $I=2$) operators in the stationary frame calculation.  With these additional operators we obtain a significant improvement in statistical precision.  We are also better able to demonstrate control over the contamination from neglected excited states and to more reliably estimate their effects.  We also applied a second approach to the analysis of our multi-operator, multi-state data, the generalized eigenvalue problem (GEVP) method.  This new method gave results consistent with those of our traditional fitting approach with similar statistical errors.

The moving frame calculation allows us to directly calculate the LL factor from our lattice QCD data, the results of which are presented in Section~\ref{sec:phase_shift:LL} and utilized in Ref~\cite{Abbott:2020hxn}.   As described in Section~\ref{sec:systematic_error} our final result from 741 configurations and a two-state fit to the $3\times3$ matrix of two-point Green's function coming from three $\pi\pi$ interpolating operators gives $\delta_0(471\textrm{MeV})=32.3(1.0)(1.4)^\circ$, which is in much better agreement with the dispersive prediction of $35.9^\circ$, a number obtained by evaluating Eqs. (17.1)-(17.3) of Ref.~\cite{Colangelo:2001df}.  This value was obtained in Ref.~\cite{Colangelo:2001df} using $M_\pi=139.6$ MeV.  However, elsewhere in the present paper, we treat the neutral pion mass of $M_\pi=135$ MeV as the ``physical'' pion mass, following our previous papers on $K\to\pi\pi$ decay and the conventions of the RBC and UKQCD collaborations.  We therefore make a correction for the difference between the pion mass used in our lattice calculation, $m_\pi = 142.3(0.7)$ MeV, and this physical 135 MeV pion mass.  We will discuss how we deal with the differences between these several different pion masses in greater detail in Sec.\ref{sec:phase_shift}. Also note that here we have corrected our result to transfer the uncertainty in the energy at which the phase shift is determined onto the phase shift itself as described in Section~\ref{sec:systematic_error} and have correspondingly evaluated the dispersive prediction at this energy rather than at the kaon mass.\par

In the next section we describe the properties of the ensemble of gauge configurations that are used for both the calculation of $\pi\pi$ scattering presented here and our companion calculation of $K\to\pi\pi$ decay and $\varepsilon'$~\cite{Bai:2015nea}, together with a brief discussion about the G-parity boundary conditions adopted in these calculations. In Section~\ref{sec:overview} we present the operators we used, the matrix of two-point Green's function we measured and the statistical methods we used. In Sections~\ref{sec:pion_2pt_function} and \ref{sec:pipi_energy} we present in detail our fitting procedures and results for single pion and $\pi\pi$ energies and two-point function amplitudes, together with a brief comparison with another data analysis method, the generalized eigenvalue problem. With these results, in Section~\ref{sec:phase_shift} we describe how we obtain the phase shift results at various center-of-mass energies using a generalized form of L\"uscher's formula. In Section~\ref{sec:systematic_error} we explain our new approach to determining the systematic uncertainties, estimate the largest systematic errors and present the resulting error budget. Finally in Section \ref{sec:conclusions} we present our conclusions. \par


\section{Description of the gauge ensemble}
\label{sec:lattice_detail}
We employ a single $32^3\times 64$ lattice with $2+1$ flavors of M\"obius domain wall fermions with $L_s=12$ and M\"obius parameters $b+c=32/12$ and $b-c=1$ and the Iwasaki+DSDR gauge action with $\beta=1.75$, corresponding to an inverse lattice spacing of $a^{-1}=1.3784(68)$\,GeV~\cite{Blum:2014tka}. Here the dislocation suppressing determinant ratio (DSDR) reduces the dislocations, or tears in the gauge field that enhance chiral symmetry breaking at coarse lattice spacings~\cite{Vranas:1999rz,Vranas:2006zk,Fukaya:2006vs,Renfrew:2009wu}. Its use enables us to work with a large, $(4.6\ {\rm fm})^3$ spatial volume and therefore have good control over finite-volume systematic errors, without a dramatic increase in computational cost, albeit at the cost of increased discretization errors. We use G-parity boundary conditions (GPBC) in three spatial directions in order to obtain physical kinematics for the $K\to\pi\pi$ decay.

The lattice parameters are equal to those of the 32ID ensemble documented in Refs.~\cite{Arthur:2012yc, Blum:2014tka}, with the addition of GPBC and a slightly lower pion mass of 142 MeV versus the 172\,MeV used previously. This enables us to take advantage of existing results such as the value of the lattice spacing and also to compute the non-perturbative renormalization factors for the $K\to\pi\pi$ matrix elements, in an environment free of the complexities associated with GPBC. 

Below we first discuss the generation of these ensembles and measured properties including the plaquette, chiral condensate and autocorrelation times. We then discuss GPBC and how they differ from the usual periodic boundary condition (PBC). A detailed discussion on GPBC can be found in Ref.~\cite{Christ:2019sah}.

\subsection{Ensemble Generation}

The ensemble used for our 2015 calculation comprised 864 gauge configurations (after thermalization), with measurements performed on every fourth configuration giving 216 in total. Following our publication an error was discovered~\cite{Bai:2016ocm} in the generation of the random numbers used to set the conjugate momentum at the start of each Monte Carlo trajectory that introduced small correlations between widely separated lattice sites. While the effects were found to be two-to-three orders of magnitude smaller than our statistical errors, we nevertheless do not include these configurations in our new calculation. 

In order to rapidly improve the statistical precision of our calculation we generated configurations via 7 independent Markov chains, each originating from widely separated configurations in our original thermalized ensemble. To compensate for any residual effects of the random number error we discarded the first 20 configurations of each stream, which is approximately 5 times the integrated autocorrelation time (see below). These configurations were generated using the hybrid Monte Carlo technique for which the Hamiltonian can be decomposed as
\begin{dmath}
H = T + S_G + S_{\rm lQ}(0.0001,1) + S_{\rm hQ}(0.045,1) + S_{\rm DSDR}\,,
\end{dmath}
where $T$ is the kinetic term and ``lQ'' and ``hQ'' denote the light and heavy quarks, respectively. The fermion actions comprise ratios of determinants
\begin{dmath}
S_X(a,b) = -\ln\left\{ {\rm det}\left( \frac{M^\dag(a)M(a)}{M^\dag(b)M(b)} \right)^{n_X/4} \right\}\,,
\end{dmath}
where $M(m)$ is the Dirac matrix with mass $m$, $n_X$ is the number of quark species of type $X$ and for $b=1$ the denominator represents the Pauli-Villars term required by the domain wall formalism. The determinants of the squared matrix are used such that the matrices which are inverted when approximating the determinant are Hermitian and positive-definite and thus suitable for the conjugate gradient (CG) algorithm and can be obtained from a convergent pseudo-fermion integration. Here the fractional power is required by the fact that for G-parity boundary conditions, the determinant of the squared matrix represents the contribution of four quark flavors~\cite{Christ:2019sah}, hence a square-root is required for the two light flavors and a fourth-root for the strange quark in order to perform a 2+1 flavor simulation. (Note that for periodic boundary conditions, the squared-matrix determinant represents the action of two quark flavors, hence for a 2+1 flavor simulation only the square root of the strange-quark determinant is typically required.) The fractional power is achieved using the rational hybrid Monte Carlo (RHMC) algorithm. The light-quark action is further decomposed into two pieces using the Hasenbusch mass splitting technique~\cite{Hasenbusch:2001ne} as follows:
\begin{dmath}
S_{\rm lQ}(0.0001,1) \to S_{\rm lQ}(0.0001,0.007) +  S_{\rm lQ}(0.007,1)\,.
\end{dmath}
We use an integration scheme comprising four levels of nested Omelyan integrator with Omelyan parameter $\lambda=0.22$, using the layout detailed in Table~\ref{tab-integrator-layout}. Over 2200 configurations were generated in these 7 streams.

\begin{table}[tb]
\centering
\begin{tabular}{cccc}
\hline\hline
Level (i) & $S_i$ & $n^S_i$ \\
\hline
1 & $S_{\rm lQ}(0.0001,0.007)$ & 1 \\
2 & $S_{\rm lQ}(0.007,1) + S_{\rm hQ}(0.045,1)$ & 2\\
3 & $S_{\rm DSDR}$ & 2\\
4 & $S_G$ & 1\\
\end{tabular}
\caption{The decomposition of the action onto the four levels of the nested integrator. The top-most integrator ($i=1$) is integrated over 16 steps of step-size $1/16$ for a trajectory length of 1 MD time unit. Each update step of the action-integrator $S_i$ of size $\tau$ is divided into $n^S_i$ equal-sized steps, with $n^S_i$ given in the third column. For a detailed discussion of the nested integrator technique we refer the reader to Appendix A of Ref.~\cite{Arthur:2012yc} 
\label{tab-integrator-layout} }
\end{table}

The use of RHMC for the light quark determinant introduces a significant cost overhead, primarily because the various mixed-precision techniques that have been developed to improve the efficiency of the standard CG algorithm are not generally applicable to the underlying multi-shift CG algorithm, which requires all the starting vectors to lie within the same Krylov space thus precluding the use of restarted methods. In addition we found that tighter stopping conditions on the inversion than are typical when applied to heavy quarks were required to ensure good acceptance and that more poles (20 in this case versus the ${\cal O}(10)$ required for a typical heavy quark calculation) were required to span the measured eigenvalue range. With some effort we were able to achieve a 70\% performance increase, as measured on the IBM BlueGene/Q machines upon which a majority of our ensemble generation was performed, by combining a ``reliable update'' step with a subsequent loop over each pole with a conventional mixed-precision restarted CG~\cite{Kelly:2014ifa}.

A more significant improvement in the configuration generation was obtained by implementing the ``exact one-flavor algorithm'' (EOFA)~\cite{Chen:2014hyy,Chen:2014bbc} formulated by the TWQCD collaboration, in which a Hermitian positive-definite action for a {\it single species} of domain wall fermion is derived. The use of the EOFA allows us to circumvent the use of RHMC in the light quark sector, opening the door for various optimizations. We determined~\cite{Jung:2017xef} that with a suitable preconditioning, the EOFA can be reformulated in a way that is not only more efficient but also allows for the re-use of the majority of our existing high-performance code for regular domain wall fermions. Coupled with algorithm and integrator tuning we achieved a $4.2\times$ reduction in the time to generate a gauge configuration on the same hardware~\cite{Jung:2017xef}. Utilizing this algorithm we extended 3 of our 7 streams by a total of nearly 3000 additional gauge configurations.

For this calculation we have measured on a subset of 741 configurations with consecutive measurements separated by 4 molecular dynamics time units (MDTU), which amounts to roughly 60\% of the available configurations given this measurement separation.

\subsection{Ensemble Properties}

\begin{figure}[tb]
\centering
\includegraphics[width=0.48\textwidth]{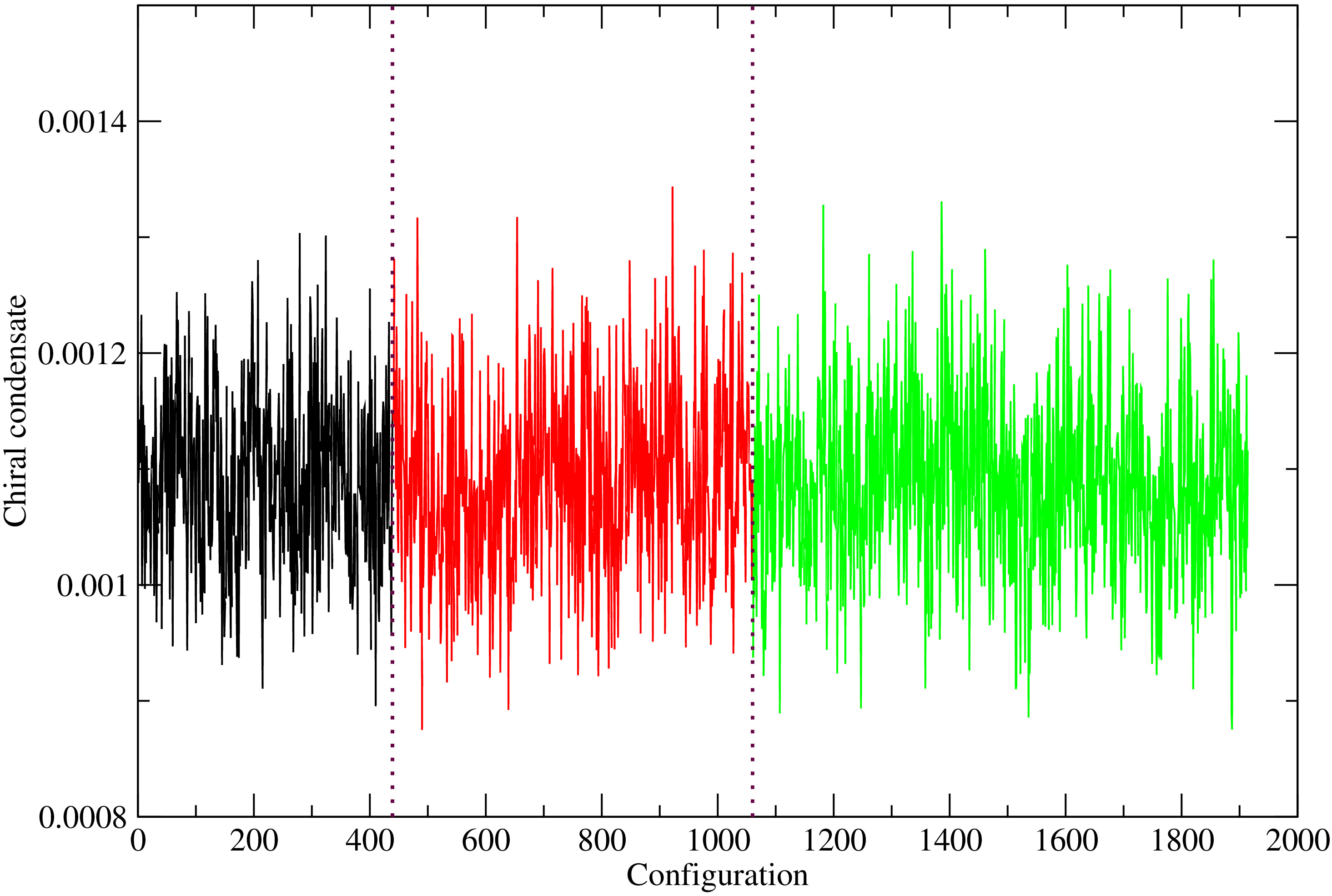}
\includegraphics[width=0.48\textwidth]{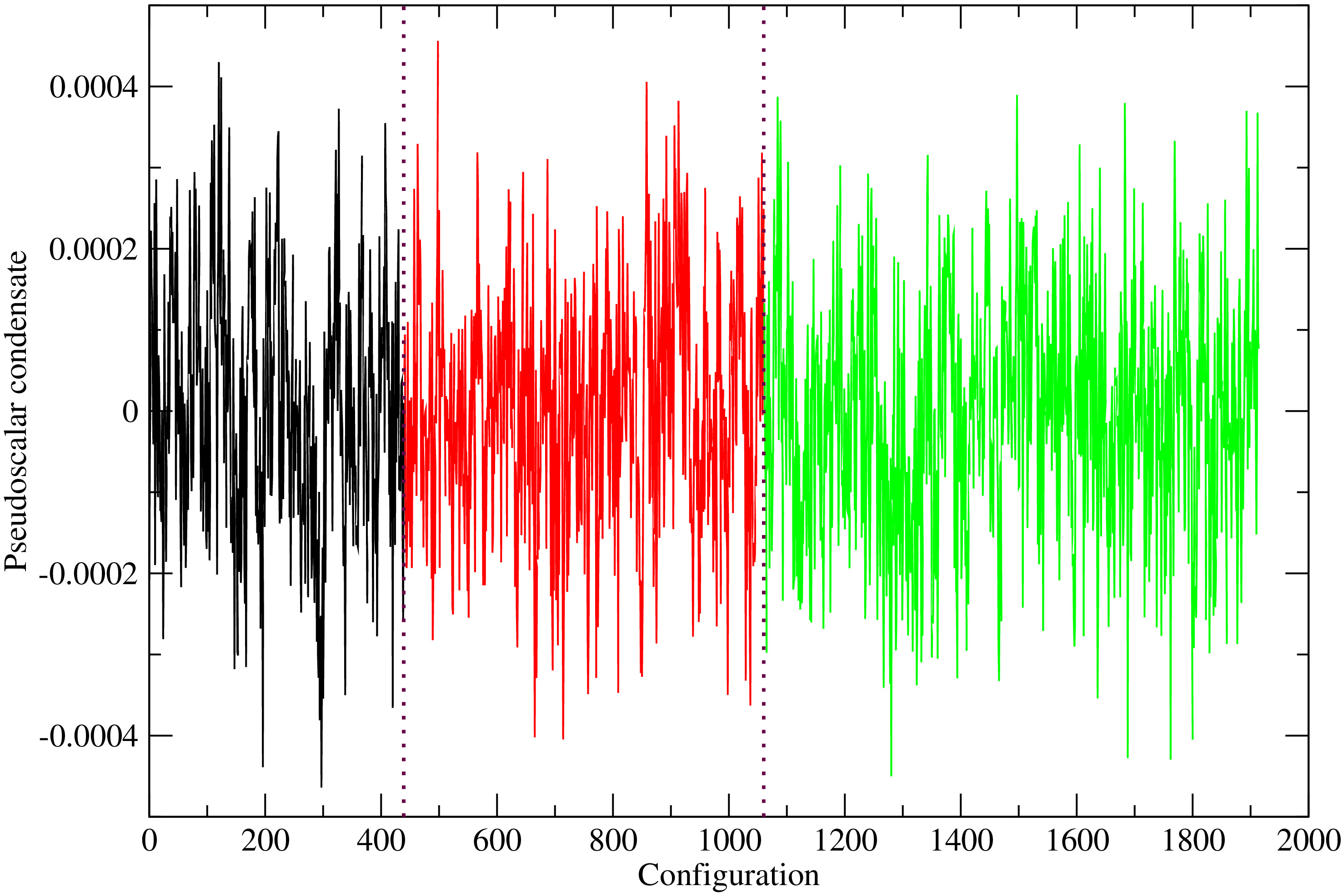}\\
\includegraphics[width=0.48\textwidth]{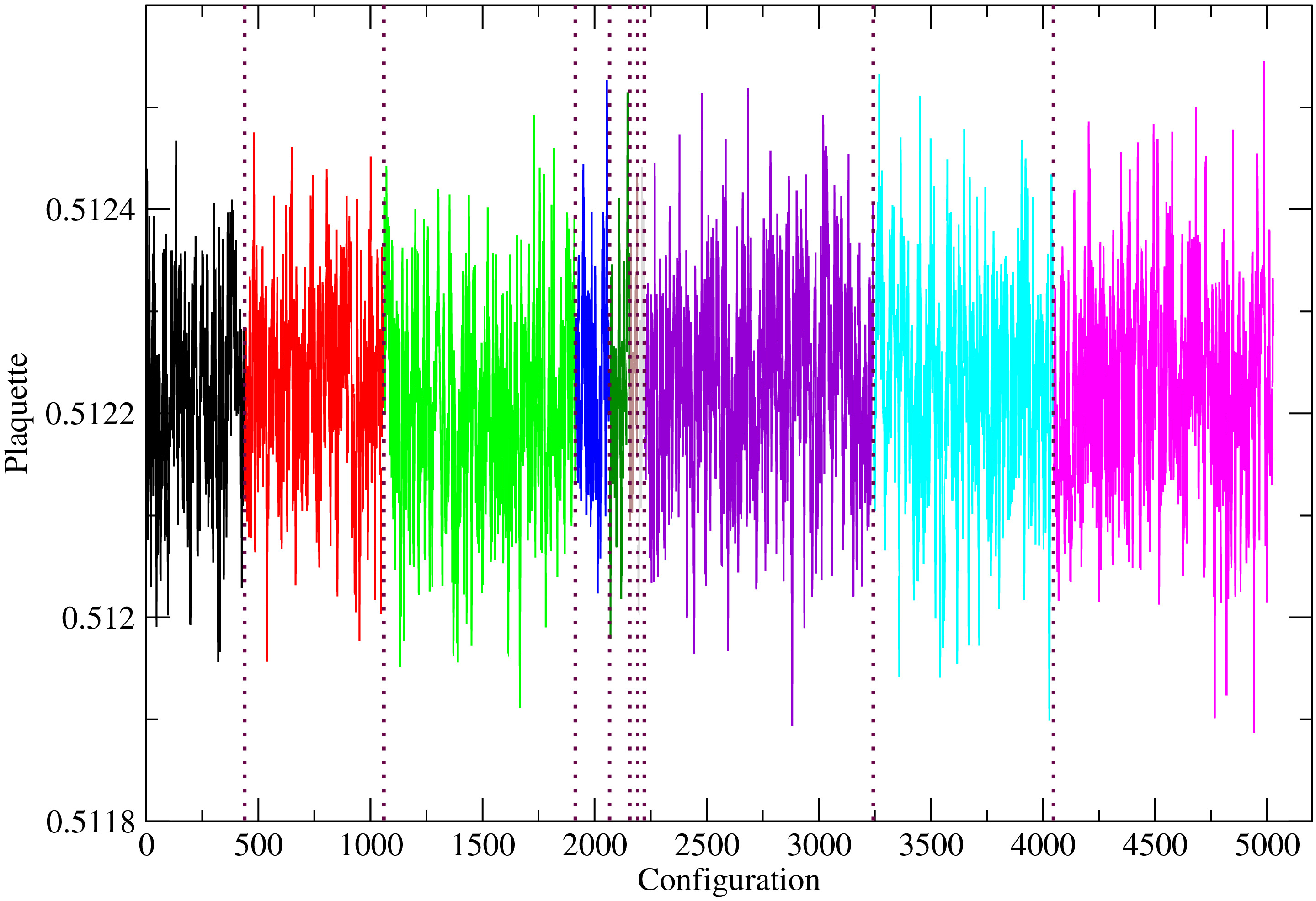}
\caption{Evolution of the chiral condensate (upper-left), pseudoscalar density (upper-right) and plaquette (lower). The different colors and vertical dotted lines indicate the boundaries of the 10 independent Markov chains. Note that measurements of the chiral condensate and pseudo-scalar density were performed only on the subset of configurations corresponding to the first three chains of the lower figure. The last three ensembles in the lower figure were generated using the EOFA technique. \label{fig-evo-plots}}
\end{figure}

\begin{table}[tb]
\centering
\begin{tabular}{c|ccc}
\hline\hline
Ensemble & $\langle P\rangle$ & $\langle \bar\psi \psi\rangle$ & $\langle \bar\psi \gamma^5\psi\rangle$\\
\hline
1 & $ 0.5122174(95)$ & $ 0.0010920(45)$ & $ 6(15)\times 10^{-6}$\\
2 & $ 0.5122277(65)$ & $ 0.0010915(39)$ & $ 1.7(9.7)\times 10^{-6}$\\
3 & $ 0.5122022(66)$ & $ 0.0010891(34)$ & $-4.7(8.4)\times 10^{-6}$\\
4 & $ 0.512231(14)$ & - & -\\
5 & $ 0.512224(22)$ & - & -\\
6 & $ 0.512214(27)$ & - & -\\
7 & $ 0.512195(70)$ & - & -\\
\hline
8 & $ 0.5122326(57)$ & - & -\\
9 & $ 0.5122254(69)$ & - & -\\
10 & $ 0.5122256(57)$ & - & -\\
\end{tabular}
\caption{Expectation values of the plaquette, chiral condensate and pseudoscalar density. The ensembles are given in the same order as in the plot shown in Figure~\ref{fig-evo-plots}. The chiral condensate and pseudoscalar density were measured only on the first three ensembles. The last three ensembles, separated by a horizontal line, were generated using the EOFA technique. The error bars were obtained using the jackknife technique applied to data that had been binned over blocks of size 12 MDTU.\label{tab-evo-avgs}}
\end{table}

\begin{figure}[tb]
\centering
\includegraphics[width=0.48\textwidth]{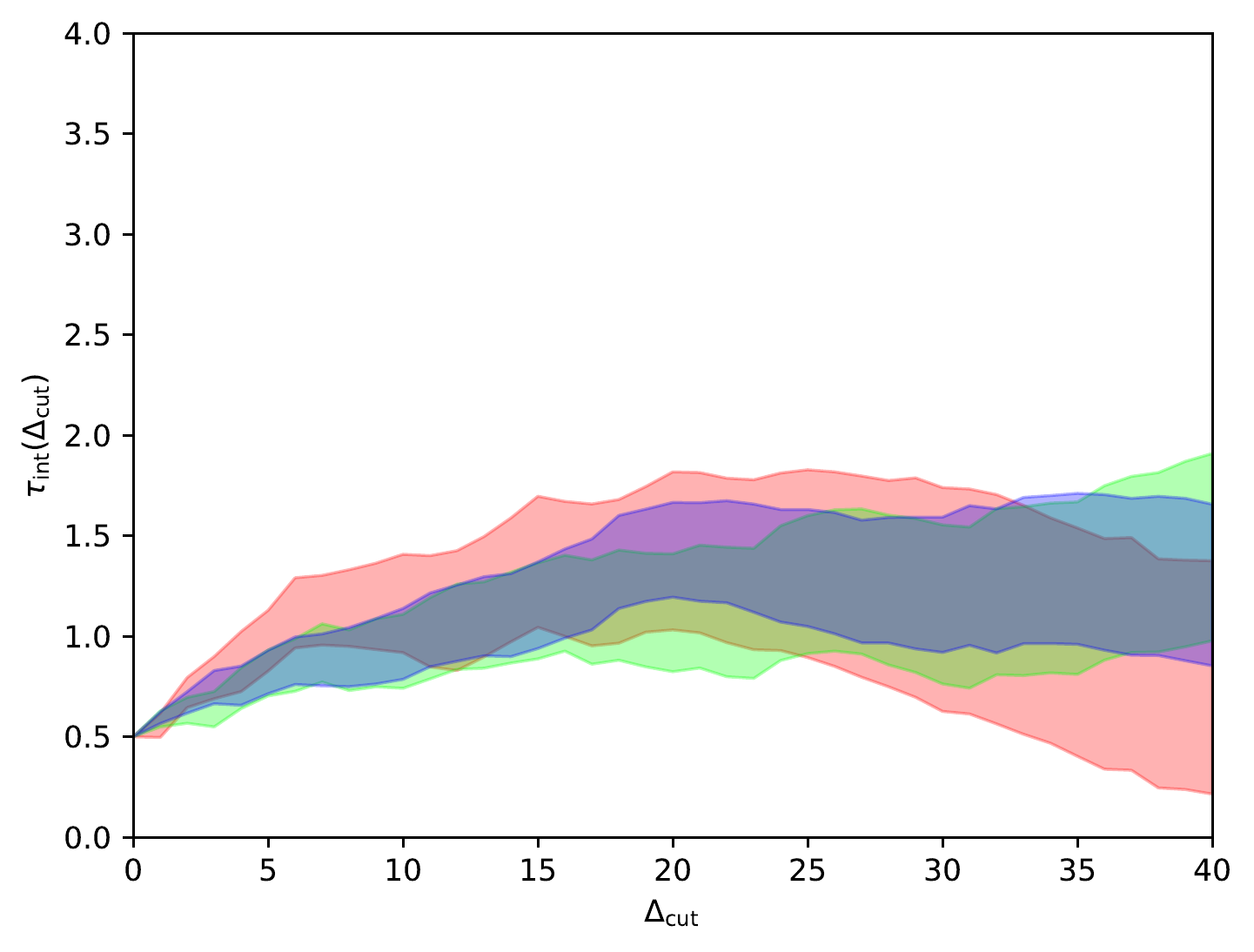}
\includegraphics[width=0.48\textwidth]{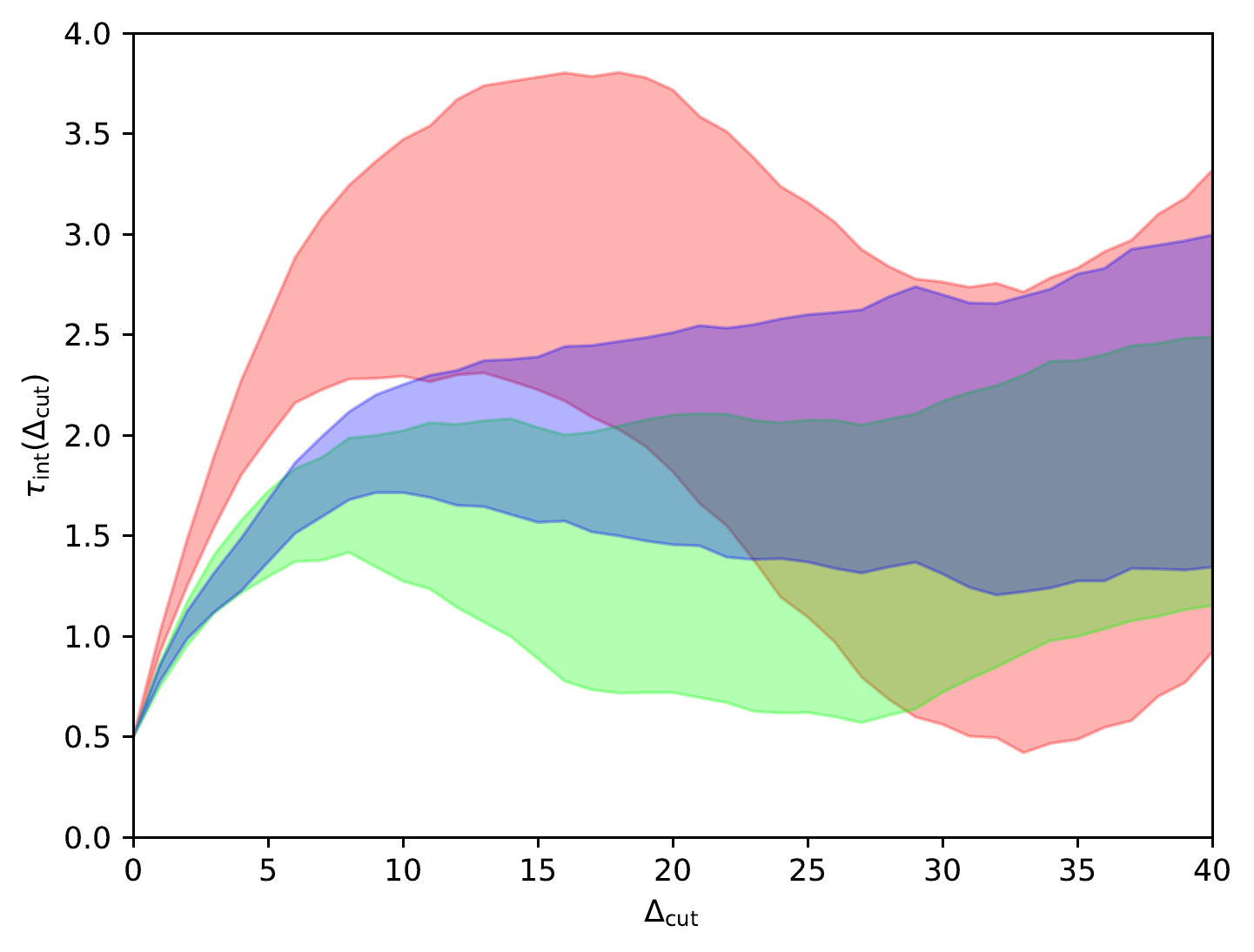}\\
\includegraphics[width=0.48\textwidth]{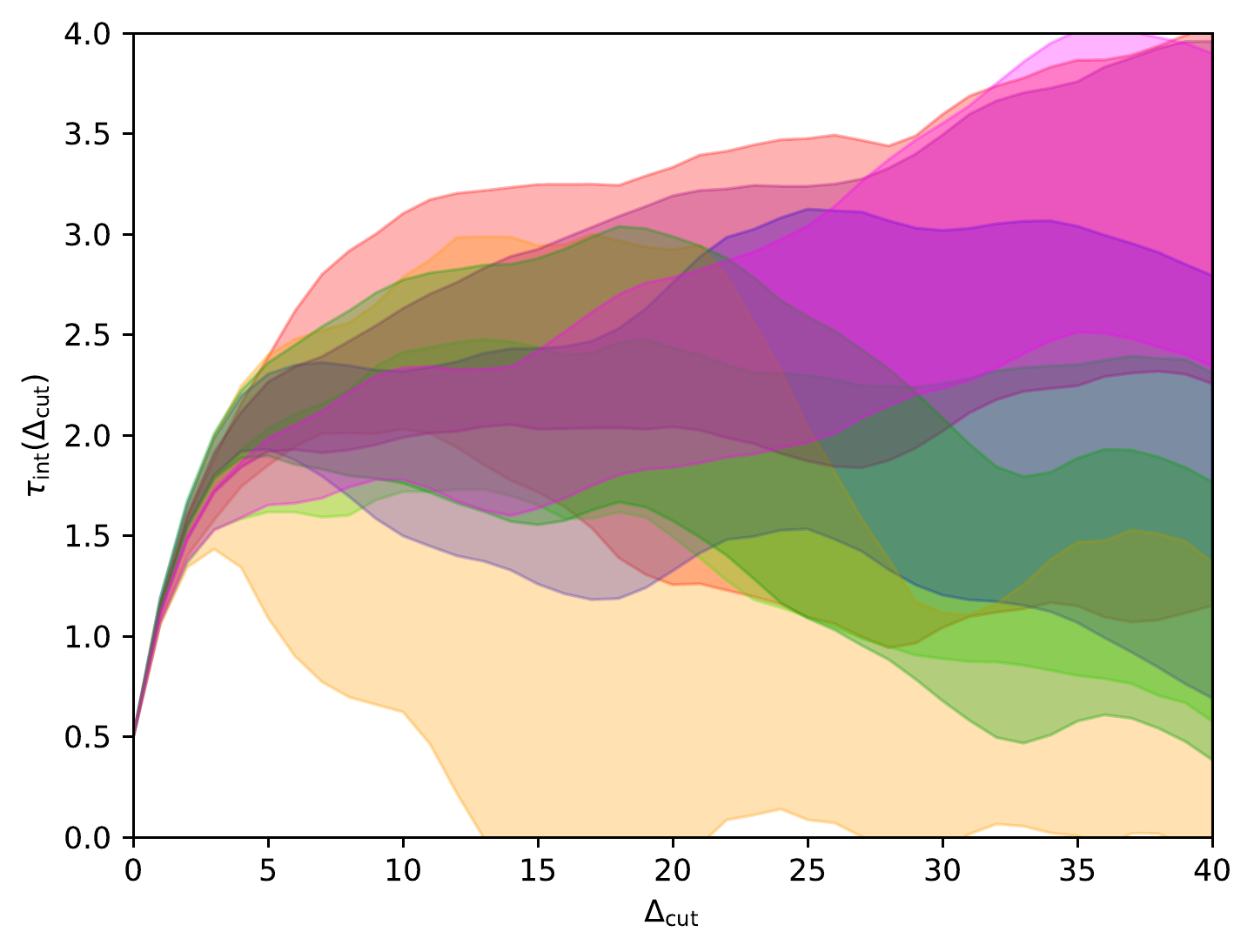}
\caption{The integrated autocorrelation time $\tau_{\rm int}(\Delta_{\rm cut})$ as a function of $\Delta_{\rm cut}$ for the chiral condensate (upper-left), pseudoscalar density (upper-right) and plaquette (lower) superimposed for the different streams. The two very short streams are not included in this analysis. The chiral condensate and pseudoscalar density were measured on only three of the ensembles. \label{fig-tau-int}}
\end{figure}

In Figure~\ref{fig-evo-plots} we plot the evolution of the chiral condensate and pseudoscalar density as well as the plaquette. We remind the reader that the chains were each generated from already-thermalized and well-separated configurations of our original ensemble and hence we expect to observe no thermalization effects in these plots. The expectation values of these quantities for each of the ensembles are listed in Table~\ref{tab-evo-avgs}. In Figure~\ref{fig-tau-int} we plot the integrated autocorrelation time
\begin{dmath}
\tau_{\rm int}(\Delta_{\rm cut})= \frac{1}{2} + \sum_{\Delta=1}^{\Delta_{\rm cut}}C(\Delta)\,,
\end{dmath}
where $C(\Delta)$ is the autocorrelation function 
\begin{dmath}
C(\Delta) = \frac{1}{N-\Delta}\sum_{i=1}^{N-\Delta } \widetilde C(i,\Delta)\,,\label{eq-autocorr-func-def}
\end{dmath}
with
\begin{dmath}
\widetilde C(i,\Delta) = \frac{( v_{i} - \bar v ) ( v_{i+\Delta} - \bar v )}{\sigma^2_v}\,.
\end{dmath}
Here $N$ is the number of samples $v_i$ of some quantity with mean $\bar v$ and standard deviation $\sigma_v$. The standard errors on $\tau_{\rm int}$ shown in the figure are estimated using a bootstrap resampling procedure applied to the quantities $\widetilde C(i,\Delta)$ prior to performing the average in Eq.~\ref{eq-autocorr-func-def}, and described in Refs.~\cite{Arthur:2012opa,Blum:2014tka}. 

Because this is a non-standard application of resampling we provide here a detailed description and justification of the method.  Since the functions $\widetilde C(i,\Delta)$ from which the integrated autocorrelation time $\tau_{\mathrm{int}}(\Delta_{\mathrm{cut}})$ is computed, depend on samples taken at different points in the Markov chain of gauge configurations, we do not attempt to obtain our error estimate by selecting samples from this Markov chain.  Instead for each $i$ we view the set of $\Delta_{\mathrm{max}}$ values, $\{\widetilde C(i,\Delta)\}_{1\le\Delta\le\Delta_{\mathrm{max}}}$ as a stochastic sample and deduce the statistical errors in our result for $\tau_{\mathrm{int}}(\Delta_{\mathrm{cut}})$ from the fluctuations among these samples. This is analogous to the usual treatment of a two-operator correlation function $M_k(t)$ computed on a configuration $k$ for a range of time separations $t$ between the two operators.  Here $\Delta_{\mathrm{max}}$ is the largest value of $\Delta$ that we include in our analysis. In this analysis we use $\Delta_{\rm max}=40$.  Thus, the number of samples is $N-\Delta_{\rm max}$ with $1 \le i \le N-\Delta_{\rm max}$.   

Consider the quantities $\widetilde C(i,\Delta)$ and $\widetilde C(j,\Delta)$ for $j\geq i$, which contain measurements from the pairs of configurations $(i,i+\Delta)$ and $(j,j+\Delta)$, respectively. Due to the autocorrelations in the underlying data these values are correlated, with the corresponding correlation function peaking when $j$ coincides with $i$ {\it or} with $i+\Delta$, and falling off exponentially in the time separation $j-i$ away from these points. The secondary peak occurring when $j$ is close to $i+\Delta$ is smaller than the primary peak at $j=i$ because only one of the two configurations involved in each pair coincide. Furthermore it is straightforward to show that the secondary peak vanishes entirely for $\Delta$ sufficiently large that the configurations $i$ and $i+\Delta$ become effectively independent. 

Thus, while the correlations between $\widetilde C(i,\Delta)$ with different values of $i$ have an unusual form, a standard binning procedure:
\begin{dmath}
\widehat C(\alpha, \Delta) = \frac{1}{B}\sum_{i=B\alpha+1}^{B(\alpha+1)} \widetilde C(i,\Delta)\label{eq-autocorr-func-bin}
\end{dmath}
is sufficient to generate values $\widehat C(\alpha, \Delta)$ that, for a large enough bin size $B$, are statistically independent in $\alpha$ and that are therefore amenable to bootstrap (or jackknife) resampling.

The full procedure is then:
\begin{enumerate}
 \item Truncate the collection of data to be analyzed to $\widetilde C(i,\Delta)$ for $1 \le i \le N-\Delta_{\rm max}$ and $1\le \Delta \le \Delta_{\rm max}$.
 \item Bin this collection of data according to Eq.~\eqref{eq-autocorr-func-bin} producing $N_{\rm bin}$ samples $\{\widehat C(\alpha)\}_{1\le \alpha \le N_{\rm bin}}$ where each sample $\widehat C(\alpha)$ represents the $\Delta_{\rm max}$ quantities $\{\widehat C(\alpha,\Delta)\}_{1\le \Delta \le \Delta_{\rm max}}$ with $N_{\rm bin} = \left\lfloor (N - \Delta_{\rm max})/B \right\rfloor$.
 \item Bootstrap resample $\{\widehat C(\alpha)\}_{1\le \alpha \le N_{\rm bin}}$ in $\alpha$ producing $N_{\rm boot}$ bootstrap ensembles $\{\widehat C_b(\alpha)\}_{1\le \alpha \le N_{\rm bin}}$, each comprising $N_{\rm bin}$ elements where $b$ is the bootstrap ensemble index. For this analysis we use $N_{\rm boot}=500$.
 \item Compute the autocorrelation function under the bootstrap,
  $$C_b(\Delta) = \frac{1}{N_{\rm bin}}\sum_{\alpha=1}^{N_{\rm bin}} \widehat C_b(\alpha, \Delta)$$
 for each $\Delta \leq \Delta_{\rm max}$.
 \item Compute the integrated autocorrelation function,
 $$\tau_{{\rm int},b}(\Delta_{\rm cut})= \frac{1}{2} + \sum_{\Delta=1}^{\Delta_{\rm cut}}C_b(\Delta)$$
 for $\Delta_{\rm cut}\leq \Delta_{\rm max}$.
\end{enumerate}
The standard deviation of the bootstrap distribution of $\tau_{{\rm int},b}(\Delta_{\rm cut})$ over $b$ provides an estimate of the standard error on $\tau_{\rm int}(\Delta_{\rm cut})$. The appropriate bin size can be found, as usual, by increasing $B$ until the error estimates stabilize; for the present analysis a bin size $B=15$ was found to be sufficient. From Figure~\ref{fig-tau-int} we estimate an integrated autocorrelation time of $\tau_{\rm int}{\sim}$3-4 MDTU, which is close to the separation of 4 MDTU between our measurements. We therefore expect minimal autocorrelation effects on our measurements, but to be certain of our error estimates we will account for any residual effects using the non-overlapping block bootstrap procedure, as we will detail in Section~\ref{sec:overview}.

\subsection{G-parity boundary conditions}
\label{sec:lattice_detail_GPBC}
The most significant difference between our calculation and those of other groups is the boundary conditions: in this work we use G-parity boundary conditions in all three spatial directions. G-parity is a symmetry of the QCD Lagrangian under charge conjugation coupled with a 180 degree isospin rotation about the y-axis. Applied as a spatial boundary condition on the quark fields this transforms a quark flavor doublet  (u,d) into $(-C\Bar{\textrm{d}}^T,C\bar{\textrm{u}}^T)$ as it passes through the boundary, where $C$ is the $4\times4$ charge conjugation matrix. \par

As with all boundary condition variants the introduction of GPBC modifies the finite-volume spectrum.  As described below, we take advantage of this change to improve the accuracy of our $K \rightarrow \pi\pi$ calculation. With GPBC applied to the up and down quarks and after introducing a fictional doublet partner $s'$ to the strange quark to which GPBC are also applied (with the additional species suitably weighted out of the path integral, cf. Ref.~\cite{Abbott:2020hxn}), a kaon state can be identified which satisfies periodic boundary condition while the pion states must satisfy anti-periodic boundary conditions (APBC).  Recall that all three pions are odd under G-parity.  This means that we can introduce a kaon ground state which is at rest, while the $\pi\pi$ ground state will be composed of two moving pions, with momenta close to $\pm \pi/L$ in each direction (the deviations from $\pm \pi/L$ being due to the $\pi\pi$ interactions we seek to measure). We can then tune the lattice parameters so that the initial kaon and the final $\pi\pi$ ground state have the same energy. This makes the $K \rightarrow \pi\pi$ calculation much easier since we can focus on the dominant ground state contribution to the $K \rightarrow \pi\pi$ matrix element. In contrast, on a lattice with PBC the $\pi\pi$ ground state will be composed with two nearly stationary pions, and we must tune the lattice spacing so that the energy of an excited $\pi\pi$ state matches the kaon mass. The matrix element of interest in this case will be a subdominant contribution to the three-point Green's function and obtaining a precise result becomes much more challenging. For more details on performing lattice simulations with G-parity boundary conditions including further discussion of the lattice symmetries and the treatment of the strange quark, we refer the reader to Ref.~\cite{Christ:2019sah}. For the remainder of this subsection we will focus specifically on how these boundary conditions affect the measurement of the two-pion system. \par

Including GPBC introduces some significant differences from a calculation with PBC.  Three significant differences might be identified.  First, the $\pi\pi$ states that can be studied with these two types of boundary condition will be different.  When non-interacting pions satisfy APBC in all three directions their allowed momenta become $(2n_1+1, 2n_2+1, 2n_3+1)\frac{\pi}{L}$, where $n_i$ are integers. These are different from those on a volume with PBC, where the allowed momenta are $(2n_1, 2n_2, 2n_3)\frac{\pi}{L}$.  However, if we take advantage of moving frames, there is still a correspondence between the states that we can construct on a PBC volume and those present for a volume obeying GPBC. For example, if we wish to work with a $\pi\pi$ state comprising two pions at rest, for a volume with PBC we can do the calculation in the stationary frame, where the two component pion operators are constructed with zero momentum.  However, for a GPBC volume the calculation can be performed in a moving frame where both pions have the same momentum, {\it e.g.} $(\pi/L,\pi/L,\pi/L)$.  With this choice, in the center-of-mass frame these two pions are at rest. \par

Since a moving frame calculation relies on a distorted volume which doesn't have cubic symmetry, there will be lower angular momentum partial waves whose phase shifts will enter the quantization condition that determines the $s$-wave phase shift, {\it e.g.} $d$-waves.  In the stationary frame the lowest partial waves that enter beyond the $s$-wave are those with with $l=4$.  Fortunately in this work the interaction energies involved in our moving frame calculations are relatively small (around the kaon mass), and those higher partial waves that enter will have a negligible effect on the $s$-wave phase shift. \par

A second troublesome aspect of G-parity is the breaking of cubic symmetry at the quark level even for a lattice with cubic symmetry.  As discussed in Ref.~\cite{Christ:2019sah} there is a sign convention that can be chosen when G-parity is imposed in one direction that can be changed by changing the relative sign of the up and down quark fields.  However, the choice of this sign in the remaining two directions is not conventional and breaks cubic symmetry by identifying one of the four diagonals connecting two corners of the cubic lattice and passing through its center.  For a cubic volume in a stationary frame, the symmetry group is broken down from $O_h$ to $D_{3d}$~\cite{atkins1970tables}. 

There are two effects of this breaking of $O_h$ symmetry that we need to consider: First, its effect on the two-pion eigenstates of the QCD transfer matrix and second its effect on the rotational properties of the quark-level interpolating operators used to create those pions.  Because of confinement the relevant degrees of freedom affected by the G-parity boundary conditions are the pions.  Since G-parity boundary conditions are translationally invariant, for the $O_h$-breaking properties of the quarks which make up the pion to have an effect, a single isolated quark must propagate across the lattice and through the boundary, a phenomenon that should be highly suppressed by effects of quark confinement. While this argument suggests that the two-pion eigenstates of the transfer matrix should fall into representations of the $O_h$ group, it is possible that the four-quark interpolating operators used to create these states will couple to more than one irreducible $O_h$ representation and care must be taken when constructing translationally covariant operators to suppress the creation of finite-volume states belonging to unwanted representations of the cubic symmetry group $O_h$.  This will be discussed when we write out the explicit form of these operators in Sec.~\ref{sec:overview} and the remaining cubic-symmetry breaking effects are discussed in Sec.~\ref{sec:systematic_error}. \par

Finally around-the-world effects in a moving frame will be different in a volume with GPBC compared to one with PBC.  When we are performing a moving frame calculation in a volume with GPBC with one of the three smallest allowed total momenta (those with $P_{\rm tot} = (\pm2,0,0)$, $(\pm2,\pm2,0)$ or $(\pm2,\pm2,\pm2)$ in units of $\pi/L$), the first-order around-the-world contribution will come from a single pion propagating from one $\pi\pi$ interpolating operator to the second (leg A) and a second single pion propagating from the second, through the time boundary to the first (leg B).  This behavior is shown schematically as part of a later more detailed discussion in Figure~\ref{fig:pipi_2nd_atw}.

For GPBC the momentum injected by each $\pi\pi$ interpolating operator can change the direction but not the magnitude of the momentum carried by the pion as it moves from leg A to leg B.  Thus, for GPBC this around-the-world pion can carry the same energy on each leg and so that its contribution behaves as a constant when the time separation between the two operators is changed.  We refer to this case where the pions in both legs carry momenta of minimum magnitude as the ``first-order'' around-the-world effect.  The case in which the pion propagating in one of the legs carries momentum greater than the minimum is termed ``second-order''.  Both cases are considered when performing the fits described in Section~\ref{sec:pipi_energy_I2_moving}.   In contrast, for the three smallest non-zero total momenta in a calculation with periodic boundary conditions all of the around-the-world terms will be time-dependent since the pions in the two legs will have different energies.


\section{Overview of the measurements}
\label{sec:overview}
In this section we describe the details of the interpolating operators used in this calculation, the two-point functions that we study and the specific contractions that are evaluated.  In the final subsection we outline the fitting methods employed and the methods used to determine a statistical error and assign a $p$-value to those fits. \par

\subsection{Interpolating Operators}
\label{sec:overview:ops}
Here we discuss the structure of the interpolating operators used in this work. There are two different types of two-pion interpolating operators.  The first type are denoted as ``$\pi\pi(\ldots)$" operators and are constructed as the product of two single-pion interpolating operators and for which the parentheses and the quantity contained within are used both to specify the pion momenta and to distinguish these labels from the general set of $\pi\pi$ interpolating operators which can produce two pions when acting on the vacuum, the set in which all of our operators reside.  The second type has the form of a quark-bilinear scalar sigma operator which shares the same quantum number as $I=0$ $\pi\pi$ state. We start by constructing the single pion and sigma interpolating operators with momentum $\Vec{P}=\Vec{p}+\Vec{q}$, where $\Vec{p}$ and $\Vec{q}$ are the momenta of the individual quarks:

\begin{small}
\begin{eqnarray}
    \pi^{+}(t,\Vec{P}) &=& \sum_{\Vec{x},\Vec{y}}e^{-i(\Vec{p}\cdot\Vec{x}+\Vec{q}\cdot\Vec{y})}h(|\Vec{x}-\Vec{y}|)\Psi^{T}(\Vec{x},t)\frac{1}{2}(1-e^{in_p\pi}\sigma_2)\frac{i}{2}\gamma^5C\sigma_1(1+e^{in_q\pi}\sigma_2)\Psi(\Vec{y},t)
\label{eq:pion-1} \\
    \pi^{-}(t,\Vec{P}) &=& \sum_{\Vec{x},\Vec{y}}e^{-i(\Vec{p}\cdot\Vec{x}+\Vec{q}\cdot\Vec{y})}h(|\Vec{x}-\Vec{y}|)\overline{\Psi}(\Vec{x},t)\frac{1}{2}(1-e^{in_p\pi}\sigma_2)\frac{-i}{2}\gamma^5C\sigma_1(1+e^{in_q\pi}\sigma_2)\overline{\Psi}^{T}(\Vec{y},t) 
\label{eq:pion-2} \\
    \pi^{0}(t,\Vec{P}) &=& \sum_{\Vec{x},\Vec{y}}e^{-i(\Vec{p}\cdot\Vec{x}+\Vec{q}\cdot\Vec{y})}h(|\Vec{x}-\Vec{y}|)\overline{\Psi}(\Vec{x},t)\frac{1}{2}(1-e^{in_p\pi}\sigma_2)\frac{-i}
    {\sqrt{2}}\gamma^5\sigma_3(1+e^{in_q\pi}\sigma_2)\Psi(\Vec{y},t) 
\label{eq:pion-3} \\
    \sigma(t,\Vec{P}) &=& \sum_{\Vec{x},\Vec{y}}e^{-i(\Vec{p}\cdot\Vec{x}+\Vec{q}\cdot\Vec{y})}h(|\Vec{x}-\Vec{y}|)\overline{\Psi}(\Vec{x},t)\frac{1}{2}(1-e^{in_p\pi}\sigma_2)\frac{1}
    {\sqrt{2}}(1+e^{in_q\pi}\sigma_2)\Psi(\Vec{y},t)
\label{eq:pion-4}
\end{eqnarray}
\end{small}
where, using the notation of Ref.~\cite{Christ:2019sah}, $\Psi$ and $\overline{\Psi}$ are the quark and anti-quark isospin doublets defined as:
\begin{equation}
\Psi = \colvectwo{d}{C\bar u^T}  \quad \mbox{and} \quad \overline{\Psi} = (\bar d, u^TC).
\end{equation}
As explained in Ref.~\cite{Christ:2019sah} the $2\times2$ flavor projection matrix $(1+e^{in_q\pi}\sigma_2)$ ensures that the quark field $\sum_{\vec x}(1+e^{in_q\pi}\sigma_2)e^{-i\vec q\cdot \vec x}\Psi(x)$ transforms as an eigenstate under translations (including positions which translate through the boundaries) if the integer $n_q= Lq_i/\pi - 1/2$ for all three components $\{q_i\}_{1 \le i \le 3}$ of the momentum vector $\Vec q$. Here $C$ is the $4\times4$ charge conjugation matrix and $h(|\Vec{x}|)$ is the meson smearing function. In this work, we choose all the smearing functions to be the 1$s$ hydrogen wave function $h(x)=e^{-x/r}$, with a radius $r = 2$ for both the pion and sigma operators. This smearing function is introduced to increase the overlap between the pion and sigma interpolating operators and the lattice pion and $\pi\pi$ ground states while at the same time reducing the overlap of the $I=0$ $\pi\pi$ operator with the vacuum state. In earlier studies this smearing was found to give a two-fold reduction in statistical errors~\cite{zhang2015kaon}. \par

With the operators constructed above, we use the all-to-all (A2A) propagator technique~\cite{Foley:2005ac} to perform the measurements. The A2A technique divides the quark propagator into an exact low mode contribution which we can calculate using the Lanczos algorithm and a high mode contribution which can be accessed using stochastic approximation. In our calculation, we choose the number of low mode eigenvectors to be 900. For the high mode contribution, we perform spin, color, flavor and time dilution ({\it i.e.} we perform a separate inversion for each of the 24 colors, spins and flavors for each time slice).  We use the same spatial field of random numbers for these 24 inversions but a different such field for each time slice~\cite{zhang2015kaon}.  We choose the number of random hits to be 1 ({\it i.e.} we use only a single random field on each time slice) since increasing it does not reduce the uncertainty~\cite{zhang2015kaon}.

We will work with two groups of pion operators.  The first is labeled as $\pi(111)$ with 8 different operators. These operators create pions carrying momenta $(\pm \pi/L,\pm \pi/L,\pm \pi/L)$. The second group is labeled as $\pi(311)$ and contains 24 different operators.  For this group one of the momentum components is replaced by $\pm 3\pi/L$. \par

We then combine two of these single-pion interpolating operators to construct $\pi\pi(\Vec{p},\Vec{q})$ operators with momenta $\Vec{P}=\Vec{p}+\Vec{q}$, where now $\Vec{p}$ and $\Vec{q}$ are the momenta of the individual pions: 
\begin{equation}
    O^{\alpha,\beta}_{\pi\pi}(t,\Vec{p},\Vec{q}) = \pi^\alpha(t+4,\Vec{p}) \pi^\beta(t,\Vec{q})\,,
\end{equation}
where $\alpha$ and $\beta$ are isospin indices. As suggested by this equation, when we construct the $\pi\pi(\Vec{p},\Vec{q})$ operators, we separate the two single-pion operators in the time direction by 4 units.  This suppresses the statistical error from the disconnected diagrams (the V diagram below) by a factor of two in the $I=0$ channel~\cite{liu2012kaon}. For consistency, when we construct the $I=2$ $\pi\pi(\ldots)$ operators, we also separate the two pion operators by 4 units in the time direction.\par

\subsubsection{Momentum decomposition}

The cubic symmetry breaking mentioned in Section~\ref{sec:lattice_detail} manifests as differences in the overlap factors between interpolating operators and finite-volume states whose momenta are related by cubic rotations (the energies themselves are not affected). In order to obtain $\pi\pi$ interpolating operators that respect the cubic symmetry and that can therefore be related to the continuum $s$-wave states, it is vital that we control this symmetry breaking. In Ref.~\cite{Christ:2019sah} it was demonstrated that the cubic symmetry breaking in the pion states can be heavily suppressed by averaging over pairs of pion interpolating operators of the same total momenta but with different assignments of quark momenta. We apply this technique for the present work and extend it to include the sigma operator. The two quark and anti-quark momentum pairs for each pion momentum are listed in Appendix~\ref{sec:appendix_quark_momentum}. In Section~\ref{sec:systematic_error} we carefully analyze our data in order to account for any residual cubic symmetry breaking effects as a systematic error. \par

In evaluating the Wick contractions it is often convenient to utilize the $\gamma^5$-hermiticity of the quark propagator ${\cal G}$:
\begin{equation} 
\gamma^5 [{\cal G}(x,y) ]^\dagger \gamma^5 = {\cal G}(y,x)\,, 
\end{equation} 
where the dagger ($\dagger$) indicates the hermitian conjugate of the matrix in its spin, color and flavor indices, in order to exchange the source and sink for a particular quark propagator. It is worth mentioning here that $\gamma^5$-hermiticity is not an exact symmetry between the A2A approximations to the quark propagators used here because of the asymmetric treatment of the source and sink in the A2A approach.  A further implication of our use of $\gamma^5$-hermiticity to combine related contractions arises from the effective exchange of the $\bar q$ and $q$ operators appearing in a meson field when $\gamma^5$-hermiticity is used on both the propagator leaving $\bar q$ and that arriving at $q$.  By symmetrizing over the assignments of momenta to the $\bar q$ and $q$ factors in each meson field, we insure that this use of $\gamma^5$-hermiticity does not result in a different amplitude. This determines the final pion interpolating operator we use: for each pion momentum we average over a total of four quark and anti-quark momentum assignments. For the sigma operator, since it satisfies PBC and has zero momentum we average over the eight different quark momentum assignments that are listed in Appendix~\ref{sec:appendix_quark_momentum} to suppress cubic symmetry breaking. (Note: this symmetrical treatment of the quark and anti-quark components of the meson field implies that the contractions presented in Refs.~\cite{ zhang2015kaon} and \cite{Abbott:2020hxn} for the case of a local pion interpolating operator can be unambiguously extended to the case of a non-local meson field.) \par

\subsubsection{Total momentum}

We perform both a stationary-frame calculation where the total two-pion momentum is zero and moving-frame calculations for which the total momentum is non-zero.  In the stationary-frame calculation we include the scalar $\sigma$ operator for the $I=0$ channel and for both isospin channels two classes of bilinear pair ``$\pi\pi(\ldots)$'' operators: One class has both pions in the group $\pi(111)$ but with opposite momenta, which we label $\pi\pi(111,111)$. The second class is made up of pions in the group $\pi(311)$, again with opposite momenta and are labeled $\pi\pi(311,311)$. \par

In the moving frame calculation we can also construct a $\pi\pi(111,311)$ operator for which the constituent pion operators belong to the two different groups described above. For the present work we did not collect data using a sigma operator with non-zero momentum; however the analysis presented in the following sections suggests the inclusion of this operator may be beneficial in future work. In summary, we therefore have three different classes of operators in the moving-frame calculation for each isospin channel, as well as in the stationary frame $I=0$ channel and only two classes of operators in the stationary-frame, $I=2$ calculation. (Note: our notation distinguishing the two-pion interpolating operators does not specify the total momentum that they carry which must be determined from the context.)  \par

\subsubsection{Angular momentum}

\begin{table}
\begin{center}
\begin{tabular}{|c|c|c|c|c|c|c|}
\hline
 Total momentum & Symmetry group & Angular momentum & Representation \\
\hline
$(0,0,0)\frac{\pi}{L}$ & $O_h$ & $l=0$ & $A_1$ \\
  \hline
$(0,0,0)\frac{\pi}{L}$ & $O_h$ & $l=2$ & $T_2$ \\
  \hline  
$(2,0,0)\frac{\pi}{L}$ & $C_{4v}$ & $l=0$ & $A_1$ \\
  \hline
$(2,2,0)\frac{\pi}{L}$ & $C_{2v}$ & $l=0$ & $A_1$ \\
  \hline
$(2,2,2)\frac{\pi}{L}$ & $C_{3v}$ & $l=0$ & $A_1$ \\
  \hline  
\end{tabular}
\end{center}
\caption{Symmetry group and the corresponding representation we used in this work for each of the different total momenta.}
\label{table:sym_group_rep}
\end{table}

After identifying numerous $\pi\pi$ operators with different total momenta, the next step is to project those $\pi\pi$ operators onto angular momentum eigenstates. In this work we are interested in only the $s$-wave phase shift but we will also use $d$-wave states to estimate the size of cubic symmetry breaking in Section~\ref{sec:systematic_error}. The angular momentum $l$ indexes the irreducible representations of the infinite volume SO(3) Lie group, but the finite-volume lattice (assuming we have successfully overcome the cubic symmetry breaking) is symmetric under only a discrete subgroup of SO(3): either the cubic group for the stationary frame or a smaller, related group in the case of the moving frame for which relativistic length contraction alters the shape of the finite volume when viewed from the perspective of the center of mass frame. In order to generate angular momentum eigenstates on the lattice we must therefore establish a mapping from the irreducible representations $\Gamma$ of the discrete lattice symmetry group $G$ to those of SO(3), from which, given a desired value of $l$, we can determine an appropriate choice of irreducible representation of $G$ in which to construct our lattice operators. In general this mapping is one-to-many such that to each representation $\Gamma$ of $G$ there corresponds a set $S(G,\Gamma)$ of values of $l$ to which it corresponds in the SO(3) group. As such there are usually several representations which satisfy this condition, and we want to choose the one that is the simplest and which couples to the fewest other values of $l$, {\it i.e.} for which the set $S(G,\Gamma)$ is the smallest.  For example, we can always use the maximally symmetric representation ($A_1$) to obtain the $s$-wave phase shift. For $d$-wave states in the stationary frame, we can use the $T_2$ representation~\cite{atkins1970tables}. The discrete symmetry groups and  representations used when constructing the two-pion interpolating operator for our various choices of center-of-mass momenta are listed in Table~\ref{table:sym_group_rep}.\par

The second step is to construct an operator in the representation $\Gamma$ by combining the operators in one of the classes described above using the characters of $\Gamma$. The detailed procedure is as follows:
\begin{equation}
    O_{\pi\pi, \Gamma, i}^{\alpha\beta}(\Vec{P},t) = \sum_{\hat{T}\in G}\chi_\Gamma(\hat{T})O_{\pi\pi,i}^{\alpha\beta}(t+4,t,\frac{\Vec{P}}{2}+\hat{T}[\Vec{p}\,], \frac{\Vec{P}}{2}-\hat{T}[\Vec{p}\,])\,.
\label{equ:project_angular_mom}
\end{equation}

Here $\hat{T}[\Vec{p}\,]$ means we apply symmetry operation $\hat{T}$ on momentum $\Vec{p}$. We sum over all elements $\hat{T}$ of the finite-volume symmetry group G, $\Vec{P}$ is the total momentum, and $\chi(\hat{T})$ is the character of each group element $\hat{T}$ in the representation $\Gamma$. We choose $\Vec{p}$ so that all the $\pi\pi$ operators appearing in the sum belong to the $i^{\rm th}$ class. After projection, for each total momentum $\Vec{P}$, instead of having three or two classes of $\pi\pi$ operators, we will only have three or two $\pi\pi$ operators, each transforming under a specific representation of $G$ and constructed from the operators within that class. Henceforth we will use the labels $\pi\pi(111,111)$, $\pi\pi(111,311)$, $\pi\pi(311,311)$ to refer to those projected operators rather than to the classes from which they were constructed.  \par

In the moving-frame calculations reported here, due to the limited number of classes (two) of single pion operators, we are only able to focus on the three sets of non-zero total momenta $\Vec P$ with the smallest individual components: $\Vec{P} = (\pm2,0,0)\frac{\pi}{L}, (\pm2,\pm2,0)\frac{\pi}{L}$ and $(\pm2,\pm2,\pm2)\frac{\pi}{L}$, so that the number of different classes of $\pi\pi$ operators we construct on the lattice is more than one (three in this work).


\subsection{Matrix of two-point correlation functions}

We begin a discussion of the correlation functions using a single operator constrained to a single timeslice (recall our $\pi\pi(\ldots)$ operators have the pion bilinears on separate timeslices). For isospin $I$ the two-point $\pi\pi$ correlation function is determined by the Euclidean Green's function
\begin{equation}
    C^{I}(t_{snk},t_{src}) = \langle O^{I\dag}_{\pi\pi}(t_{snk})O^{I}_{\pi\pi}(t_{src})\rangle\,,
\label{eq:two-point}
\end{equation} 
where $\langle\ldots\rangle$ indicates the expectation value from a Euclidean-space Feynman path integral, performed in a finite spatial volume of side $L$ and time extent $T$, obeying periodic boundary conditions for the gauge field but anti-periodic boundary conditions for the fermion fields in the time direction and $G$-parity boundary conditions in the three spatial directions.  

Here and in our two earlier papers~\cite{Christ:2019sah, Abbott:2020hxn} the hermitian conjugate which appears on the left-hand operator in Green's functions such as shown in Eq.~\eqref{eq:two-point} requires some explanation.  For the case that the operator involves Euclidean fields evaluated at a single time, the hermitian conjugate represents a combination of path integral field variables which corresponds to the Hermitian conjugate of the indicated operator in the time-independent Schr\"odinger picture which is subsequently transformed to the time-dependent Heisenberg picture operator whose expectation values are described by a Euclidean path integral. For the case that the operator $O^I_{\pi\pi}$ is itself the product of two such operators evaluated at different times, each operator is to be interpreted in this fashion. In this case the two operators appearing in this pair are always symmetrized to insure that the resulting two-point functions are positive as this notation suggests in spite of the fact that their order is not exchanged by this prescription.

By inserting two complete sets of intermediate states, we can rewrite this two-point function as
\begin{equation}
\begin{aligned}
    C^{I}(t_{snk},t_{src}) &= \langle\pi|O^{I\dag}_{\pi\pi}|\pi\rangle\langle\pi|O^{I}_{\pi\pi}|\pi\rangle e^{-t E_{\pi,in}} e^{-(T-t) E_{\pi,out}} \\
    & +  \langle0|O^{I\dag}_{\pi\pi}|\pi\pi\rangle\langle\pi\pi|O^{I}_{\pi\pi}|0\rangle e^{-t E_{\pi\pi}} \\
    & +\langle\pi\pi|O^{I\dag}_{\pi\pi}|0\rangle\langle0|O^{I}_{\pi\pi}|\pi\pi\rangle e^{-(T-t) E_{\pi\pi}}\\
    & + \langle0|O^{I\dag}_{\pi\pi}|0\rangle\langle0|O^{I}_{\pi\pi}|0\rangle \delta_{I,0}
\end{aligned}
\label{equ:single_state_correlator}
\end{equation}
in the limit where both $t\equiv t_{snk}-t_{src}$ and $T-t$ are large so that we can neglect the contribution from excited intermediate states. Notice the first term describes the ``around-the-world effect'', which is exponentially suppressed in $T$. Here $E_{\pi,in}$ and $E_{\pi,out}$ are the energies of the pions propagating from the source along the positive and negative time directions, respectively. These two energies should be the same in a stationary frame calculation but they may be different for a moving frame. The second and third terms, which can be combined together into a cosh function of the time separation t, describe the ground state $\pi\pi$ scattering, one for the forward propagating $\pi\pi$ along the time direction and the other for the backward propagating case. The last term, which describes the contribution of the vacuum intermediate state, appears only in the $I=0$ channel and does not describe the physics of $\pi\pi$ scattering. This term is the largest source of statistical error because it is time-independent and therefore results in a decreasing signal-to-noise ratio as we increase the time separation $t$ to suppress excited state contamination. \par

In practice, due to the rapid reduction in signal-to-noise ratio and the finite temporal extent of the lattice it is necessary to include data in the region where $t$ or $T-t$ is not very large. By including data from smaller time separations our results will be affected by contamination from excited-states. One way to suppress these errors is to expand the sum over intermediate states in  Eq.~\eqref{equ:single_state_correlator} to include not only the ground state but also one or more excited states and then to fit using this more complicated expression. However, even if we only include one more state, performing such a multi-state fit may be difficult using a single interpolating operator since we are attempting to determine an increasing number of parameters purely from the time dependence of data with a rapidly falling single-to-noise. \par

While increasing statistics will ultimately allow the various states to be isolated, a far more powerful technique is to introduce additional interpolating operators which all share the same quantum numbers and therefore project onto the same set of states, albeit with different coefficients. While naively equivalent to increasing statistics, the additional operators actually introduce a wealth of new information that helps constrain the fit. This additional information can also be exploited more directly using the GEVP technique (described in more detail in Section~\ref{sec:pipi_energy}) whereby the energies of $N$ states can be obtained from Green's functions comprising $N$ operators using only three timeslices. A simpler method which allows for the detection of the presence of excited states using data from only a single timeslice by looking at the ``normalized determinant'' will be discussed in Section~\ref{sec:pipi_energy}.

In order to perform a stable fit where both ground and excited states are included, we introduce additional interpolating operators which all share the same quantum numbers so that the number of operators can be larger than or equal to the number of states included in the fit.  Thus, we consider the matrix of two-point correlation functions:
\begin{equation}
    C^{I}_{ij}(t_{snk},t_{src}) = \langle O^{I\dag}_{i}(t_{snk})O^{I}_{j}(t_{src})\rangle\,,
\label{eqn:correlator_matrix}
\end{equation}
where indices $i$ and $j$ distinguish the operators. We can then expand Eq.~\eqref{eqn:correlator_matrix} to include excited-state contributions:
\begin{equation}
\begin{aligned}
    C^{I}_{ij}(t_{snk},t_{src}) &= \langle\pi|O^{I\dag}_{i}|\pi\rangle\langle\pi|O^{I}_{j}|\pi\rangle e^{-t E_{\pi,in}} e^{-(T-t) E_{\pi,out}} \\
    & \hskip -3mm + \sum_{n=0}^{m} \left\{\langle0|O^{I\dag}_{i}|n\rangle\langle n|O^{I}_{j}|0\rangle e^{-t E_{n}} + \langle n|O^{I\dag}_{i}|0\rangle\langle0|O^{I}_{j}|n\rangle e^{-(T-t) E_{n}}\right\} \\ 
    & \hskip -3mm + \langle0|O^{I\dag}_{\pi\pi}|0\rangle\langle0|O^{I}_{\pi\pi}|0\rangle \delta_{I,0}\,.
\end{aligned}
\end{equation}
Now the excited state contamination error has been reduced since the lightest state that we neglect is the one with energy $E_{m+1}$, which is higher than $E_{1}$, the energy of the first state that we neglected in Eq.~\eqref{equ:single_state_correlator}. For simplicity in the discussion above we have identified a single time $t_{snk/src}$ that is associated with each two-pion operator.   However, these operators are constructed from two, single-pion operators evaluated at the times $t_{snk/src}+4$ and $t_{snk/src}$ as shown in Eq.~\eqref{equ:project_angular_mom}.  In the remainder of this paper we will use the variable $t = t_{snk}-t_{src} -4$ to describe the separation between the two operators which indicates a minimum distance of propagation needed to connect the two, two-pion operators.\par

Assuming that the fit is able to reliably obtain the parameters then clearly the larger number of states that are included in the fit, the smaller the resulting excited state contamination. However, given the added computational cost and resulting fit complexity, we should be careful to include only operators which help to distinguish the relevant excited states.  An important criterion, discussed later, is the degree to which the operators introduced overlap with the state being studied or a common set of excited states. \par

\subsection{Contraction diagrams}

We are interested in the scattering process for specific isospin channels. The $I=0$ and $I=2$ $\pi\pi$ state with $I_z = 2$ can be constructed from $\pi^{+}$, $\pi^{-}$, $\pi^{0}$ states as below:
\begin{equation}
    |I=2,I_z=2\rangle =|\pi^{+}\rangle |\pi^{+}\rangle
\end{equation}
\begin{equation}
    |I=0,I_z=0\rangle = \frac{1}{\sqrt{3}}\left\{|\pi^{+}\rangle |\pi^{-}\rangle - |\pi^{0}\rangle |\pi^{0}\rangle + |\pi^{-}\rangle |\pi^{+}\rangle\right\}\,.
\end{equation}

The matrix of two-point correlation functions for the $\pi\pi$ and $\sigma$ operators can be obtained from a linear combination of eight different diagrams, labeled as $C$, $D$, $R$, $V$, $C_{\sigma\pi\pi}$, $V_{\sigma\pi\pi}$, $C_{\sigma\sigma}$ and $V_{\sigma\sigma}$, each corresponding to a particular Wick contraction that is identified in Fig.~\ref{fig:diagram_definition}. Their definition in terms of quark propagator is given in  Appendix~\ref{sec:appendix_contraction}.  They can be combined to obtain the two-point correlation functions as follows:
\begin{equation}
\begin{aligned}
    \langle\pi\pi(t)\pi\pi(0)\rangle^{I=2} &= 2D-2C \\
    \langle\pi\pi(t)\pi\pi(0)\rangle^{I=0} &= 2D+C-6R+3V
\end{aligned}
\label{equ:pipi_contraction}
\end{equation}

\begin{equation}
\begin{aligned}
    \langle\sigma(t)\sigma(0)\rangle &= \frac{1}{2} V_{\sigma\sigma} - \frac{1}{2} C_{\sigma\sigma} \\
    \langle\sigma(t)\pi\pi(0)\rangle &= \frac{\sqrt{6}}{4}V_{\sigma\pi\pi} - \frac{\sqrt{6}}{2}C_{\sigma\pi\pi}\,.
\end{aligned}
\label{equ:pipi_sigma_contraction}
\end{equation}

\begin{figure}[tb]
\centering
\fbox{\includegraphics[width=0.2\textwidth]{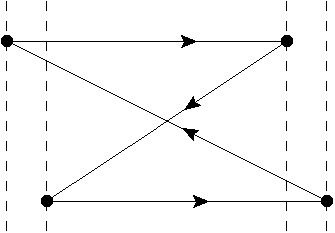}}
\fbox{\includegraphics[width=0.2\textwidth]{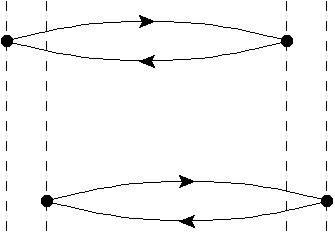}}
\fbox{\includegraphics[width=0.2\textwidth]{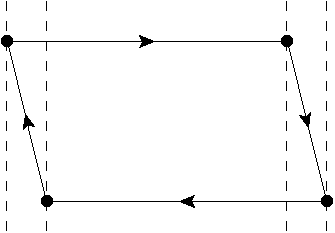}}
\fbox{\includegraphics[width=0.2\textwidth]{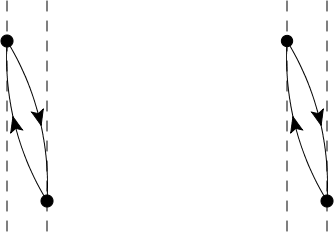}}\\
\fbox{\includegraphics[width=0.2\textwidth]{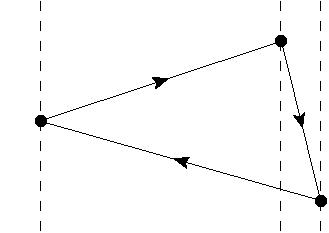}}
\fbox{\includegraphics[width=0.2\textwidth]{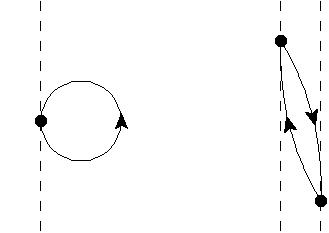}}
\fbox{\includegraphics[width=0.2\textwidth]{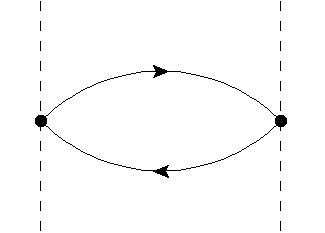}}
\fbox{\includegraphics[width=0.2\textwidth]{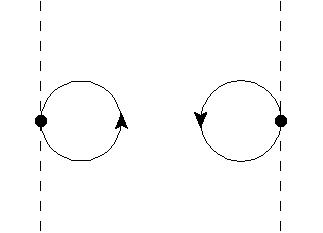}}
\caption{Diagrams showing the contractions which contribute to the two-point functions involving the $\pi\pi(\ldots)$ and $\sigma$ operators.  The solid dots indicate the positions of the pion two-quark operators and the dotted vertical lines passing through these points indicate the separate 3-dimensional time slices on which these operators are placed, with the nearby pairs of lines separated by four time units as described in Eq.~\eqref{equ:project_angular_mom}. Identical diagrams appear for the $\sigma$ operator only with a single vertical line at the source and/or sink, with the dots now representing the scalar bilinear. The top 4 diagrams are labeled by $C$, $D$, $R$ and $V$ diagrams from left to right, and the lower 4 diagrams are labeled by $C_{\sigma\pi\pi}$, $V_{\sigma\pi\pi}$, $C_{\sigma\sigma}$ and $V_{\sigma\sigma}$ from left to right.}
\label{fig:diagram_definition}
\end{figure}

If we were to perform the contractions for each of the different total momenta by substituting Eq.~\eqref{equ:project_angular_mom} into Eqs.~\eqref{equ:pipi_contraction} and \eqref{equ:pipi_sigma_contraction}, the number of different contractions to be evaluated for each gauge configuration would be 7848, which is unnecessarily large. The technique which we employ to reduce the number of momentum combinations takes advantage of three kinds of symmetry in $\pi\pi$ scattering: parity symmetry, which corresponds to changing each momentum from $\Vec{p}$ to $-\Vec{p}$\,, axis permutation symmetry, which permutes the three coordinate axes and an ``auxiliary-diagram'' symmetry, which relies on the combination of $\gamma_5$ hermiticity and the ``around-the-world'' contraction to show that two diagrams whose source and sink momenta satisfy a special relation are identical (For more details about the ``auxiliary-diagram'' symmetry, we refer readers to Ref.~\cite{Kelly:2019yxg}).  Using a subset of the gauge configurations in this study, we have found that excluding all but one of the momentum combinations that are related by these three symmetries does not increase the statistical error for the measured $\pi\pi$ energy.  This strategy substantially reduces the number of momentum combinations from 7848 to 1037~\cite{Kelly:2019yxg}. \par

\subsection{Estimating statistical errors and goodness of fit}

In this paper we use multi-state correlated fits to determine the energies of each state and the overlap amplitudes between the different states and operators. The fitting procedure is flexible, e.g. we can perform a fit where the number of operators and states are different and we can perform a ``frozen fit'' where some of the parameters are held fixed during the fit, which is useful in the excited-state error analysis.  An important benefit of our fitting procedure is our ability to calculate a $p$-value, which is a measure of how well our data matches with our theoretical expectation for the time dependence of the two-point function being analyzed. \par

However, the determination of statistical errors and the calculation of a $p$-value are not straightforward.  Not only are we performing a correlated fit where the covariance matrix is itself determined by the data and therefore has its own, often substantial uncertainties, but there are autocorrelations between configurations, since the sampling interval between neighboring configurations used in our analysis is comparable to or smaller than the autocorrelation time which separates truly independent samples.  While our number of samples, 741, is relatively large compared to many lattice calculations, if we group these samples into bins of two or four and thereby reduce the autocorrelations between these binned samples, the resulting decrease in the effective number of samples loses significant information about the fluctuations which is required for adequate control of the covariance matrix upon which our correlated fits are based.  \par

Fortunately, we have developed methods to solve both of these issues.  These methods are based on a combination of the jackknife and the non-overlapping blocked-bootstrap resampling techniques~\cite{Kelly:2019wfj}.  The bootstrap technique uses uncorrelated, non-overlapping blocks of data for its samples and gives statistical errors unaffected by the autocorrelation between our 741 samples.  However, the inner jackknife resampling introduced to calculate the covariance matrix for each outer bootstrap sample is applied to the unbinned data obtained as a union of all of the blocks in a given jackknife sample. In this paper the block size is chosen to be 8 to suppress the effects of autocorrelation.  Finally the distribution of bootstrap means about the mean for the entire sample, determines the proper $\chi^2$ distribution that can be used to correctly determine the $p$-value for the fit.  (Recall that the usual standard $\chi^2$ distribution is not accurate when $\chi^2$ is determined using an uncertain covariance matrix in the presence of autocorrelations.)  More details of this method can be found in Ref.~\cite{Kelly:2019wfj}.  \par


\section{Single pion energies and mass}
\label{sec:pion_2pt_function}
In order to determine the pion energy and mass, we calculate a two-point function using the neutral pion operator:
\begin{equation}
    C(\Vec{p},t_{snk},t_{src}) = \left\langle \pi^0(t_{snk},-\Vec{p})  \pi^0(t_{src},\Vec{p}) \right\rangle
\end{equation}
for all possible values of $t_{src}$ and $t_{snk}$ and then we average over $t_{src}$ while keeping $t = t_{snk} - t_{src}$ fixed. We have in total 32 different pion momenta, 8 from the $\pi(111)$ group of operators and the other 24 from the $\pi(311)$ group. Up to the effects of the cubic symmetry breaking induced by the boundary conditions, which are heavily suppressed by the procedure discussed in Section~\ref{sec:overview} and the residual effects shown to be negligible in Section~\ref{sec:systematic_error}, the two point functions within each group are related by cubic rotations hence we average the two-point functions within each group. This leaves us with two correlation functions, $C^i_{\pi}(t)$, where $i\in\{(111),(311)\}$ represents the momentum of the pion without specifying its direction. \par

\begin{figure}
\centering
\begin{minipage}{0.45\textwidth}
  \centering
  \includegraphics[width=1.0\linewidth]{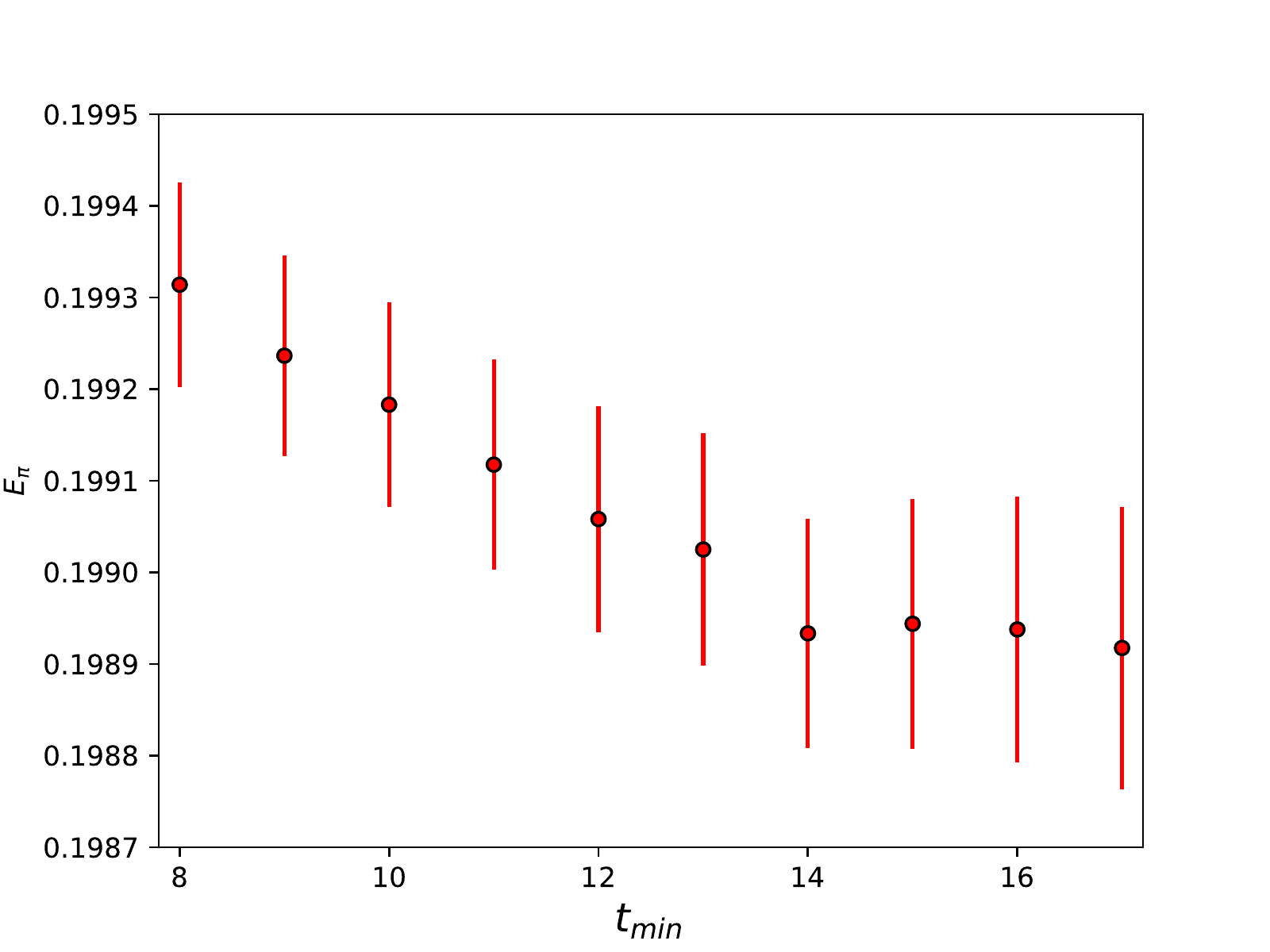}
\end{minipage}%
\begin{minipage}{0.45\textwidth}
  \centering
  \includegraphics[width=1.0\linewidth]{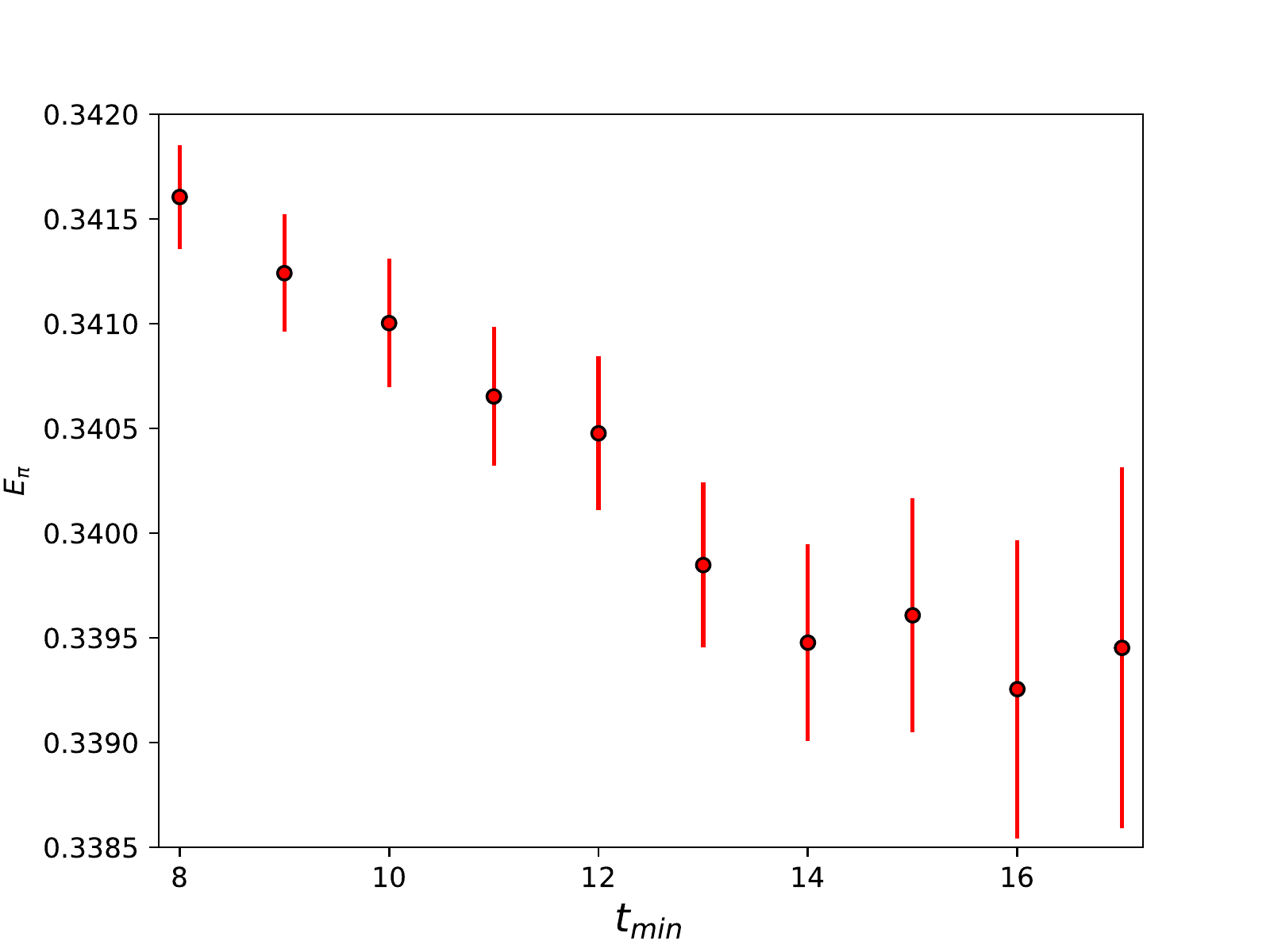}
\end{minipage}
\caption{The $t_{\mathrm{min}}$ dependence of the fitted energy $E_\pi$ for the $\pi(111)$(left) and $\pi(311)$(right) cases.  Here $E_\pi$ is shown in lattice units with $t_{\mathrm{max}}$ fixed to be 29.
\label{fig:single_pion}}
\end{figure}

We then perform correlated fits of each correlation function to the form
\begin{equation}
    C^i_{\pi}(t) = A^i_{\pi}(e^{-E^i_{\pi}t} + e^{-E^i_{\pi}(T - t)})
\end{equation}
using various fit ranges, all of which share the same upper limit $t_{\mathrm{max}} = 29$. Here $A^i_{\pi}$ is related to the normalization of the operator $O_\pi^i$ while $E^i_\pi$ is the energy of a moving pion state with momentum $(1,1,1)\frac{\pi}{L}$ or $(3,1,1)\frac{\pi}{L}$.  The fitted results for $E^i_\pi$ plotted as a function of $t_{\mathrm{min}}$ are shown in Figure~\ref{fig:single_pion}. From both plots we can see a clear plateau starting from $t_{\mathrm{min}}=14$. The result for $E_\pi$ is insensitive to $t_{\mathrm{max}}$ so we make the same choice of $t_{\mathrm{max}}=29$ as was made in Ref.~ \cite{Bai:2015nea}. For those reasons we choose the fit range to be $14-29$ and the fit results for that choice are listed in Table~\ref{table:pion}. The good $p$-values for both fits suggest that our data is well described by this single-state model. \par

Knowledge of the mass of the pion is required for the determination of the $\pi\pi$ phase shifts via the L\"uscher procedure. Unfortunately, with GPBC we are unable to measure this mass directly and must instead infer it from the energy of a moving state with a suitable choice of dispersion relation. In Table~\ref{table:pion} we give the results of applying the continuum dispersion relation to the (111) and (311) moving pion energies, which are labeled as $m_{\pi,\textrm{CD}}$. We can see that the resulting masses are inconsistent, which we interpret as the result of discretization effects on the dispersion relation. We also calculate the pion mass using the dispersion relation obeyed by a free particle on our discrete lattice
\begin{equation}
     \textrm{cosh}(E_{\pi}) = \textrm{cosh}(m_{\pi}) + \sum_{i=1}^3 (1-\textrm{cos}(p_i))\,,
    \label{equ:lat_disp}
\end{equation}
where the pion mass is identified as the energy of a pion with zero-momentum.  The results are listed in Table~\ref{table:pion} as $m_{\pi,\textrm{LD}}$ and are consistent between the two momenta. \par

The large discrepancy between the two pion masses calculated using different dispersion relations suggests that when we calculate the pion mass using the larger-momenta $\pi(311)$ operators the result has not only a statistical error that is 3 times larger than that from the $\pi(111)$ operators, but also a large systematic error. For the remainder of this paper, we will use $m_{\pi,\textrm{CD}} = 142.3(0.7)$  MeV calculated from the $\pi(111)$ operators using the continuum dispersion relation as the pion mass.  This $142.3(0.7)$  MeV value differs from the physical pion mass of 135 MeV by 7 MeV. This introduces an ``unphysical pion mass'' error into our results which will be discussed in Sections~\ref{sec:phase_shift} and \ref{sec:systematic_error}.  We will neglect the discretization error that remains in our determination of the pion mass since the 1 MeV discrepancy between the $m_{\pi,\textrm{CD}}$ and $m_{\pi,\textrm{LD}}$ in Table~\ref{table:pion} is small compared to the 7 MeV ``unphysical pion mass'' error identified above.\par

\begin{table}
\begin{center}
\begin{tabular}{|c|c|c|c|c|c|c|}
\hline
 State & Fit range & $A_{\pi}$ & $E_{\pi}$ & $p$-value & $m_{\pi,\textrm{CD}}$(MeV) & $m_{\pi,\textrm{LD}}$(MeV)\\
\hline
$\pi(111)$ & 14-29 & $6.194(11)\times10^6$ & 0.19893(13) & 0.99 & 142.3(0.7) & 143.3(0.7)\\
  \hline
$\pi(311)$ & 14-29 & $3.138(18)\times10^6$ & 0.33948(47) & 0.64 & 132.4(2.4) & 144.3(2.3)\\
  \hline
\end{tabular}
\end{center}
\caption{Results for the fitted energies for the pion states with momenta in the groups $(111)$ and $(311)$.  The right-most two columns show the pion masses calculated from those energies using the continuum ($m_{\pi,\textrm{CD}}$) and free-particle lattice ($m_{\pi,\textrm{LD}}$) dispersion relations.  We have converted to units of MeV by using the inverse lattice spacing for this ensemble, $1/a = 1.3784(68)$ GeV, where the error on $a$ also has been propagated into the errors on the energies given here.}
\label{table:pion}
\end{table}


\section{Finite-volume \texorpdfstring{$\pi\pi$}{pp} energies}
\label{sec:pipi_energy}
In this section we describe our multi-state, multi-operator fitting strategies and the resulting fit parameters for both the stationary frame and the moving frame calculations and for both the $I=0$ and $I=2$ channels.  Since these four situations are different, we will discuss them separately.  At the end of this section we briefly discuss results obtained from another data analysis technique, the GEVP.  This both provides alternative results for these quantities and an opportunity to compare these two methods.  Because the primary focus of this paper is on the properties of the $\pi\pi$ ground state, this discussion of the GEVP method is limited to the ground state energies which it determines. \par

\subsection{Stationary frame}
\subsubsection{\texorpdfstring{$I=2$}{I=2} Channel}

In the stationary $I=2$ channel, we have two classes of operators, $\pi\pi(111,111)$ and $\pi\pi(311,311)$. We project them onto the trivial $A_1$ representation  of the cubic symmetry group, which is the approximate symmetry group of a finite-volume lattice.   (A discussion of possible cubic symmetry breaking effects resulting from our G-parity boundary conditions will be presented in Section~\ref{sec:systematic_error}.)  This projection results in two different $\pi\pi$ operators, $O_a = \pi\pi_{A_1}(111,111)$ and $O_b = \pi\pi_{A_1}(311,311)$.  We then calculate the matrix of two-point functions constructed from these two operators by measuring
\begin{equation}
   C_{ij}(t_{snk}, t=t_{snk}-t_{src}-\Delta) = \langle O_i^\dagger(t_{snk})O_j(t_{src}) \rangle, 
\label{eq:delta-conventions}
\end{equation}
where $\Delta = 4$ is the time-separation between two pion fields used to construct each $\pi\pi$ operator. We average over all values of $t_{src}$ while fixing $t$ and then average the data at $t$ with that at $t = T - t - 2\Delta$ to improve the statistics.  (The individual single-pion operators at the times $t_{snk/src}+\Delta$ and $t_{snk/src}$ that make up each two-pion operator are constructed to be identical so when taking this second average we are combining equivalent physical quantities.)  We then try two different fitting strategies: \par

1) Fit the single two-point function $C_{aa}(t)$ assuming a single intermediate state and an around-the-world constant using the form
\begin{equation}
    C_{aa}(t) = A\left(e^{-E_{\pi\pi}t} + e^{-E_{\pi\pi}(T - t - 2\Delta)}\right) + B,
\label{eq:fit-form-single-op}
\end{equation}
where $A$ describes the normalization of the operator, $E_{\pi\pi}$ is the energy of the finite-volume $\pi\pi$ ground state and $B$ is the around-the-world constant. Thus, a total of three fit parameters are required.   We neglect all data related to the second operator $O_b$ so this is a one-operator, one-state fit. \par
2) Fit the upper triangular component of the $2\times 2$ matrix of two-point functions $C_{ab}$ using two intermediate states and three different around-the-world constants using the form
\begin{equation}
    C_{ij}(t) = \sum_{x=1}^2 A_{ix}  A_{jx} \left(e^{-E_xt} + e^{-E_x(T - t - 2\Delta)}\right) + B_{ij},
\label{eq:fit-form}
\end{equation}
where $A_{ix}$ is the overlap between the $i^{th}$ operator and the $x^{th}$ state; $E_x$ is the energy of the $x^{th}$ state and $B_{ij}$ is the around-the-world constant constructed from operators $O_i$ and $O_j$ for a total of 9 real fit parameters. Note that, as the lower triangular component of the matrix is related to the upper triangular component by the time-translational symmetry, we did not measure these terms in order to reduce the computational cost.

\begin{figure}
\centering
\begin{minipage}{0.45\textwidth}
  \centering
  \includegraphics[width=1.0\linewidth]{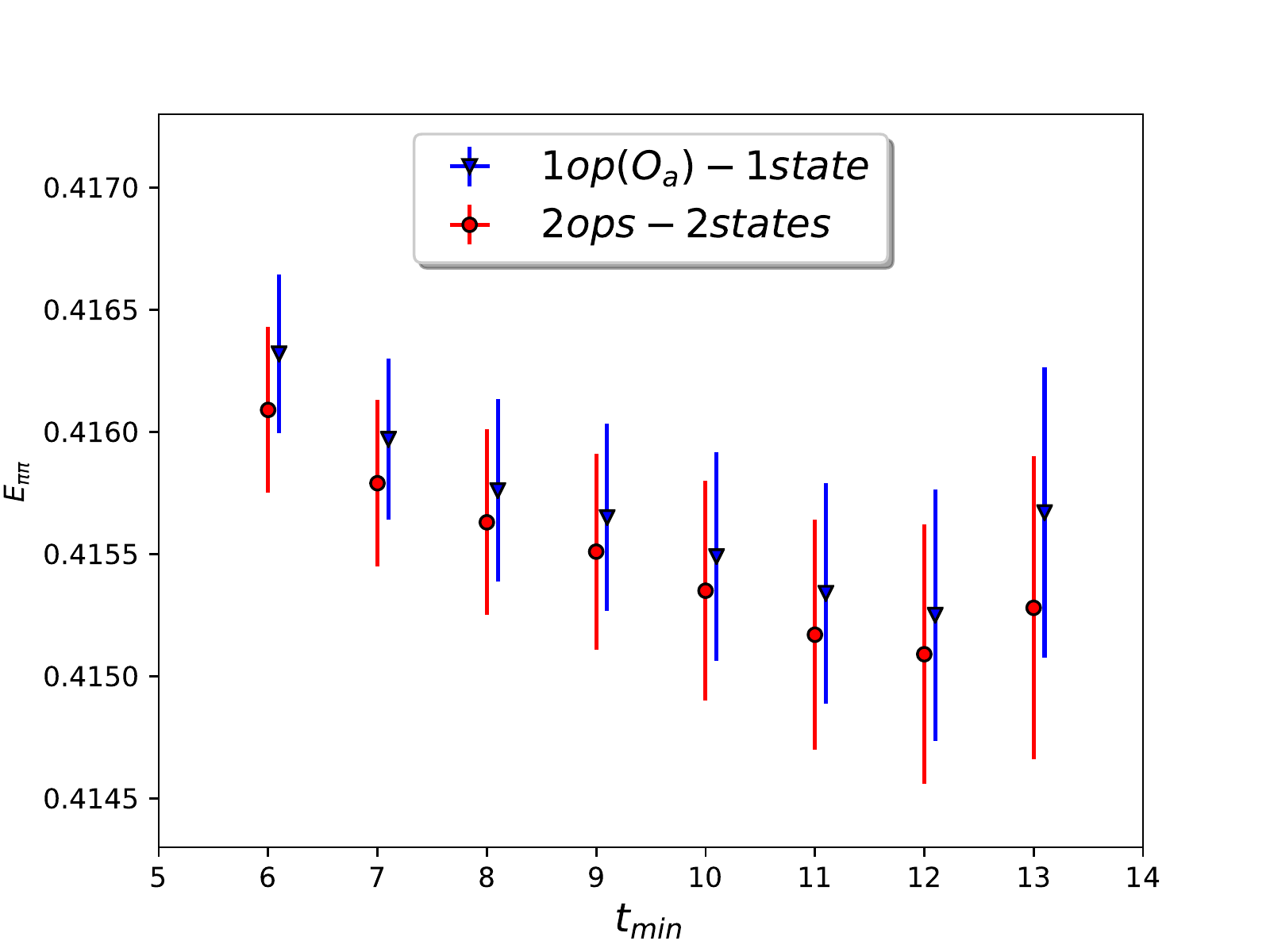}
  \label{fig:stationary_I2}
\end{minipage}%
\begin{minipage}{0.45\textwidth}
  \centering
  \includegraphics[width=1.0\linewidth]{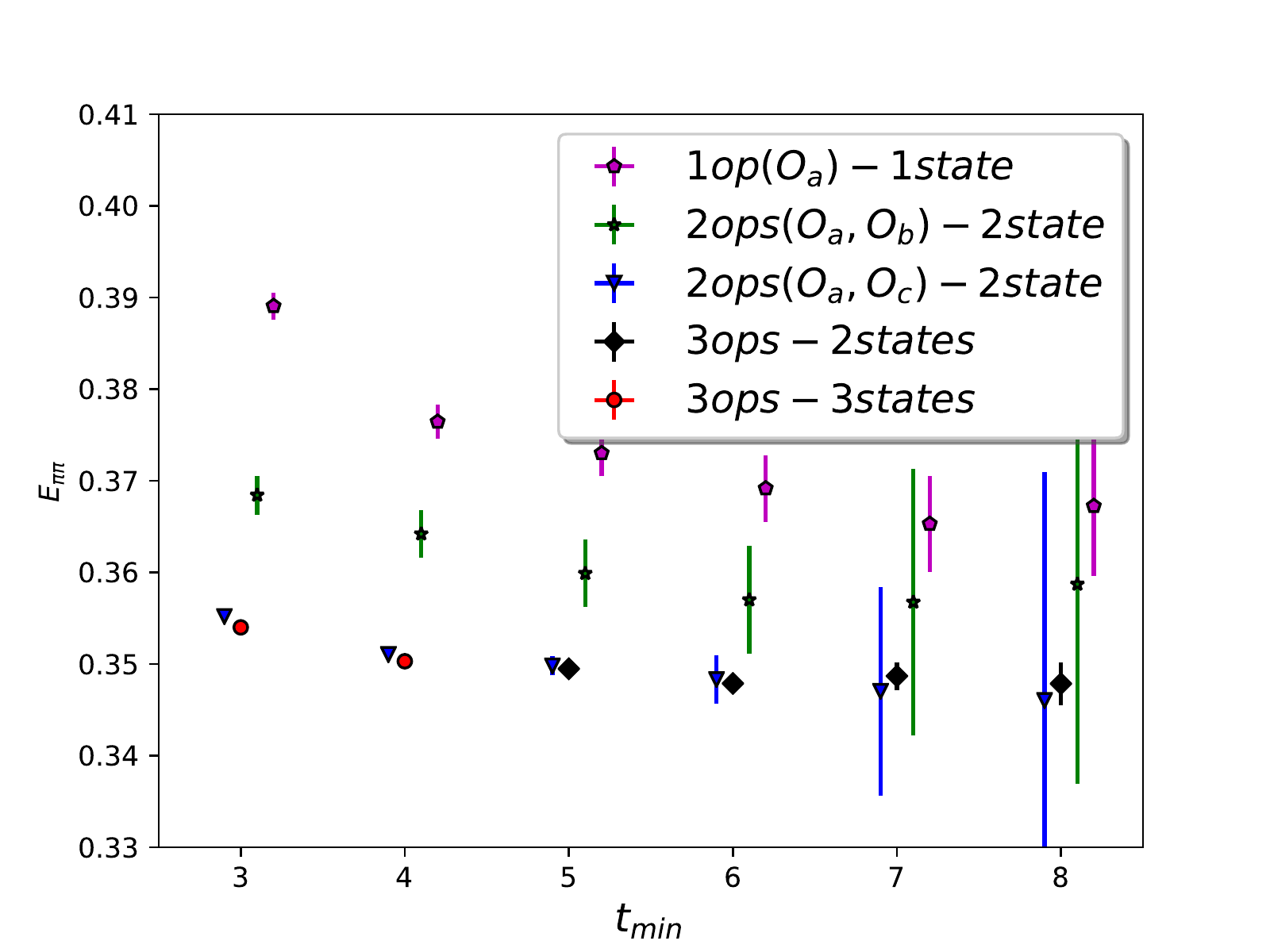}
  \label{fig:stationary_I0}
\end{minipage}
\vspace{-1cm}
\caption{The $t_{\mathrm{min}}$ dependence of fitted ground state energy for the stationary $\pi\pi_{I=2}$ channel (left) with $t_{\mathrm{max}}=25$ and the stationary $\pi\pi_{I=0}$ channel (right) with $t_{\mathrm{max}}=15$. Left: The circles represent the two-operator, two-state fit and downward pointing triangles the one-operator, one-state fit. Right: The pentagons represent the one-operator, one-state fit.  The stars and downward pointing triangles show the results from the two two-operator, two-state fits.  Finally the circles show the three-operator, three-state fit for $t_{\mathrm{min}}=3$ and 4 while the diamonds show the three-operator, two-state fit for $t_{\mathrm{min}}=5-8$.  For the $I=0$ channel, including additional operators (especially the $\sigma$) substantially improves the determination of the ground state energy.}
\label{Fig:stationary}
\end{figure}

For each case, we perform correlated fits with various choices for $t_{\mathrm{min}}$ and set $t_{\mathrm{max}}=25$. We plot the resulting ground state energy as a function of $t_{\mathrm{min}}$ in the left panel of Figure~\ref{Fig:stationary}. As we can see from the plot, the introduction of the second operator does not noticeably improve the fit result, as the ground state energies given by both fitting strategies are statistically consistent for all $t_{\mathrm{min}}$ and the statistical errors are also consistent.  As we increase $t_{\mathrm{min}}$, the ground state energy first decreases, which suggests a non-negligible excited state contamination for small $t_{\mathrm{min}}$ and then reaches a plateau for $t_{\mathrm{min}}\approx10$.  We adopt the 2-operator, 2-state fit with the fitting range of $10-25$ for our final result. In Table~\ref{tab:I=2} we list the $p$-values and the final parameters obtained from that approach. We observe an excellent $p$-value indicating a strong consistency between the data and our model.

The fact that $B_{aa}$ is $60\sigma$ resolved from zero suggests the importance of including these around-the-world constants in our fits.  This conclusion can also be reached by performing a similar fit in which the only change is that these constants are excluded.  These fits give $p$-values that are consistent with zero, suggesting that these constants are required. \par

We also observe that the matrix of overlap amplitudes $A_{ix}$ is nearly diagonal, where the operator $O_a$ predominantly couples to the ground state and the operator $O_b$ couples almost exclusively with the first excited state. The overlap factors between the operators and excited states is essential to exploiting the power of the multi-operator technique; without it one is merely performing several independent fits simultaneously. The fact that the amplitude matrix is near diagonal therefore likely explains the lack of improvement of the fit to the ground state energy when the second operator is introduced. The reason why this matrix is so diagonal can be intuitively explained by the weak strength of the $\pi\pi$ interaction potential in the $I=2$ channel as indicated by the small phase shifts. Such an interaction is required for the pions to exchange momentum and thus transform into other $\pi\pi$ states. \par

\begin{table}
\begin{center}
\begin{tabular}{ | c | c | c | c | c | }
  \hline
   $I=2$ channel& (2,2,2) & (2,2,0) & (2,0,0) & (0,0,0) \\
  \hline
  Fit range & 10-25 & 12-25 & 11-25 & 10-25\\
  \hline
  Fit strategy & 3op-3state & 3op-3state & 3op-3state & 2op-2state\\
  \hline
  $A_{a0}$ &  0.3941(6) &  0.2770(5) & 0.1933(3) &  0.4214(9) \\
  \hline
  $A_{a1}$ &  0.004684(565) &  0.007011(548) &  0.009301(455) & 0.012(10) \\
  \hline
  $A_{a2}$ &  0.001209(1890) &  0.005350(1812) &  0.005249(1482) & - \\ 
  \hline
  $A_{b0}$ & $-2665(31)\times 10^{-6}$ & $-4632(27)\times 10^{-6}$ & -0.007711(43) & -0.01164(10) \\
  \hline
  $A_{b1}$ &  0.08800(29) &  0.07457(39) &  0.07485(34) & 0.0696(60) \\
  \hline
  $A_{b2}$ &  0.003506(901) &  0.001437(1382) &  0.0050(13) & - \\ 
  \hline
  $A_{c0}$ & $-9626(124)\times 10^{-7}$ & $-1522(11)\times 10^{-6}$ & $-2327(14)\times 10^{-6}$ & - \\ 
  \hline
  $A_{c1}$ & $-3319(114)\times 10^{-6}$ & $-3914(162)\times 10^{-6}$ & $-4637(145)\times 10^{-6}$ & - \\ 
  \hline
  $A_{c2}$ &  0.04690(66) &  0.04592(111) &  0.03940(103) & - \\ 
  \hline
  $E_0$ & 0.3984(3) &  0.4001(3) &  0.4045(3) &  0.41535(45) \\
  \hline
  $E_1$ &  0.5453(7) &  0.5480(10) &  0.5514(9) &  0.713(17) \\
  \hline
  $E_2$ &  0.6902(28) &  0.6874(40) &  0.6916(48) & - \\ 
  \hline
  $B_{aa}$ & $8097(68)\times 10^{-9}$ & $4034(35)\times 10^{-9}$ & $1979(19)\times 10^{-9}$ & $940(16)\times 10^{-8}$\\
  \hline
  $B_{ab}$ & $-5748(3888)\times 10^{-12}$ & $-1388(169)\times 10^{-11}$ & $-1865(157)\times 10^{-11}$ & $134(350)\times 10^{-11}$\\
  \hline
  $B_{ac}$ & $-1025(178)\times 10^{-11}$ & $-8835(934)\times 10^{-12}$ & $-9792(986)\times 10^{-12}$ & -\\
  \hline
  $B_{bb}$ & $1136(154)\times 10^{-11}$ & $9200(1074)\times 10^{-12}$ & $9798(1111)\times 10^{-12}$ & $-101(16)\times 10^{-9}$\\
  \hline
  $B_{bc}$ & $-2642(4459)\times 10^{-13}$ & $206(3020)\times 10^{-13}$ & $-2617(3372)\times 10^{-13}$ & -\\
  \hline
  $B_{cc}$ & $2967(3749)\times 10^{-13}$ & $1084(2540)\times 10^{-13}$ & $2153(2980)\times 10^{-13}$ & -\\
  \hline
  $p$-value & 0.477  &  0.641 & 0.293 & 0.159 \\ 
  \hline
 \end{tabular}
\end{center}
\caption{Final fitting results for the $I=2$, $\pi\pi$ channel.  The right-most column lists the parameters obtained from a two-operator, two-state fit to the $\pi\pi$-$\pi\pi$ correlation function in the case of total momentum $(0,0,0)$ that is discussed in this subsection.  The next three columns from the right show the parameters obtained from three-operator, three-state fits for three non-zero values of the total momentum.}
\label{tab:I=2}
\end{table}

\subsubsection{\texorpdfstring{$I=0$}{I=0} Channel}
In the stationary $I=0$ channel, we have three classes of interpolating operator, two of which are constructed from two-pion interpolating operators and the other is the stationary $\sigma$ operator.  After projecting the $\pi\pi$ operators onto the $A_1$ representation, we obtain three different operators: $O_a = \pi\pi_{A_1}(111,111)$, $O_b = \pi\pi_{A_1}(311,311)$ and $O_c = \sigma$ and calculate the matrix of two-point functions
\begin{equation}
    C_{ij}(t_{snk},t=t_{snk}-t_{src}-\Delta_j) = \langle O_i^\dagger(t_{snk})O_j(t_{src}) \rangle - \langle 0 | O_i(t_{snk}) | 0\rangle\langle 0 |O_j(t_{src}) | 0\rangle,
\end{equation}
where the second term represents the vacuum subtraction which removes the disconnected piece in Eq.~\eqref{equ:single_state_correlator}, since it does not contribute to $\pi\pi$ scattering. We then average over all $t_{src}$ while fixing $t = t_{snk} - t_{src}$ and average the data at $t$ with that at $t = T - t - \Delta_i - \Delta_j$.   Here $\Delta_a=\Delta_b=4$ while $\Delta_c = 0$.  We then explore three different fitting strategies: \par

1) Fit $C_{aa}(t)$ using a single state and the equation
\begin{equation}
    C_{aa}(t) = A\left(e^{-E_{\pi\pi}t} + e^{-E_{\pi\pi}(T - t - 2\Delta_a)}\right),
\end{equation}
where $A$ and $E$ have the same physical meaning as in the stationary $I=2$ fit. This is a one-operator, one-state fit and we have only two fit parameters in total. In contrast with the stationary $I=2$ fit, here we neglect the around-the-world constant since an estimate of the size of the dominant contribution resulting from a single pion propagating through the temporal boundary gives a value which is approximately ten times smaller than the statistical error on these noisier $I=0$ channel data.  Note, if fit as a free parameter, the result for this around-the-world constant is consistent with zero and gives a ground-state energy consistent with the result obtained when this constant is excluded, but with a statistical error that is $50\%$ larger.  \par

2) Fit the upper triangular components of the $2\times 2$ submatrix spanned by $O_a$ and one of the other two operators using two states and the equation
\begin{equation}
    C_{ij}(t) = \sum_{x=1}^N A_{ix}A_{jx}\left(e^{-E_xt} + e^{-E_x(T - t - \Delta_i - \Delta_j)}\right),
\label{eq:multistate}
\end{equation}
where $N=2$, $A_{ix}$ is the overlap amplitude between the $i^{th}$ operator and the $x^{th}$ state, $E_x$ is the energy of the $x^{th}$ finite-volume state and $(i,j)$ takes values from either $\{a,b\}$ or $\{a,c\}$. Thus, this is a six-parameter fit.  An analysis similar to that mentioned in 1) above shows that the three around-the-world constants should be excluded. \par

3) Fit the upper triangular component of the entire $3\times3$ matrix of two-point functions using two or three states and the fitting form given in Eq.~\eqref{eq:multistate} where $N= 2$ or 3 is the number of states we include in the fit. We neglect the around-the-world constants for the same reasons as above, resulting in 12 (N=3) or 8 (N=2) fit parameters in total. \par

For each fitting strategy, we perform correlated fits with various values of $t_{\mathrm{min}}$ and set $t_{\mathrm{max}} = 15$.  We do not extend $t_{\mathrm{max}}$ to 25 as we did for the $I=2$ channel since the data for $t>15$ have larger statistical errors than in the $I=2$ case, so including them will not benefit our fit.  However, adding more fit points will destabilize the correlation matrix inversion procedure because of its increased dimension. We also risk introducing data for which the neglected around-the-world contribution may be a dominant component of the large-time data that has been introduced. This behavior is suggested because although the around-the-world constants remain statistically consistent with zero the $p$-value does fall as $t_{\mathrm{max}}$ is increased.  Note that we do not observe any corresponding statistically significant effects on the amplitudes and energies as $t_{\rm max}$ is increased suggesting that our fits remain robust even in the presence of around-the-world contributions.  A similar issue is encountered for the moving frame $I=0$ fits and is discussed in greater detail in Section~\ref{sec:pipi_energy_moving_atw}.  \par

We plot the ground state energies from these fits as a function of $t_{\mathrm{min}}$ in the right panel of Fig.~\ref{Fig:stationary}. For the three-operator case, we perform the three-state fit for $t_{\mathrm{min}}$ $\le 4$ while for $t_{\mathrm{min}} \ge 5$ we use the two-state fit as we observed that the three-state fits with $t_{\rm min}\geq 5$ were unstable and did not converge for many bootstrap samples, indicating that the third state can no longer be reliably resolved in the data. As we increase the number of operators, the ground state energy at fixed $t_{\mathrm{min}}$ becomes significantly lower and the plateau region becomes more clear and begins earlier.  We conclude that in contrast with the $I=2$ channel, the introduction of the two extra operators, especially the $\sigma$ interpolating operator, substantially reduces not only the statistical error but also the systematic error resulting from excited state contamination.  \par

Since the plateau region for the three-operator fit starts at $t=6$, we choose the three-operator, two-state fit with a fitting range of $6-15$ to determine our final results.  In the right-hand column of Table~\ref{tab:I=0} we list the $p$-value and final parameters for that fit. We can see that especially for the $\pi\pi(111,111)$ (a) and $\sigma$ (c) the overlap amplitudes between a given operator and the two states are of comparable size, which explains the effectiveness of the multiple operators that we included.  This large overlap factors between operators and states is consistent with the fact that the phase shift and hence the $\pi\pi$ interaction strength, is considerably larger than in the $I=2$ case.  Hence the exchange of momentum between the two pions required for the mixing between states is enhanced.  For $I=2$ the two $\pi\pi$ operators assign momenta with different magnitudes to the pions and would each couple to a different $\pi\pi$ energy eigenstate if the pions were non-interacting.

The fact that the overlap between operator $O_a$ and the first excited state is about half of the overlap of that operator with the ground state also provides a strong indication that there is likely to be non-negligible excited-state contamination in a single-operator, single-state fit.  This explains the substantial discrepancy between the phase shift at an energy near the kaon mass that we published in Ref.~\cite{Bai:2015nea} and both the results presented here and those from the earlier dispersive prediction~\cite{Colangelo:2001df}. This can also be seen in the right panel of Fig.~\ref{Fig:stationary}, where the single-operator fit reaches an apparent plateau at around $t=6$ or 7 with an energy that is consistent with our previously published value but which is substantially larger than the ground-state revealed by the introduction of the additional operators.\par

\begin{table}
\begin{center}
\begin{tabular}{ | c | c | c | c | c | }
  \hline
   $I=0$ channel& (2,2,2) & (2,2,0) & (2,0,0) & (0,0,0) \\
  \hline
  Fit range & 6-10 & 8-15 & 7-15 & 6-15\\
  \hline
  Fit strategy & 3op-3state & 3op-3state & 3op-3state & 3op-2state\\
  \hline
  $A_{a0}$ & 0.3873(7) &   0.2626(31) &  0.1772(26) &  0.3682(31) \\
  \hline
  $A_{a1}$ & -0.02647(391) & -0.05371(1262) &  -0.05431(776) & -0.1712(91) \\
  \hline
  $A_{a2}$ & -0.01354(312) & -0.03438(559) &  -0.02450(274) & - \\ 
  \hline
  $A_{b0}$ & $-1298(439)\times 10^{-6}$ & 0.002231(1392) & 0.005861(1306) & 0.0038(3) \\
  \hline
  $A_{b1}$ & 0.08361(100) & 0.06894(318) &  0.06781(261) & 0.0513(27) \\
  \hline
  $A_{b2}$ & -0.01121(395) &  -0.01277(940) & -0.02008(636) & - \\ 
  \hline
  $A_{c0}$ & $-8172(1223)\times 10^{-7}$ & $-1920(3981)\times 10^{-7}$ & $4871(4239)\times 10^{-7}$ & -0.000431(4) \\ 
  \hline
  $A_{c1}$ & 0.000837(1050) & 0.001713(2049) & 0.003439(1464) & -0.000314(17) \\ 
  \hline
  $A_{c2}$ & 0.04786(126) & 0.04602(456) &  0.03735(263) & - \\ 
  \hline
  $E_0$ & 0.3972(4) &  0.3895(17) &  0.3774(23) &  0.3479(11) \\
  \hline
  $E_1$ &  0.5264(37) &  0.5129(100) &  0.5032(75) &  0.569(13) \\
  \hline
  $E_2$ &  0.6881(93) &  0.6758(243) &  0.6514(183) & - \\ 
  \hline
  $p$-value & 0.094  & 0.016  &  0.635 & 0.314 \\ 
  \hline
 \end{tabular}
\end{center}
\caption{Table giving our final fitting results for $I=0$, $\pi\pi$ channel. The right-most column lists the parameters obtained from a three-operator, two-state fit to the $\pi\pi$-$\pi\pi$ correlation function in the case of total momentum $(0,0,0)$ that is discussed in this subsection.  The next three columns from the right show the parameters obtained from three-operator, three-state fits for three non-zero values of the total momentum.}
\label{tab:I=0}
\end{table}

\subsection{Moving frame}
\begin{figure}[tb]
\centering
\fbox{\includegraphics[width=0.45\textwidth]{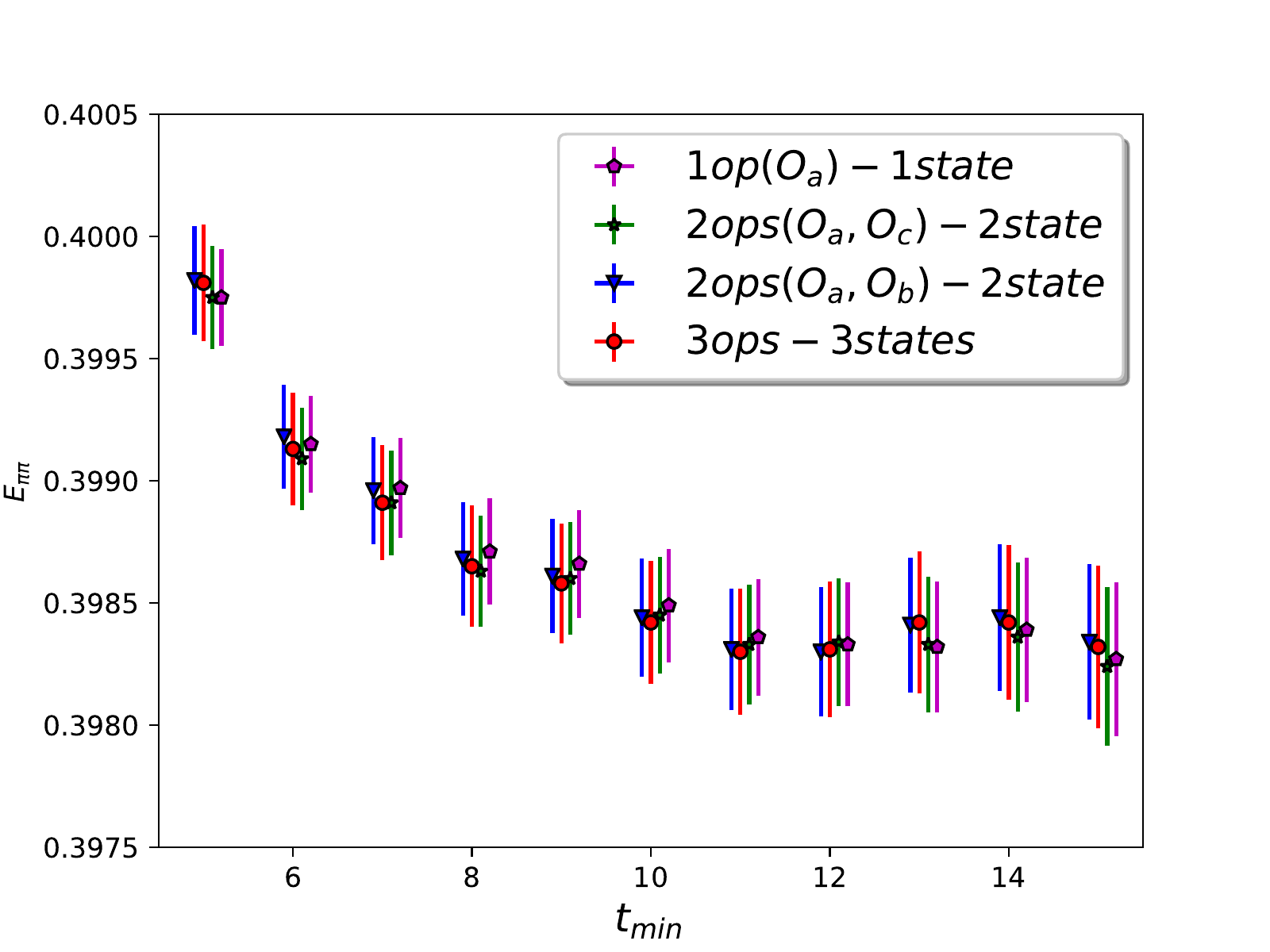}}
\fbox{\includegraphics[width=0.45\textwidth]{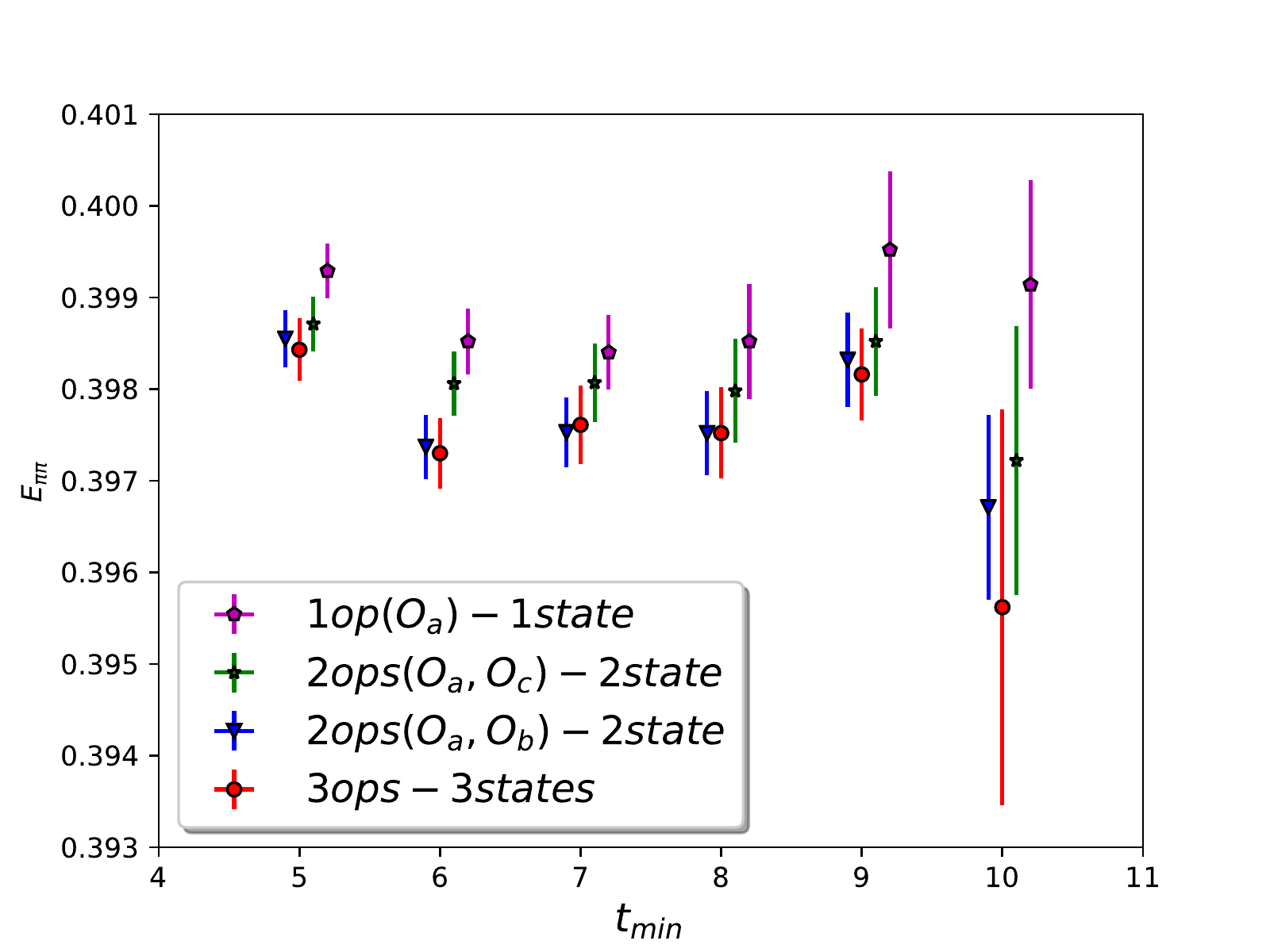}}\\
\fbox{\includegraphics[width=0.45\textwidth]{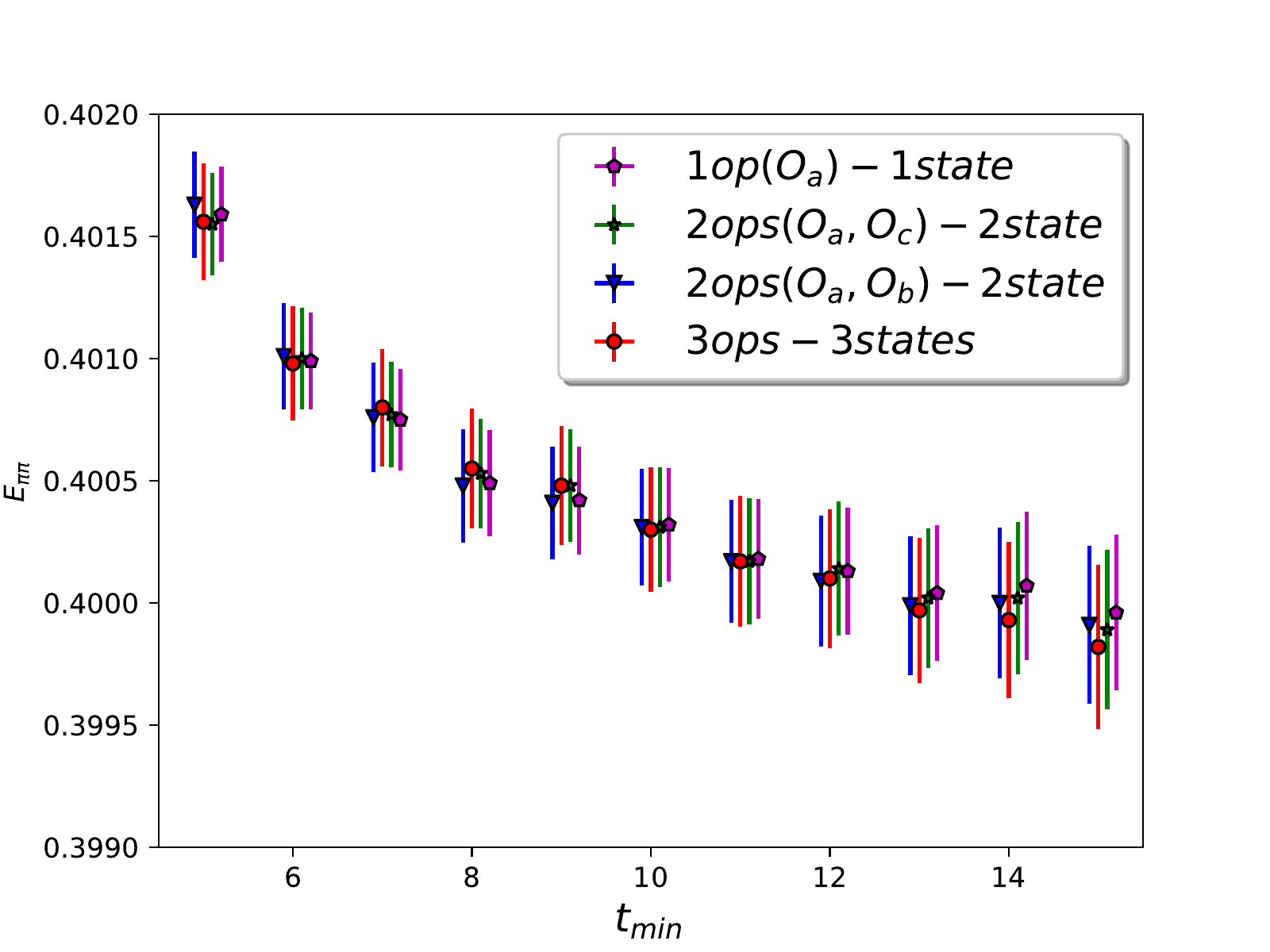}}
\fbox{\includegraphics[width=0.45\textwidth]{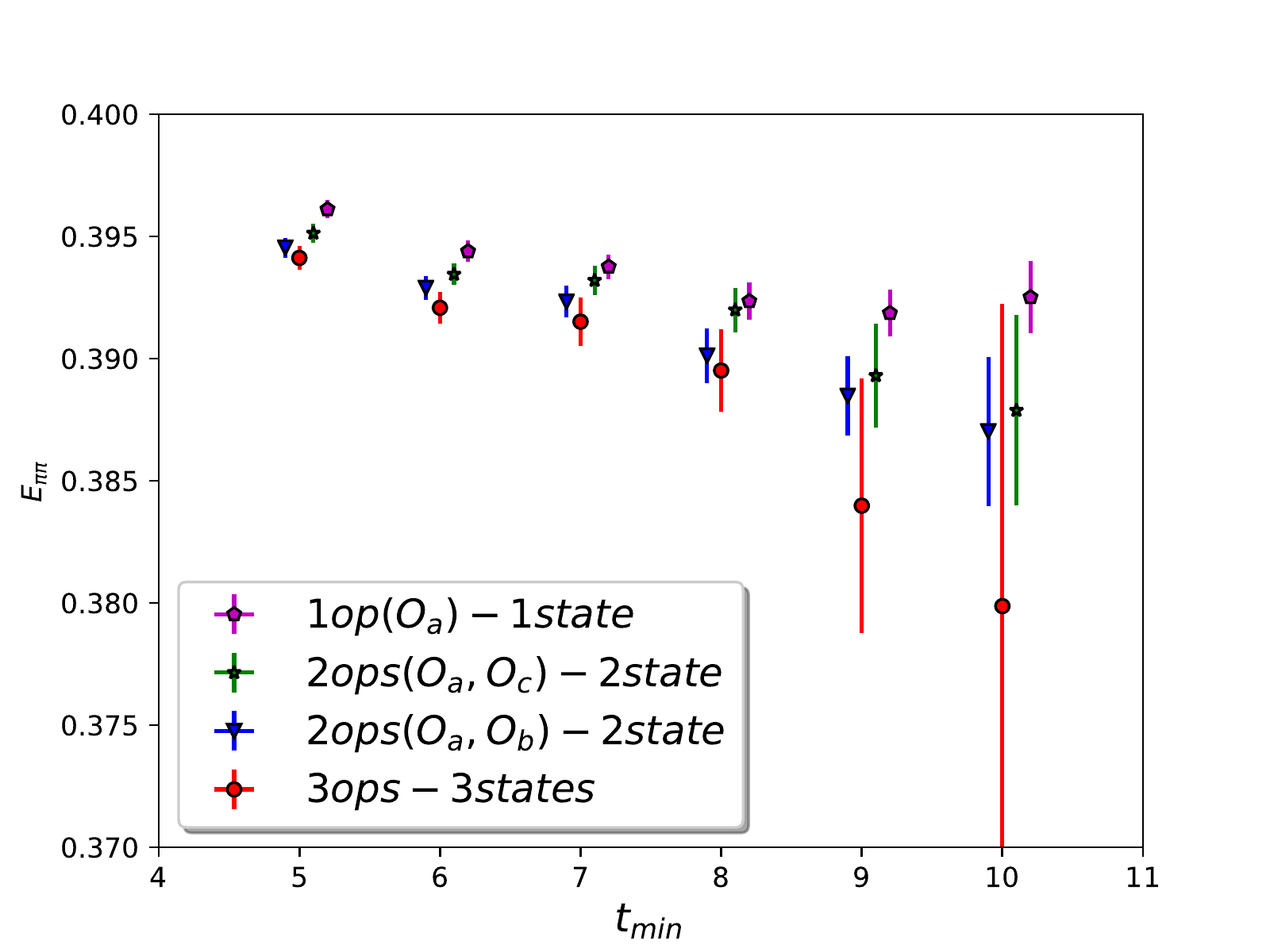}}\\
\fbox{\includegraphics[width=0.45\textwidth]{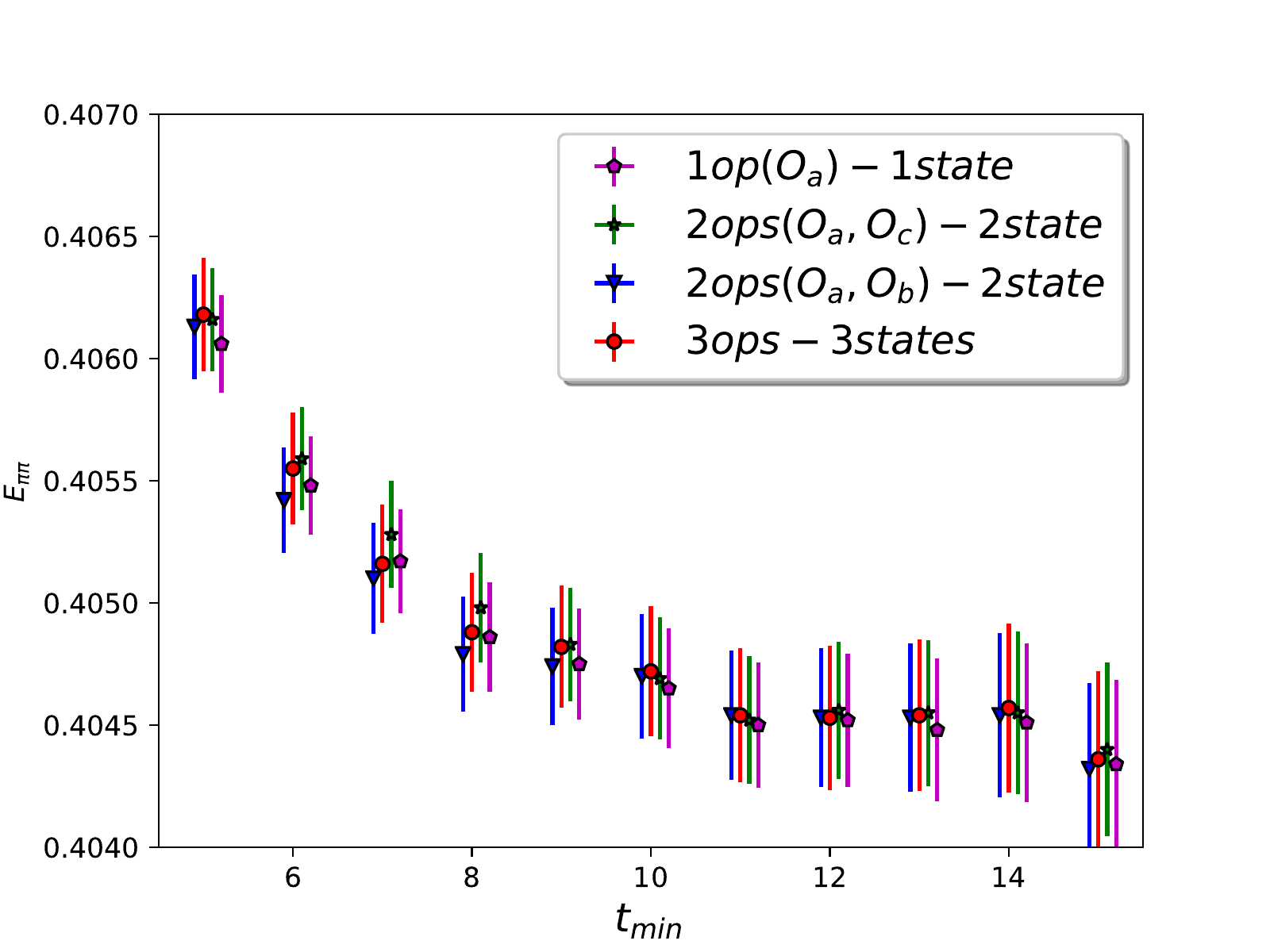}}
\fbox{\includegraphics[width=0.45\textwidth]{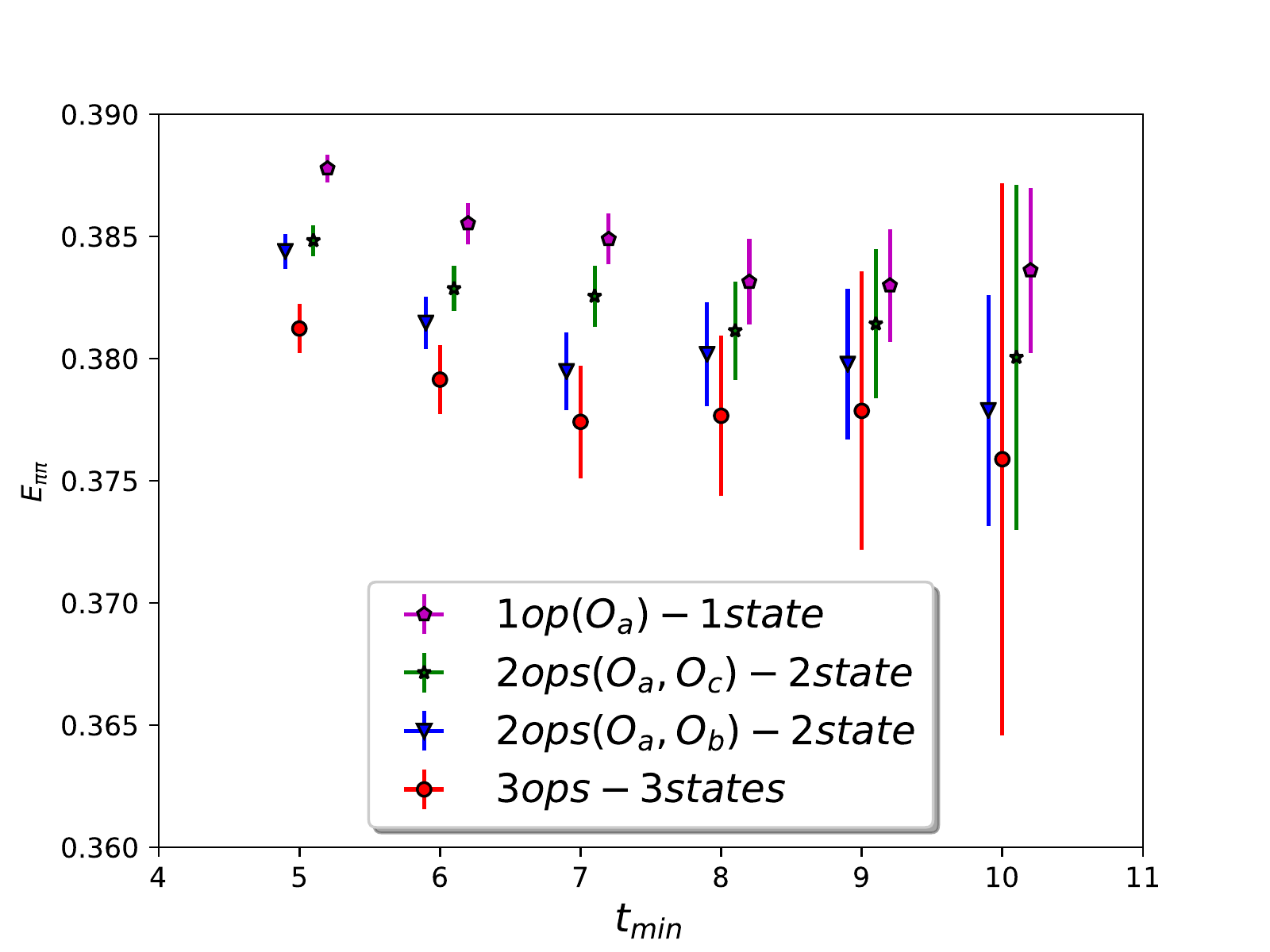}}
\caption{The $t_{\mathrm{min}}$ dependence of the fitted ground state energy for the moving $\pi\pi_{I=2}$ channel (left) and moving $\pi\pi_{I=0}$ channel (right) with $t_{\mathrm{max}}=25$ ($I=2$) and 15 ($I=0$). The upper, middle and lower panels are for total momenta $(2,2,2)\frac{\pi}{L}$, $(2,2,0)\frac{\pi}{L}$ and $(2,0,0)\frac{\pi}{L}$, respectively.  Our final results were obtained from three-operator, three-state fits. Reading from top to bottom the values of  $t_{\mathrm{min}}$ for these results were for $I=2$ $t_{\mathrm{min}}=10$, 12, 11 and  for $I=0$ $t_{\mathrm{min}}= 6$, 8 and 7.}
\label{Fig:moving}
\end{figure}

\subsubsection{\texorpdfstring{$I=2$}{} Channel}
\label{sec:pipi_energy_I2_moving}
In the moving $I=2$ channel, we have three classes of operators, $\pi\pi(111,111)$, $\pi\pi(111,311)$ and $\pi\pi(311,311)$. We project them onto the trivial representation of the little group of the cubic symmetry group which leaves the total momentum unchanged.  These little groups are $C_{4v}$ for $\Vec{P}_{tot} = (\pm2,0,0)\frac{\pi}{L}$, $C_{2v}$ for $\Vec{P}_{tot} = (\pm2,\pm2,0)\frac{\pi}{L}$ and $C_{3v}$ for $\Vec{P}_{tot} = (\pm2,\pm2,\pm2)\frac{\pi}{L}$.  For each choice of $\Vec{P}_{tot}$, this gives us three different operators, $O_a = \pi\pi_{A_1}(111,111)$, $O_b = \pi\pi_{A_1}(111,311)$ and $O_c = \pi\pi_{A_1}(311,311)$. We calculate the matrix of two-point functions constructed from these three operators, $C_{ij}(t_{snk}, t_{src})$ and combine the various values of $t_{src}$ and $t_{snk}$ in the same way as was done for the stationary $I=2$ calculation, except for an extra step where for each value of $|\Vec{P}_{tot}|$, we also average over all of the possible total momentum directions. This leaves us with three correlation matrices, one for each $|\Vec{P}_{tot}|$. We then try three different fitting strategies for each $|\Vec{P}_{tot}|$: \par

1) Fit $C_{aa}$ alone with a single state and an around-the-world constant, as we did in the stationary $I=2$ calculation. \par

2) Fit the upper triangular component of the $2\times 2$ submatrix spanned by $O_a$ and one of the other two operators using two states and three different around-the-world constants using the equation
\begin{equation}
    C_{ij}(t) = \sum_{x=1}^N A_{ix}A_{jx}\left(e^{-E_xt} + e^{-E_x(T - t - 2\Delta)}\right) + B_{ij},
\label{eq:moving-matrix}
\end{equation}
where the definitions of $A_{ix}$, $E_x$ and $B_{ij}$ are the same as the stationary frame, $N=2$ and $(i,j)$ takes value from either $\{a,b\}$ or $\{a,c\}$, giving nine fit parameters for either fit. \par

3) Fit the upper triangular component of the entire $3\times3$ matrix of two-point functions using three states, six around-the-world constants and Eq.~\eqref{eq:moving-matrix} with $N=3$.  In this case there are a total of 18 fit parameters. \par

For each value of $|\Vec{P}_{tot}|$ and fitting strategy, we perform correlated fits with $t_{\mathrm{max}}=25$, vary the value of $t_{\mathrm{min}}$ and plot the fitted ground state energy as a function of $t_{\mathrm{min}}$ in Figure~\ref{Fig:moving}, as in the stationary $I=2$ calculation. \par

Similar to the stationary $I=2$ calculation, for all three values of $|\Vec{P}_{tot}|$, the introduction of the two extra operators has little impact on the ground state energy. As we increase $t_{\mathrm{min}}$, the ground state energy first decreases, suggesting a non-negligible excited state error for small $t_{\mathrm{min}}$ and then reaches the plateau region. This plateau starts at $t_{\mathrm{min}}=11$ for $P_{tot} = (\pm2,0,0)\frac{\pi}{L}$, $t_{\mathrm{min}}=12$ for $P_{tot} = (\pm2,\pm2,0)\frac{\pi}{L}$ and $t_{\mathrm{min}}=10$ for $P_{tot} = (\pm2,\pm2,\pm2)\frac{\pi}{L}$. We choose the three-operator, three-state fit with tmax=25 and tmin fixed to the start of the plateau region identified above.\par

In Table~\ref{tab:I=2} we list the $p$-value and the final parameters for each choice of $P_{tot}$. With the chosen fit ranges we observe excellent $p$-values for all values of the total momentum. The fact that for each $P_{tot}$, $B_{aa}$ is $100\sigma$ resolved from zero suggests the importance of including these around-the-world constants in the fitting. The overlap matrices are all nearly diagonal as in the stationary $I=2$ calculation so that each operator is dominated by a different one of the three states.  Thus, as was the case for the stationary frame $I=2$ calculation, this explains why the introduction of these two additional operators does not improve the determination of the ground state energy. \par

\begin{figure}
\centering
\includegraphics[width=0.45\textwidth]{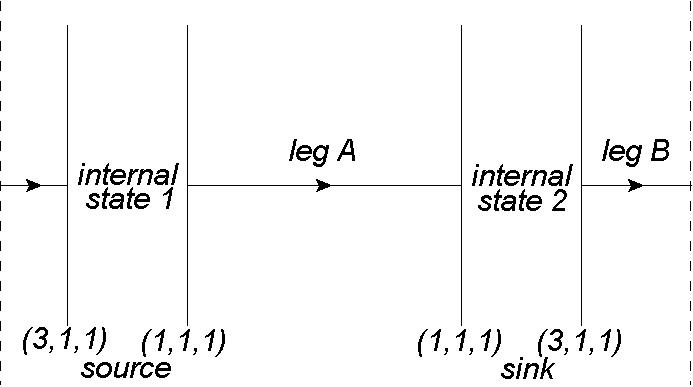}
\caption{A typical diagram for the decomposition of the ATW effect when the Green's function is constructed from two $O_b$ operators. For the first-order ATW effect, both legs are pions with momentum $(1,1,1)\pi/L$, which means if one of the internal states is the vacuum, e.g., internal state 1, then the other internal state cannot be the vacuum. For the second order ATW effect, we can choose leg A to be $(1,1,1)\pi/L$ and leg B to be $(3,1,1)\pi/L$, while keeping both internal states to be the vacuum state.}
\label{fig:pipi_2nd_atw}
\end{figure}

It is also worth mentioning that the constant terms we include in the fit only describe the lowest-order around-the-world (ATW) effect mentioned in Sec.~\ref{sec:lattice_detail_GPBC}, where both the pions on leg A (direct propagation between the two single-pion operators) and leg B (propagation through the temporal boundary) carry a minimum momenta with components $\pm \pi/L$.  Here we refer to segments of an around-the-world propagation path identified in Fig.~\ref{fig:pipi_2nd_atw}.  In contrast to the stationary case, the higher-order ATW terms in the moving frame need not be described by a constant term in the Green's function. For example, one of the pions on leg A or leg B could be replaced by a pion one of whose components has the larger $\pm3\pi/L$ value.  This possibility still conserves momentum and will show an exponential time dependence. 

When compared with the first-order ATW effect, this second-order ATW effect is exponentially suppressed by the energy difference between a pion with three $\pm\pi/L$ momentum components and a pion with one component increased to $\pm3\pi/L$.   However, in our calculation, due to the time separation $\Delta$ between the two single-pion operators that make up our $\pi\pi$ operator, this second-order effect can be enhanced in some cases. For example, we can look at the Green's function constructed from two $O_b$ operators. We define the state that propagates between the two temporally-separated pion operators in our $\pi\pi$ operator as the ``internal state". Notice we have two internal states here, since we have two $\pi\pi$ operators. For the first-order case, the two internal states cannot both be the vacuum while conserving momentum, but for the second-order effect they can.  This is illustrated in Figure~\ref{fig:pipi_2nd_atw}.  Thus, in this example the second-order effect is enhanced at least by a factor of $e^{E\Delta}=4.2$, where $E$ is the lowest energy of the internal state which we  approximate by the $I=0$ $\pi\pi$ energy.\par

In order to investigate the size of the higher-order ATW terms we perform a fit to the $I=2$ data. It can be easily shown that third- and higher-order ATW effects are always exponentially suppressed when compared with the first-order and second-order effects.  This means we can perform a fit which includes some extra parameters which represent the second-order ATW effect and neglect third- and higher-order effects. Here we fit the matrix of correlation functions with the following fit function:
\begin{equation}
    C_{ij}(t) = \sum_{x=1}^N A_{ix}A_{jx}\left(e^{-E_xt} + e^{-E_x(T - t - 2\Delta)}\right) + B_{ij} + D_{ij} \left( e^{-(E_1^{\pi}-E_0^{\pi})t} +  e^{-(E_1^{\pi}-E_0^{\pi})(T - t - 2\Delta)} \right)\,.
\end{equation}

Compared with Eq.~\eqref{eq:moving-matrix}, the extra term with coefficient $D_{ij}$ describes the second order ATW effect. Here $E_0^{\pi}$ and $E_1^{\pi}$ are the energies of moving pions with momenta $(1,1,1)\pi/L$ and $(3,1,1)\pi/L$, respectively. Their values can be obtained from Table~\ref{table:pion}. The fitting results for the ground state $\pi\pi$ energy and the sample-by-sample difference between the results with and without the second order ATW effect are shown in Table~\ref{tab:2nd_order_atw_I2}. Since the difference is negligible and statistically consistent with zero, we conclude that we need not include the second- or higher-order ATW effects in our fits. \par

\begin{table}
\begin{center}
\begin{tabular}{ | c | c | c | c |}
  \hline
   $I=2$ channel& (2,2,2) & (2,2,0) & (2,0,0)\\
  \hline
  Fit range & 10-25 & 12-25 & 11-25\\
  \hline
  $E_{ATW,2nd}$ & 0.3985(3)  &  0.4002(3) & 0.4045(3)  \\ 
  \hline
  $\delta E$ & $411(1167)\times 10^{-7}$  &  $867(1981)\times 10^{-7}$ & $-731(1700)\times 10^{-7}$ \\ 
  \hline
 \end{tabular}
\end{center}
\caption{Fit results used for estimating the size of the second-order ATW effect in the $I=2$ results given in Sec.~\ref{sec:pipi_energy_I2_moving}. Here $E_{ATW,2nd}$ is the ground state energy when the second order ATW term is included in the fit and $\delta E$ is the energy difference between $E_{ATW,2nd}$ and $E_0$ given in Table~\ref{tab:I=2}, which is significantly smaller that the statistical error on $E_0$ given in Table~\ref{tab:I=2}, in all three cases.}
\label{tab:2nd_order_atw_I2}
\end{table}

\subsubsection{\texorpdfstring{$I=0$}{} Channel}
\label{sec:pipi_energy_moving_atw}
As in the case of the moving $I=2$ channel, we have three classes of operators, defined as $O_a = \pi\pi_{A_1}(111,111)$, $O_b = \pi\pi_{A_1}(111,311)$ and $O_c = \pi\pi_{A_1}(311,311)$ which are projected onto the trivial representation of the corresponding little group.  We calculate the $3\times3$ matrix of two-point functions constructed from these operators for each of the three values of $|\Vec{P}_{tot}|$ in the same way as was done for the $I=2$ case.  We fit the data using three fitting strategies that are similar to the three used for the moving $I=2$ case, except that we exclude the around-the-world constants from all the fits. The effect of these constants will be discussed below. We then perform correlated fits with $t_{\mathrm{max}}=15$, vary $t_{\mathrm{min}}$ and plot the ground state energy as a function of $t_{\mathrm{min}}$ in Figure~\ref{Fig:moving}.  \par

Figure~\ref{Fig:moving} suggests that the introduction of the two extra operators does improve the fit result, since the ground state energy from the one-operator, one-state fit is always $2\sigma$ higher than its value from the three-operator, three-state fit, suggestive of remnant excited state contamination in the one state fit.  The consistency of the ground state energy between the two-operator ($O_a$, $O_b$), two-state fit and the three-operator, three-state fit in the plateau region suggests that operator $O_c$ may not be very useful. This is similar to the stationary $I=0$ calculation, where the operator constructed from the two $\pi(311)$ operators plays little role in controlling the excited state error.  \par

Another interesting feature is seen in the errors of the fitted parameters when we perform a single-operator, single-state fit using only the $\pi\pi(111,111)$ operator. Consider how the sizes of either the relative error of the amplitude, or the absolute error of the ground state energy change as we decrease the total momentum from $(2,2,2)\frac{2\pi}{L}$ to $(0,0,0)\frac{\pi}{L}$, when the fit range is fixed (e.g., $6-15$). The pattern is that these errors {\it increase} as the total momentum decreases, as can be seen in Table~\ref{tab:1124_effect}! This behavior conflicts with the expectation that these errors would be approximately the same based on the Lepage argument~\cite{Lepage:1989hd}. For our kinematics, the non-zero total momentum is created by reversing some of the momentum components of one of the pions.  Thus, if the modest $\pi\pi$ interactions are ignored, the four-pion states with zero total momentum which can contribute to the error will have approximately the same energy as the states which contribute to the signal. \par

\begin{table}
\begin{center}
\begin{tabular}{ | c | c | c | c | c |}
  \hline
   $P_{tot}$ & (2,2,2) & (2,2,0) & (2,0,0) & (0,0,0)\\
  \hline
  $E_0$ & 0.39852(36)  & 0.39439(44) & 0.38553(85) & 0.36917(364)  \\ 
  \hline
  $A_0$ & 0.15152(50)  &  0.07300(23) & 0.03454(15) & 0.01611(26) \\ 
  \hline
  $\frac{\delta A_0}{\bar{A_0}}$ & 0.0033  &  0.0032 & 0.0044 & 0.016  \\ 
  \hline
 \end{tabular}
\end{center}
\caption{Single operator ($O_a$) single state fit result with fit range $6-15$ for the $I=0$ channel. The absolute error of the ground state energy and relative error of the amplitude are approximately the same when $P_{tot} = (2,2,2)\frac{\pi}{L}$ and $(2,2,0)\frac{\pi}{L}$ and increase as we further decrease the total momentum. This effect can be partially understood by comparing the number of momentum matched D-type diagrams that dominate the central value of the Green's function to the total number of D-type diagrams.}
\label{tab:1124_effect}
\end{table}

This unexpected phenomenon can be understood by comparing the contributions to the central values of E0 and A0 and the corresponding errors obtained from the I=0 Green's functions..  From Eq.~\eqref{equ:pipi_contraction}, there are four types of diagram that contribute to the $I=0$ scattering. With $\sqrt{s} \le m_K$, the interaction between the pions is small, and the Green's function is dominated by the D-type diagrams when $t \le 10$ because in the non-interacting limit, the D-type diagrams represent products of two separate single-pion Green's functions. The V-type diagrams contain, in the stationary case, a vacuum contribution that is explicitly subtracted and for all four choices of $P_{tot}$ contributions in which gluons propagate between the disconnected components. The error on these diagrams does not decrease with increasing operator separation and becomes dominant when $t \ge 4$.  Given that the V-type diagrams are by far the dominant contribution to the error within our fit ranges, the size of the error on our fit results will depend primarily on the relative size of the V-diagram contribution to the overall Green's function, which, due to the dominance of the D-diagrams in the signal, is closely related to the relative size of the V and D-diagram contributions. Assuming that the errors on the amplitudes and the energies are uncorrelated, the pattern of these ratios as the total momentum varies (our four cases) should then be reflected in the errors on the fitted energies and amplitudes. \par

Notice that according to Eq.~\eqref{equ:project_angular_mom}, the $\pi\pi$ operator with definite momentum $\Vec{P}$ will contain  1, 2, 4 and 8 terms for the four cases above with total momentum containing three, two, one or zero non-zero components, respectively.  Since the number of two-point function contractions that we must evaluate grows like the product of the numbers of terms in its constituent operators, there will be 1, 4, 16 and 64 different contractions needed for each type with total momentum $(2,2,2)\frac{\pi}{L}$, $(2,2,0)\frac{\pi}{L}$, $(2,0,0)\frac{\pi}{L}$ and $(0,0,0)\frac{\pi}{L}$, respectively.  All these terms contribute V-type diagrams and hence to the error of the Green's function, with each of approximately the same size.  \par

However, not all of these terms contribute D-type diagrams (and hence to the central value of the Green's function), because of a mismatch between the momenta. This can be understood by looking at the non-interacting limit in which a D-type diagram will only be non-zero when the two independent single-pion Green's functions are non-zero.  This happens only when each of the two pions in the source $\pi\pi$ operator has the opposite momentum to that carried by one of the pions in the sink $\pi\pi$ operator, which we call momentum matching.  Counting these terms gives the numbers of momentum-matched D-type diagrams for these four possible total momenta: 2, 4, 8 and 16, respectively. After dropping a common factor of two in this counting of D-type diagrams, these estimates suggest that the proportions between the relative errors for these four total momenta become $\frac{1}{1}:\frac{4}{2}:\frac{16}{4}:\frac{64}{8} = 1:2:4:8$.  For a cosh fit, if the error of the amplitude and the error of the energy are uncorrelated, we can see that the relative error of the amplitude and the absolute error of the energy should be proportional to the relative error of the Green's function, which partially explains how the size of the errors on the ground state energy and amplitudes shown in Table~\ref{tab:I=0} changes as we decrease the total momentum.  A similar analysis can be applied to the $I=2$ channel, which suggests that these relative errors should be approximately the same, independent of the total momentum, which is consistent with what is shown in Table~\ref{tab:I=2}.\par

Similar to the moving $I=2$ channel, as we increase $t_{\mathrm{min}}$, the ground state energy first decreases, which suggests a non-negligible excited-state contamination for small $t_{\mathrm{min}}$.  The ground state energy then reaches a plateau region which starts with $t_{\mathrm{min}}=7$ for $P_{tot} = (\pm2,0,0)\frac{\pi}{L}$, $t_{\mathrm{min}}=8$ for $P_{tot} = (\pm2,\pm2,0)\frac{\pi}{L}$ and $t_{\mathrm{min}}=6$ for $P_{tot} = (\pm2,\pm2,\pm2)\frac{\pi}{L}$. For our final result we choose the three-operator, three-state fit with $t_{\mathrm{min}}$ equal to the beginning of the plateau region identified above. The choice of $t_{\mathrm{max}}$ is more subtle and will be discussed together with the effect of the neglected around-the-world constants below.  In Table~\ref{tab:I=0} we list for each $P_{tot}$ the fit range, fit procedure, as well as the resulting $p$-value and final parameters. We also observe in this table a trend towards smaller overlap factor between the operators and states, {\it i.e.}~a more diagonal amplitude matrix $A_{ix}$, as we increase the total momentum and thus decrease the center-of-mass energy. This is again consistent with our understanding of the relation between this overlap factor and the strength of the $\pi\pi$ interaction, which also decreases as the center-of-mass energy is decreased. This also explains why the additional operators appear to have the largest impact on the ground-state energy for the moving frame, $P_{\rm tot}=(2,0,0)\pi/L$ fits in Figure~\ref{Fig:moving}. \par

Next we discuss our treatment of the around-the-world constants in the fit. There are two potential sources of systematic error in our results that must be treated carefully: the excited state contamination and the around-the-world contributions.  The first error is expected to be much more significant and is discussed in Section~\ref{sec:systematic_error}.  To leading exponential order in the time extent of the lattice volume, the around-the-world contributions are time-independent constants even in this moving frame calculation because of our G-parity boundary conditions.  We observe that fitting with these around-the-world constants as free parameters results in good $p$-values for $t_{\rm min}\geq 6$ but gives results for the constants that are either statistically consistent with 0 ($P_{tot} = (2,2,2)$ and $(2,0,0)$), or which have an unphysical, negative sign ($P_{tot} = (2,2,0)$).  For all three cases either their errors when the constants are unresolved or the non-zero fitted values when these constants can be resolved are ten times larger than the expected size, that of the I=2 around-the-world constant.

For the case of $P_{tot} = (2,2,2)$ or $(2,0,0)$, we can neglect these constants in the fit since there is no statistical inconsistency between the fitted energy with and without these constants and we expect that the effects of the true around-the-world constants will be approximately ten times smaller than these sub-statistical effects.  Note that excluding theses constants from the fit gives us an improvement in the statistical error of the ground state energy by a factor of $1.2-1.5$. For the second  $P_{tot} = (2,2,0)$ case, the most likely explanation is that the constants are acting as ``nuisance parameters'' that help to partially account for the excited state contamination but do not reflect true around-the-world behavior. Rather than leaving the constants as free parameters and using an unphysical model to describe our data we choose to fix the constants to zero and to account for the systematic, excited-state contamination errors separately. \par 

The model with zero around-the-world contributions should be a good description of the data in the window $[t_{\rm min}, t_{\rm max}]$ for which $t_{\rm min}$ is large enough that excited state effects are small and $t_{\rm max}$ small enough that the contribution of these constants is small relative to the size of the data. For $t_{\rm max}=15$ we observe very poor $p$-values even for large $t_{\rm min}\leq 10$.  Reducing $t_{\rm max}$ from 15 to 10 we observe a significant improvement in the goodness-of-fit, finding acceptable $p$-values for $t_{\rm min}\geq 6$. This behavior is consistent with the effects of around-the-world contributions, although the excited state contributions may also play a role. Note however that, despite the dramatic improvement in $p$-value observed when reducing $t_{\rm max}$ from 15 to 10, we observe consistency in the ground-state fit parameters and no loss of precision, suggesting that the around-the-world systematic error is negligible and that the fits are under good control.  In Figure~\ref{Fig:moving} we use $t_{\mathrm{max}} = 15$ to show that reasonable behavior is seen when the around-the-world constants are omitted even for this large value of $t_{\mathrm{max}}$.  Further evidence that supports the argument that for $P_{tot} = (2,2,0)$ these constants are ``nuisance parameters'' can be found by including them in the fit, fixing $t_{\mathrm{max}}$ and increasing $t_{\mathrm{min}}$.  The resulting around-the-world constants monotonically decrease with increasing $t_{\mathrm{min}}$, which suggests that they likely result from excited state contamination, which is expected to decrease as $t_{\mathrm{min}}$ is increased, rather than representing the effects of single-pion around-the-world propagation. For uniformity, in Table~\ref{tab:I=0} we choose to list the results for the three smallest total momenta with the same value of $t_{\mathrm{max}}=15$.  This results in the small $p$-value of 0.016 for the $(2,2,0)$ case.  However, had we used $t_{\mathrm{max}}=12$ we would have obtained equivalent results with a $p$-value of 0.205. \par

\subsection{Normalized determinant}
It is important to emphasize that the introduction of these additional $\pi\pi(\ldots)$ operators (in all cases) and $\sigma$ operators (in the stationary $I=0$ case) offers something more than a simple statistical improvement but gives new information about the underlying energy eigenstates.  The two-point Green's functions $C_{xy}$ for $x\ne a$ and/or $y\ne a$ typically have larger statistical uncertainties than $C_{aa}$ at the same $t$, suggesting that including these additional operators may lead to only a small reduction in the statistical errors of the fitting parameters. However, in some cases including these operators significantly improves the statistical error of the ground state energy, ({\it e.g.} the stationary $I=0$ case shown in Figure~\ref{Fig:stationary}). We also observe in several cases a significant reduction in the energy of the apparent plateau as well as an earlier onset of the plateau region, suggesting that the extra operators are dramatically improving our ability to resolve nearby excited states which may be very difficult to distinguish from the ground state when we have only a single operator, even with large statistics. \par

Some insight into how this improvement comes about can be gained by considering the ``normalized determinant'' of the $N\times N$ matrix of Green's functions, $\mathcal{N}(t)$, defined as
\begin{equation}
    \mathcal{N}(t) = \frac{\textrm{Det}(C(t))}{\prod_{i=1}^N C_{ii}(t)}\,,
\label{eq:Norm_Det}
\end{equation}
where $C(t)$ is the matrix of Green's functions. We normalize the determinant using the product of the diagonal elements of the matrix so that this quantity does not depend on the scale of the interpolating operators. In fact, it can be shown that $0 \le | \mathcal{N}(t)| \le 1$.  If the number of intermediate states that contribute to $C(t)$, $N'$, is smaller than $N$ then $\mathcal{N}(t)=0$ (since the $N'$,  $N$-component vectors constructed from the matrix elements of the $N$ operators between these $N'$ states and the vacuum, which determine the $N\times N$ matrix $C_{ij}$ do not span the entire $N$ dimensional space on which $C_{ij}$ acts).  Thus, if at a given time $t$ we find $\mathcal{N}(t)\ne0$, then we can be certain that at least $N$ distinct states are contributing to $C(t)$. However, if we find $\mathcal{N}(t)=0$ we cannot conclude that there are fewer than $N$ states in the Hilbert space that contribute to the correlation matrix $C(t)$.  Thus, we cannot use $\mathcal{N}(t)$ to tell us if a sufficient number of operators has been used to distinguish all of the states that contribute to $C(t)$. When $\mathcal{N}(t) \approx 1$, it suggests that these operators create states from the vacuum which are orthogonal to each other.

\begin{figure}
\centering
\includegraphics[width=0.8\linewidth]{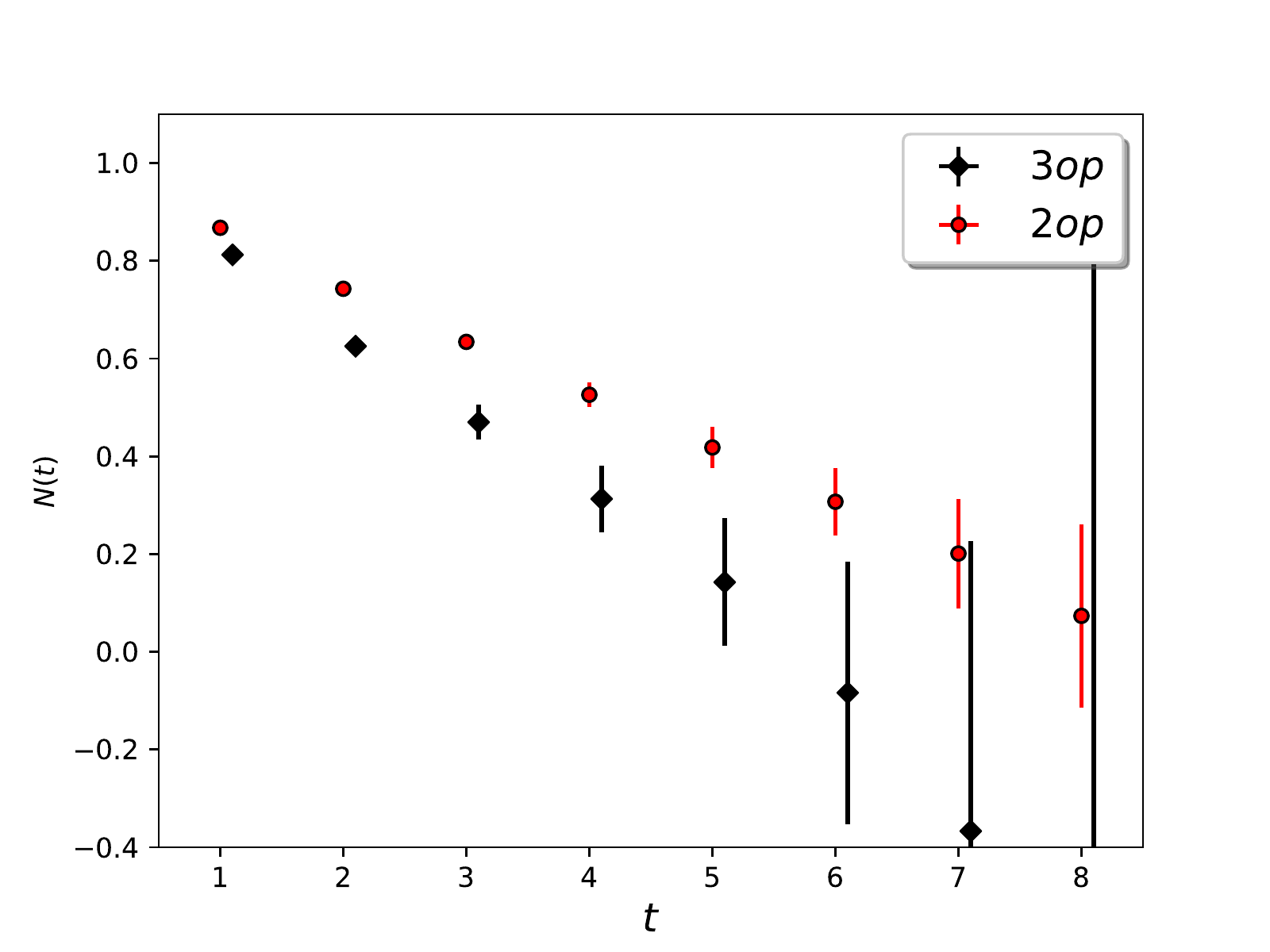}
\caption{The t dependence of $\mathcal{N}(t)$, defined in Eq.~\eqref{eq:Norm_Det} for the three dimensional (black, which include all operators) and two dimensional (red, only include $O_a$ and $O_c$ operator) matrix of Green's functions for the stationary $I=0$ case.}
\label{Fig:nor_det}
\end{figure}

In Figure~\ref{Fig:nor_det} we plot $\mathcal{N}$ as a function of $t$ for both the $2\times 2$ matrix of stationary $I=0$ two-point functions comprising $\pi\pi(111,111)$ and $\sigma$ operators and the $3\times 3$ matrix of Green's functions constructed from all three operators. For the two-dimensional matrix case we find at $t = t_{\textrm{min}} = 6$,  $\mathcal{N}(t) = 0.31(7)$ giving unambiguous proof that more than one state must be present, while for the three-dimensional matrix case, $\mathcal{N}(t)$ is relatively suppressed and takes value of 0.14(13) at $t=5$ and consistent with zero at $t\ge 6$. The observation that the third state can no longer be distinguished from the noise for $t\geq 5$ explains why we were unable to perform reliable 3-operator, 3-state fits to the $I=0$ stationary two-point functions with $t_{\rm min}\geq 5$ earlier in this section. This is closely related to the discussion of the size of the excited state systematic error, as will be explained in Section~\ref{sec:systematic_error}.  We emphasize that the determinant is computed from two-point function measurements {\it at a single time separation} and provides information beyond that which can be obtained from the time dependence of a single operator.  The slow decrease of $\mathcal{N}(t)$ as a function of $t$ throughout our fitting range suggests that there are states with similar energies, which can be distinguished even at a single time separation by the multiple operators in our fitting procedure.  Note, according to Table~\ref{tab:I=0} for the $I=0$ channel these three nearby energies expressed in units of MeV are:  $E_0=547.5(6)$, $E_1=725(5)$ and $E_2=948(13)$.  While the differences between these energies are sufficient to easily see the time dependence of shown in Figure~\ref{Fig:nor_det}, they are insufficient to be resolved in a single-operator fit, even with the statistical precision achieved with 741 configurations.

 \par

\subsection{Comparison of multi-operator multi-state fits with the GEVP method}

Multi-parameter fitting is a straightforward method to analyze the correlation functions between pairs of interpolating operators to determine the energies of finite-volume states which these operators create and the overlap amplitudes between these operators and states.   A second approach to analyze such data is the generalized eigenvalue problem (GEVP) approach~\cite{Luscher:1990ck, Blossier:2009kd}.  The GEVP can be viewed as a generalization of the concept of effective mass, from single-operator to multiple-operator Green's functions.  In principle, this approach has good control over the systematic error resulting from the excited states that are not included in the analysis.  Following the notation of Ref.~\cite{Blossier:2009kd}, the $N$-dimension GEVP can be defined as
\begin{equation}
\begin{aligned}
C(t)v_{n}(t,t_0) = \lambda_n(t,t_0) C(t_0)v_{n}(t,t_0) \quad 1\le n\le N, t_0 \le t\,,
\end{aligned}
\end{equation}
where $C(t)$ is the N-dimensional matrix of two-point functions, $v_{n}$, $1 \le n \le N$ are the eigenvectors and $\lambda_n(t,t_0)$ are the corresponding generalized eigenvalues.  (In this section only, we follow the conventions of Ref.~\cite{Blossier:2009kd}, and construct the correlation function $C_{ij}(t)$ from the product $O_i(t)O_j(0)^\dagger$.)  In the limit where the lattice temporal extent, $T$, is large, the energy of the $n^{th}$ state is related to $\lambda_n$ by
\begin{equation}
\begin{aligned}
 E_n &= \lim_{t\rightarrow \infty} E_n^{\textrm{eff}}(t,t_0) \\
 E_n^{\textrm{eff}}(t,t_0) &= \textrm{log}(\lambda_n(t,t_0)) - \textrm{log}(\lambda_n(t+1,t_0))\,.
\end{aligned}
\label{equ:GEVP_energy}
\end{equation}
\par

The GEVP approach can also be used to construct an operator $A^\dagger_n$ which creates the normalized lattice energy eigenstate: 
\begin{equation}
\begin{aligned}
 A_n(t,t_0) &= e^{-Ht}Q_n(t,t_0) \\
 Q_n(t,t_0) &= R_n(t,t_0) \sum_{i=1}^N O_i v^{n\ast}_i(t,t_0) \\
 R_n(t,t_0) &= \left(\sum_{i,j=1}^N v^{n\ast}_i(t,t_0) C_{ij}(t)v^{n}_j(t,t_0)\right)^{-0.5}\frac{\lambda_n(t_0+t/2,t_0)}{\lambda(t_0+t,t_0)}\,.
\end{aligned}
\end{equation}
It has been shown that in the region where $t_0 > t/2$, the systematic error in the energy of the $ i^{th}$ state resulting from states omitted from the analysis is constrained by~\cite{Blossier:2009kd}
\begin{equation}
\begin{aligned}
\Delta E = O\left(e^{-(E_{N+1}-E_i)t}\right).
\label{eq:GEVP-error}
\end{aligned}
\end{equation}
\par

If $T$ is not sufficiently large, we need to consider two complications to the GEVP procedures described above.  The first is around-the-world propagation, which introduces time-independent constants into the correlation functions for both isospin channels for each of our four values of total momentum. One way to eliminate this effect is to introduce a ``subtracted matrix of two-point functions'' $D(t)$ defined as $D(t) = C(t) - C(t+\delta t)$ and use this $D$ matrix in the GEVP calculation~\cite{Dudek:2012gj}. Notice that this step will not affect the formula for the energy, but a modification is needed for the operator $A_n$ which is now given by
\begin{equation}
\begin{aligned}
A_n(t,t_0) &= e^{-Ht}Q_n(t,t_0) / \sqrt{1-e^{-E_{n}\delta t}}\,.
\end{aligned}
\end{equation}
\par

The second complication comes from backward propagating states.  One way to accommodate this effect is to modify the relation between the eigenvalue and the corresponding energy. 
For more detail, see Ref.~\cite{Irges:2006hg}. We will not use this method here but instead work with a smaller time range where the effect of backward propagating states can be neglected. \\

In this paper we will compare only the ground state energies obtained in our fitting and GEVP analyses.  Specifically we compare the energies obtained from our multi-operator simultaneous fits to the GEVP effective energy defined in Eq.~\eqref{equ:GEVP_energy}.  Our comparison of the fitting and GEVP approaches begins by comparing the fitting results with those obtained from the GEVP at a fixed time $t$ chosen to be the same as the value of $t_{\mathrm{min}}$ used in the fit while the value of the GEVP quantity $t_0$ is chosen as $\left \lceil{\frac{t}{2}}\right \rceil $.  We find that the GEVP energy is statistically consistent with the fit result but its statistical error is about five times larger.  Actually the result of this direct comparison should not be surprising, since much information is lost when the GEVP method is applied to a single $(t,t_0)$ pair.  \par

An improvement to the GEVP method proposed in Ref.~\cite{Dudek:2012gj} is to fit the set of generalized eigenvalues $\lambda_0(t,t_0)$, as a function of $t$ with $t_0$ fixed and to include in that fit possible correction terms from omitted excited states.  This addresses the statistical noise problem identified above by including more of the correlation function data in the GEVP analysis.  We adopt a simple version of this fitting approach and perform a correlated fit to the GEVP eigenvalues of the form
\begin{equation}
    \lambda_0(t,t_0) = e^{-E_0(t-t_0)},
\label{eq:GEVP-fit}
\end{equation}
fitting all of the data from $t=t_0+1$ to a largest value $t_{\mathrm{max}}$.  As shown in Ref.~\cite{Blossier:2009kd}, this functional form for $ \lambda_0(t,t_0)$ will contain errors from neglected higher energy states which are bounded by
\begin{equation}
\begin{aligned}
\Delta E = O\left(e^{-(E_{N+1}-E_0)t_0}\right).
\label{eq:GEVP-error-new}
\end{aligned}
\end{equation}
\par

By choosing the smallest value of $t$ used in this GEVP fit to be one time unit above $t_0$ we are minimizing the statistical error in our result for $E_0$ for a given choice of $t_0$.  We then treat $t_0$ in the same spirit as  $t_{\mathrm{min}}$ in our previous multi-parameter fitting of the matrix of Green's functions (which we will call the ``usual fit'' in the remainder of this section).  Thus, we vary $t_0$ searching for a plateau region for sufficient large $t_0$ and then adopt as the result of this GEVP fitting that value obtained for $E_0$ from the smallest value of $t_0$ within that plateau region.

While this procedure is similar to what is used in our usual fit, we have not carried out the detailed discussion of the residual systematic errors coming from excited state contamination that is attempted in Section~\ref{sec:systematic_error} for our usual fit.  Thus, we are unable to say if choosing a larger value of $t$ than $t_0+1$ would have resulted in a sufficiently reduced systematic error to give a reduction in the total error, overcoming the increase in statistical error that would result from increasing $t-t_0$ above one.

In addition to examining the dependence of the GEVP result for $E_0$ on $t_0$, we must also make sure that our choice of $t_{\mathrm{max}}$ is appropriate: if $t_{\mathrm{max}}$ is too large, then neglecting the backward propagating state will introduce an error; if $t_{\mathrm{max}}$ is too small, then we will have a small number of input data points which makes our fit less reliable. We will choose $t_{\mathrm{max}}$ to be no larger than that used in our usual fit, so that we can use those earlier results to determine how we should treat the around-the-world effects.  Thus, based on the results obtained from our usual fit, we neglect the around the world effects for the $I=0$ channel and work directly with the correlation matrix $C(t)$ of two-point functions.  For $I=2$ these effects were found to be important so in that case we analyze the subtracted matrix $D(t)$ defined above, using $\delta t=1$.  For the $I=2$ channel, where we perform this matrix subtraction process, $t_{\mathrm{max}}$ is taken to be 20, smaller than the $t_{\mathrm{max}}=25$ in the usual fit, due to the increased noise resulting from the construction of the subtracted matrix of two-point functions $D(t)$.  The value of $t_{\mathrm{max}}$ in the $I=0$ channel is chosen to be 15, the largest used in the usual fit.  We can then plot the ground state energy from this GEVP fitting as a function of $t_0$ and look for the beginning of the plateau region and also the $p$-value in order to determine $t_0$.

\begin{figure}[tb]
\centering
\fbox{\includegraphics[width=0.40\textwidth]{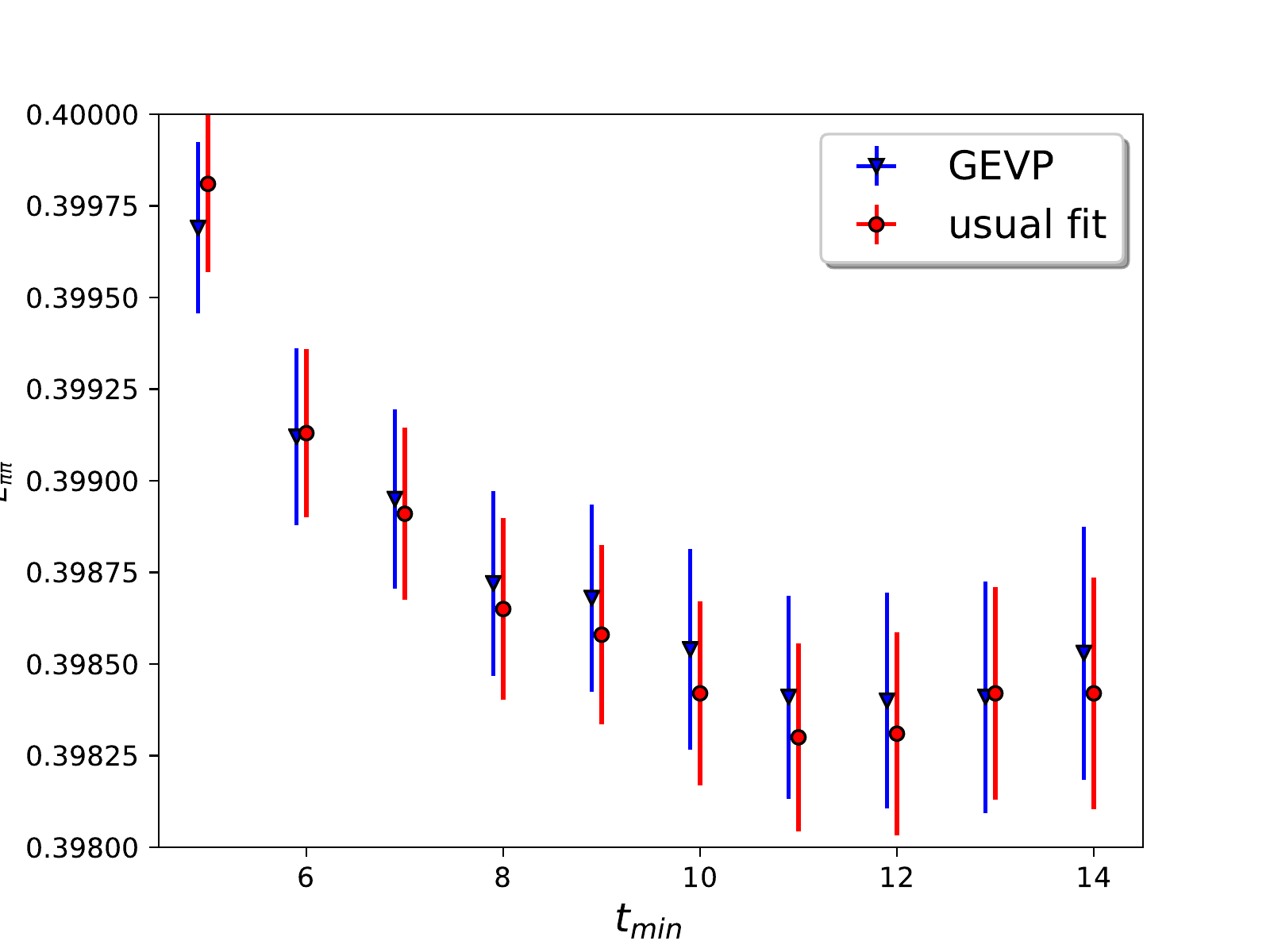}}
\fbox{\includegraphics[width=0.40\textwidth]{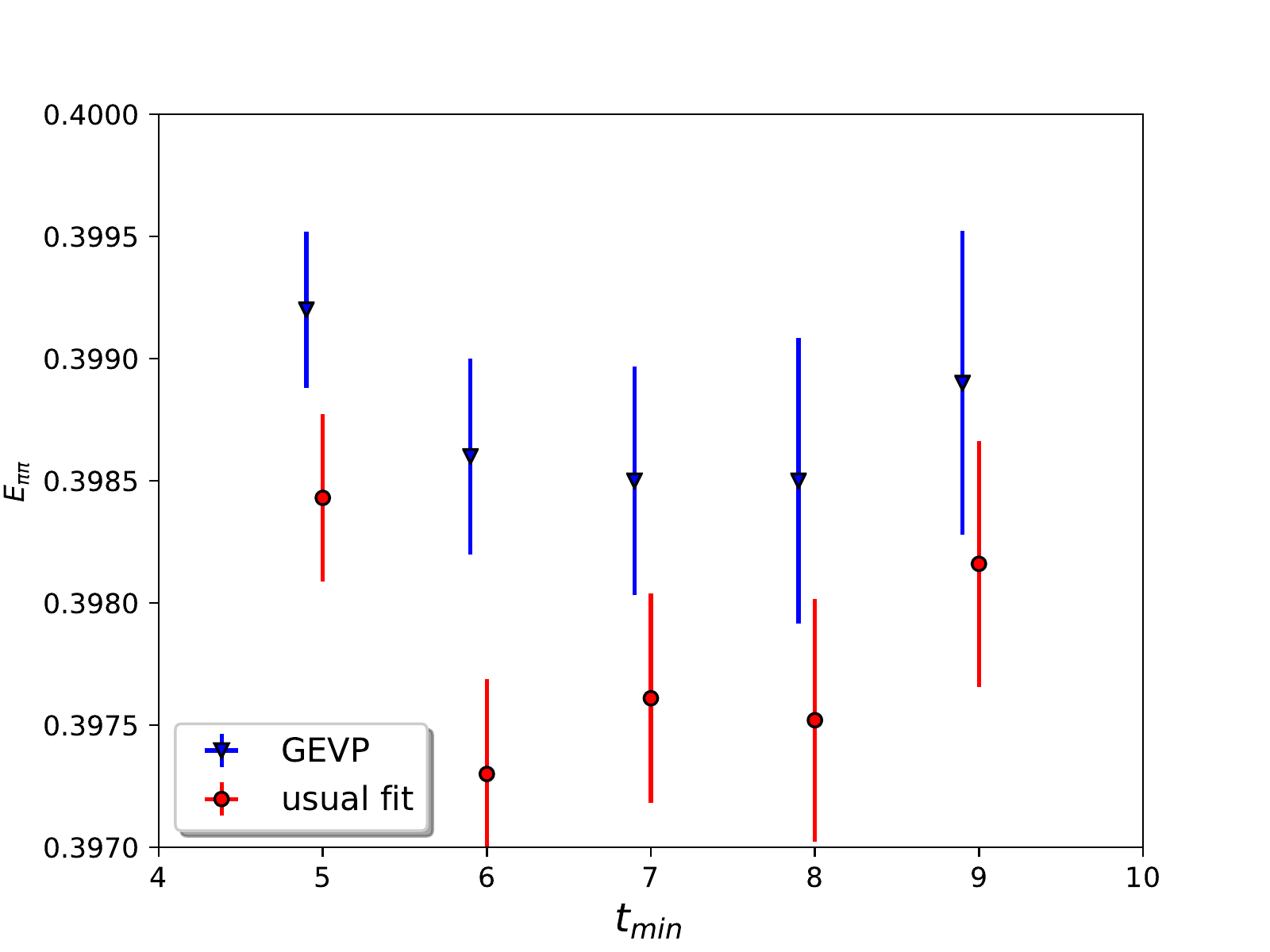}}\\
\fbox{\includegraphics[width=0.40\textwidth]{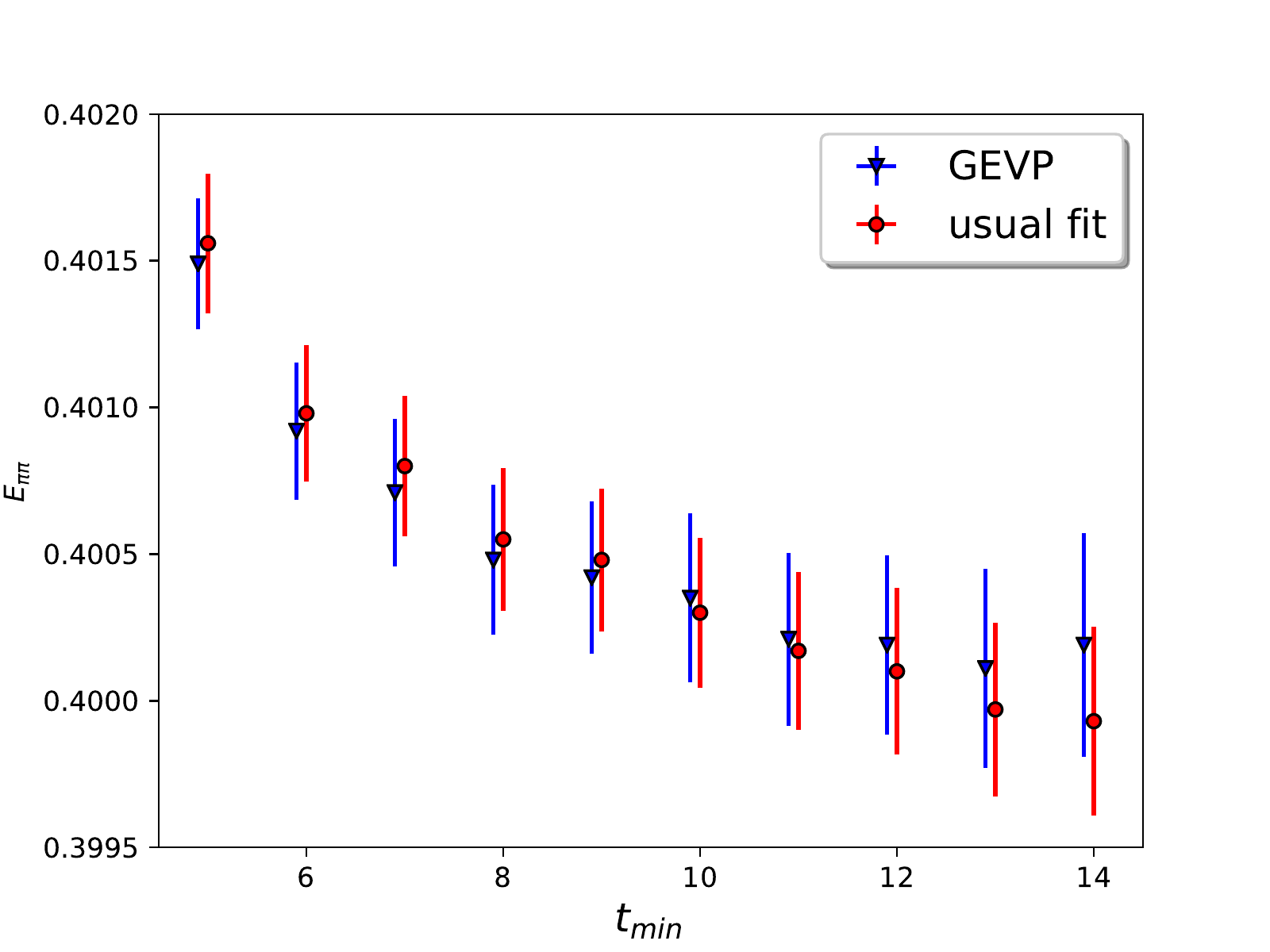}}
\fbox{\includegraphics[width=0.40\textwidth]{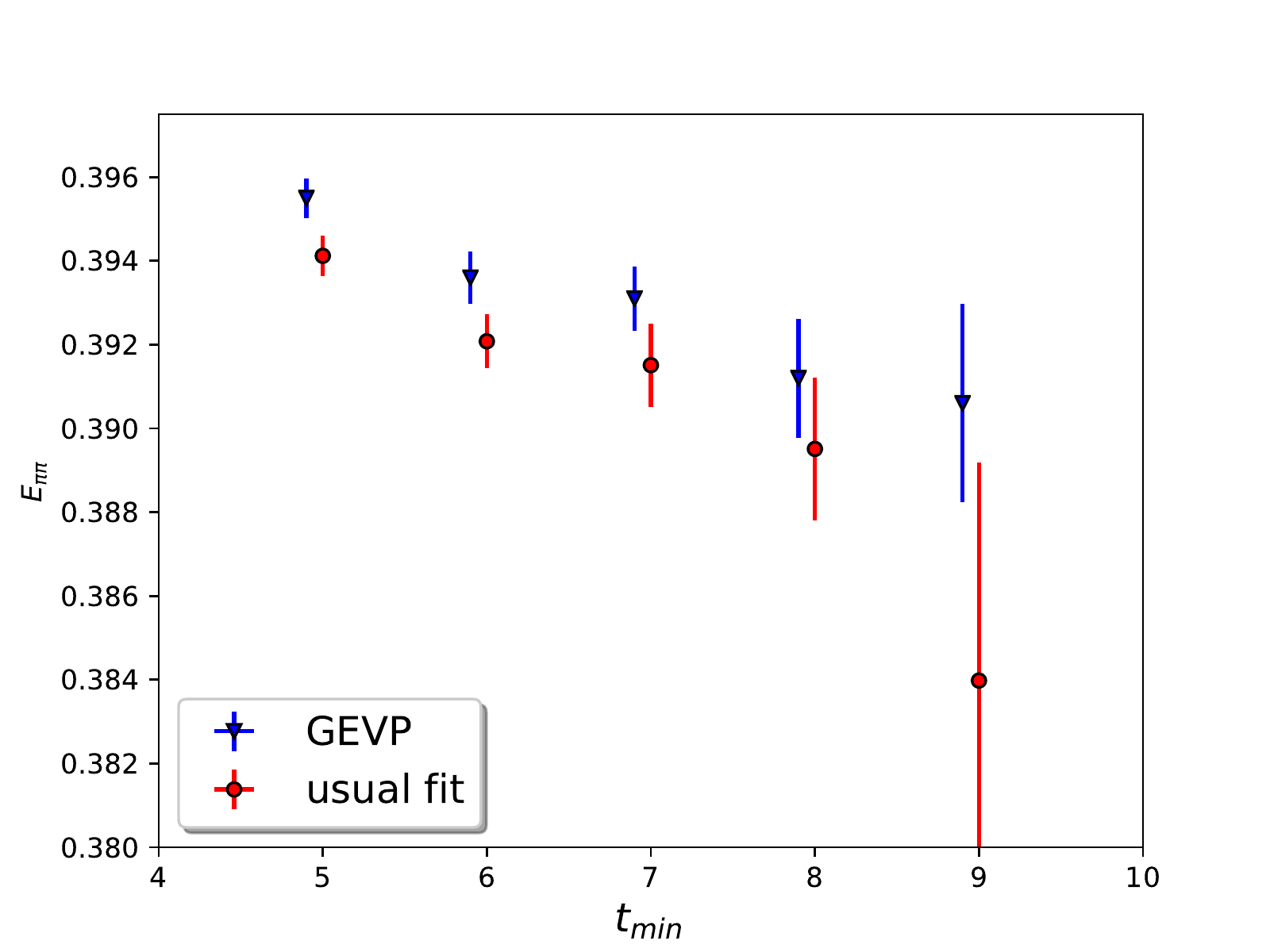}}\\
\fbox{\includegraphics[width=0.40\textwidth]{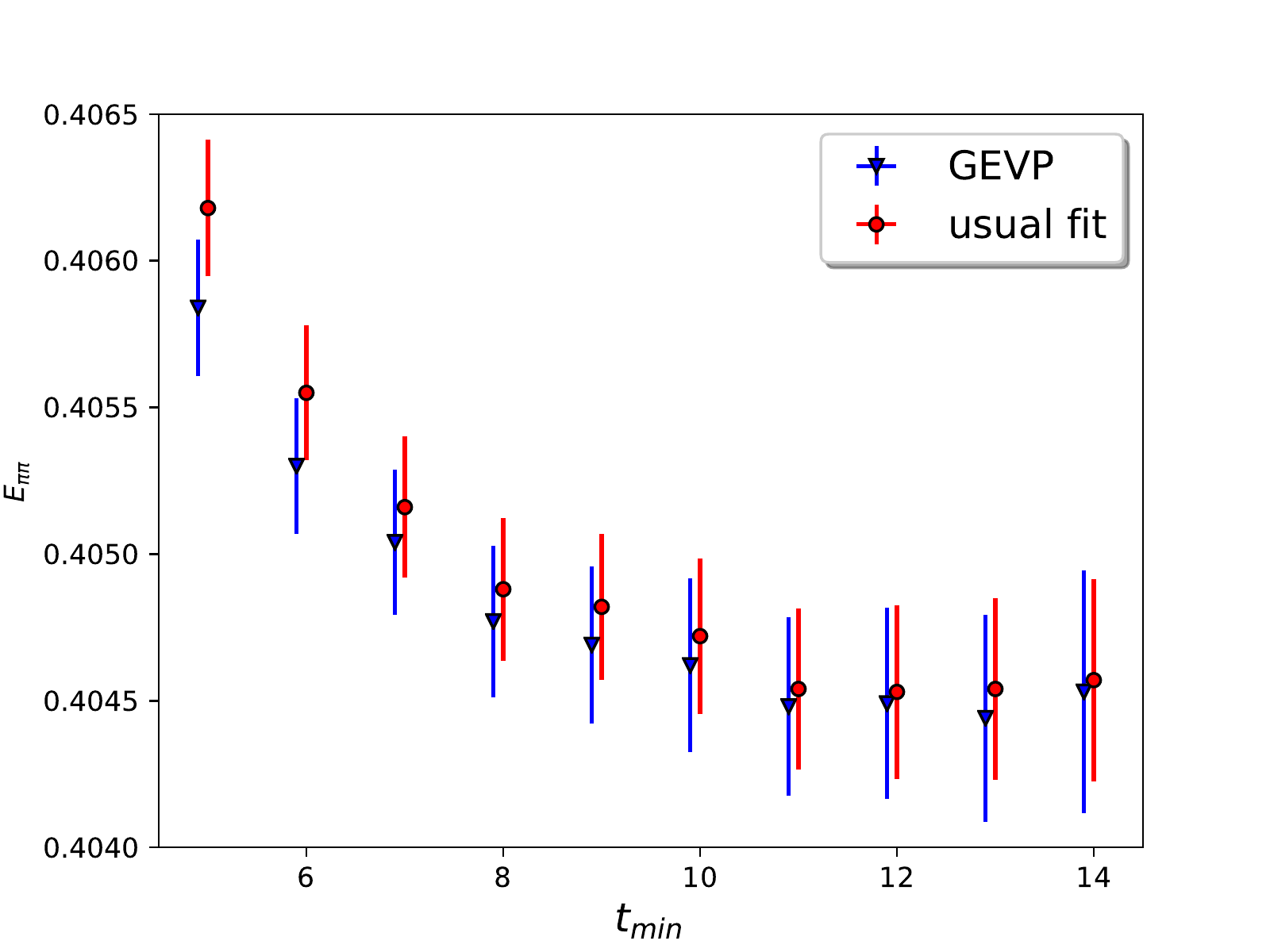}}
\fbox{\includegraphics[width=0.40\textwidth]{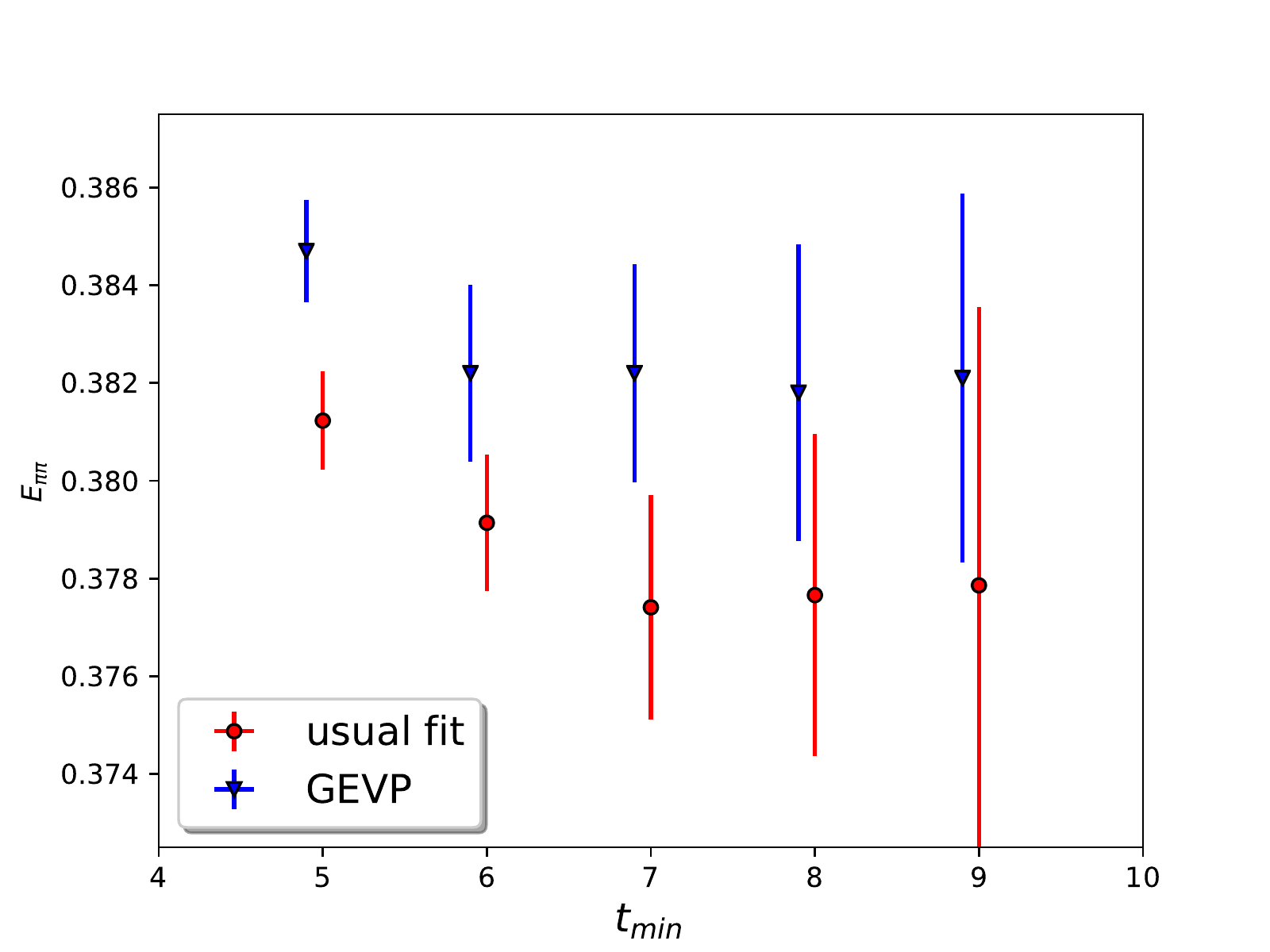}}\\
\fbox{\includegraphics[width=0.40\textwidth]{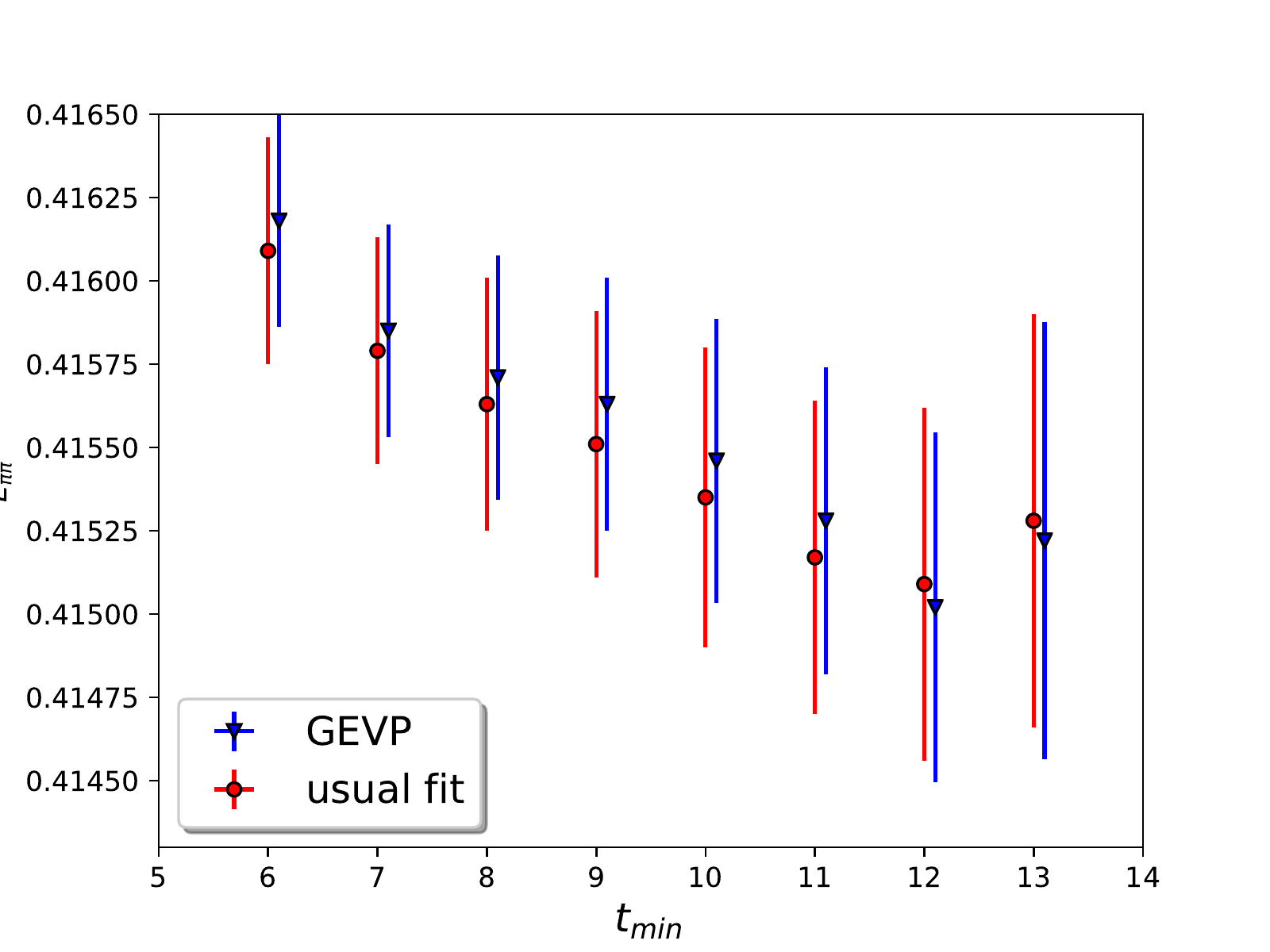}}
\fbox{\includegraphics[width=0.40\textwidth]{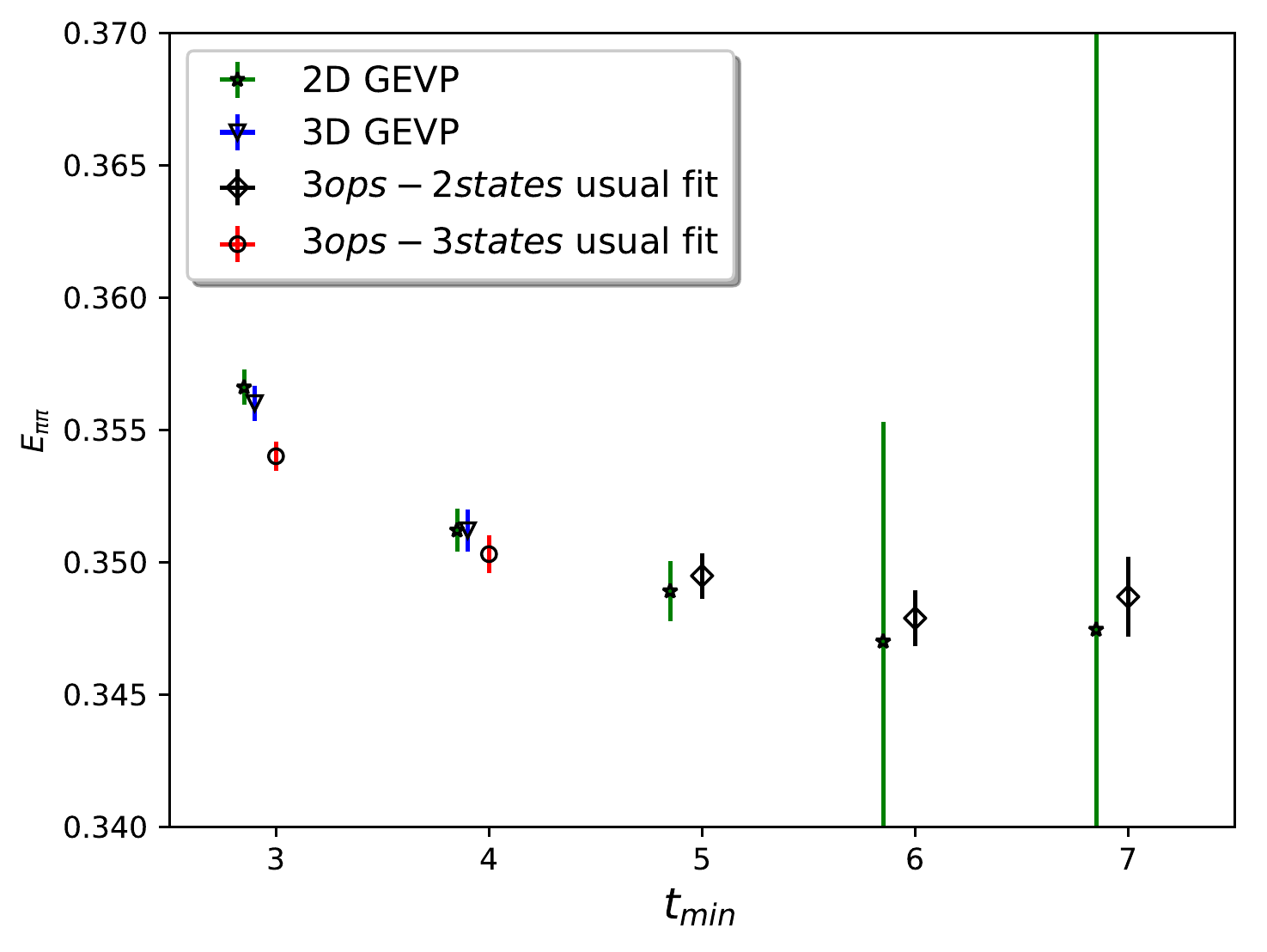}}\\
\caption{The $t_{\mathrm{min}}$ or $t_0$ dependence of the fitted ground state energy from the GEVP and the usual fit for the $\pi\pi_{I=2}$ (left) and $\pi\pi_{I=0}$ (right) channel with $t_{\mathrm{max}}=20$ ($I=2$) and 15 ($I=0$). The total momenta from the top down are $(2,2,2)\frac{\pi}{L}$, $(2,2,0)\frac{\pi}{L}$, $(2,0,0)\frac{\pi}{L}$ and 0. Here the $x$-axis represents $t_{\mathrm{min}}$ for the usual fit, and $t_0$ for the GEVP fit.  In the legend of the lower right panel, 2D and 3D indicate a $2\times2$ and $3\times3$ GEVP matrix.}
\label{fig:gevp_vs_fit}
\end{figure}

Plots that include both the GEVP fit and the usual fit results are shown in Fig.~\ref{fig:gevp_vs_fit}. Notice the x-axis represents $t_{\mathrm{min}}$ for the usual fit, and $t_0$ for the GEVP method.  This choice for plotting was made to best align both the central values and statistical errors from the two methods.  For all but the stationary $I=0$ case, both the usual fit and the GEVP fit are performed with all operators and the number of states the same as the number of operators. For the stationary $I=0$ case, the usual fit includes all the operators, and the number of states is 3 when $t_{\mathrm{min}} < 5$ and 2 when $t_{\mathrm{min}} \ge 5$.  For the GEVP fit, the statistical error blows up when $t_{\mathrm{min}} > 5$ if we include all operators, so we have used only two operators, $O_a$ and $O_c$, when $t_{\mathrm{min}} > 5$. It can be seen that for the $I=2$ channel, the two results are not only consistent with each other, but also show similar time-dependence.  For the $I=0$ channel, the two results are not consistent, but this inconsistency may come from the excited-state error, which will be discussed below.\\

\begin{table}[tb]
\begin{center}
\begin{tabular}{ | c | c | c | c | c | c | c | }
  \hline
  $P_{tot,I}$ & sub & $t_0$ & fitting range & $E_0$ (GEVP) & $E_0$ (usual fit) & $p$-value\\
  \hline
  $(222)_{I=2}$ & y & 10 & 11-20 & 0.39854(27) & 0.39842(26) & 0.385\\
  \hline
  $(220)_{I=2}$ & y & 11 & 12-20 & 0.40021(29) & 0.40010(32) & 0.306 \\
  \hline
  $(200)_{I=2}$ & y & 11 & 12-20 & 0.40447(31) & 0.40454(30) & 0.294 \\
  \hline
  $(000)_{I=2}$ & y & 11 & 12-20 & 0.41528(46) & 0.41535(45) &  0.665\\
  \hline
  $(222)_{I=0}$ & n & 6 & 7-15 & 0.3986(4) & 0.3973(4)\textcolor{red}{[17]} & 0.28 \\
  \hline
  $(220)_{I=0}$ & n & 8 & 9-15 & 0.3907(13) & 0.3895(17)\textcolor{red}{[19]} & 0.622 \\
  \hline
  $(200)_{I=0}$ & n & 6 & 7-15 & 0.3823(18) & 0.3774(23)\textcolor{red}{[52]} & 0.983 \\
  \hline
  $(000)_{I=0}$ & n & 5 & 6-15 & 0.3489(11) & 0.3479(11)\textcolor{red}{[10]} & 0.142\\
  \hline
\end{tabular}
\caption{Comparison between the ground state energy $E_0$ obtained from the GEVP fit (GEVP) and the direct matrix of two-point functions fit (usual fit) given in Tables~\ref{tab:I=2} and \ref{tab:I=0}, repeated here for convenience. The $t_0$ used for the GEVP fit is obtained from Fig.~\ref{fig:gevp_vs_fit} by recognizing the beginning of the plateau region and the resulting $t_{\mathrm{min}}=t_0+1$ is shown in the fitting range above. In the “sub” column we indicate whether (y) or not (n) we are using the subtracted matrix of two-point functions $D(t)$ to perform the GEVP calculation, removing the around-the-world effects. The statistical error for both fits are shown in parenthesis while the systematic errors from excited state contamination for the $I=0$ channel, estimated in Section~\ref{sec:systematic_error}, are shown in square brackets.}
\label{table:GEVP_comparison}
\end{center}
\end{table}

The final results and the choices of fitting setup are shown in Table~\ref{table:GEVP_comparison}.  The $p$-values shown suggest that the quality of all the GEVP fits are relatively good.  From the table we can see that for the $I=0$ channel the GEVP fit results are approximately $2\sigma$ larger than the results from the usual fit if we only include the statistical error.   However, the two are consistent if the systematic errors arising from excited state contamination in the usual fit are included.   These excited-state error estimates were obtained by independent methods as described in Section~\ref{sec:systematic_error} and are also listed in Table~\ref{tab:I=0_exc_err} and match surprisingly well the differences between the results obtained from our usual and GEVP fits. 

For the $I=2$ channel the two results are consistent with each other with comparable statistical errors which are much smaller than in the $I=0$ case. Also notice that in all cases, the GEVP method gives statistical errors that are no larger than the usual fit method, which suggests that the GEVP fitting method is a useful tool for analyzing the matrix of Green's functions for the scattering considered here. We also expect that the usual fits may be less successful than the GEVP method when we increase the number of operators in the fit due to instabilities that will likely result from the larger number of fit parameters. As used here, the GEVP fit is far simpler, being a correlated fit to a single one-parameter function of the time.  The multi-operator fits will further suffer from the quadratic increase in the number of elements in the covariance matrix, the inversion of which may become unstable once it becomes too large. In the GEVP case, the size of the covariance matrix depends only on the size of the fit window.

Nevertheless, for the problem at hand, fitting the matrix of two-point functions is more direct than fitting the GEVP eigenvalues and, as we will show in Section~\ref{sec:systematic_error}, allows considerable flexibility in estimating the size of the systematic error arising from omitted excited states. In the traditional GEVP method~\cite{Blossier:2009kd} the number of operators must be larger than or equal to the number of states, while in the multi-parameter fitting approach more states than operators can be easily accommodated.  Currently, this is a crucial step in estimating the excited state error as will be seen in Section~\ref{sec:systematic_error} and in identifying a plateau in the stationary I = 0 case as can be seen from the lower right panel of Fig.~\ref{fig:gevp_vs_fit}. However, for the case studied here we have not attempted to make similar estimates of the systematic errors in our GEVP results and effective methods for estimating such errors may well be possible there.


\section{Determination of the phase shift}
\label{sec:phase_shift}
In this section we discuss in detail how we determine the $\pi\pi$ phase shifts from the finite-volume $\pi\pi$ energies.  We begin with L\"uscher's formula generalized to the case of anti-periodic boundary conditions\footnote{In this section we will focus on the case of anti-periodic boundary conditions obeyed by the pions, which result from the G-parity boundary conditions obeyed by the quarks.}  with a general total momentum.  Next we discuss the strategy of working with energy differences to reduce discretization errors, especially for the moving frame calculations. We then calculate the $\pi\pi$ scattering phase shifts at various center-of-mass energies for both isospin channels using this technique.  We also describe our method for specifying the energy at which these phase shifts have been determined in order to reduce the effects of the slightly unphysical pion mass used in our lattice calculation. Finally we calculate the Lellouch-L\"uscher factor~\cite{Lellouch:2000pv} that is needed to interpret the finite-volume $K \rightarrow \pi\pi$ calculation~\cite{Abbott:2020hxn}.

\subsection{L\"uscher's quantization condition for non-zero total momentum and anti-periodic boundary conditions}
\label{sec:Luscher_formula_derivation}
Euclidean-space lattice QCD calculations determine finite-volume $\pi\pi$ energies from which the infinite-volume scattering phase shifts can be obtained using an approach developed by L\"uscher~\cite{Luscher:1990ux}.    While  initially derived for the case of a stationary frame and periodic boundary conditions, this approach was later generalized to non-zero total momentum \cite{Rummukainen:1995vs, Christ:2005gi, Kim:2005gf} and later this moving frame result was generalized to the case with general twisted boundary conditions\cite{Briceno:2014oea}. In this section, we will write down this result for our GPBC lattice where pions satisfy the anti-periodic boundary conditions. In particular, the $s$-wave phase shift, $\delta(s)$ can be determined from the relation: $\delta(s) + \phi^{\Vec{d},\gamma}(s) = n\pi$ where $n$ is an integer and typically allows for more than one solution to this quantization condition, resulting in a series of energy eigenstates in a single volume.  The function $\phi^{\Vec{d},\gamma}(s)$ is defined by

\begin{equation}
\begin{aligned}
\tan\left(\phi^{\Vec{d},\gamma}(s)\right) = \frac{\gamma\pi^{3/2}q}{Z_{00}^{\Vec{d},\gamma}(1,q^2)}.
\end{aligned}
\label{equ:luscher_phase_shift}
\end{equation}
Here $s$ is the square of the invariant mass of the two-pion system, $\gamma$ is the Lorentz factor which boosts the laboratory frame to the CM frame and $q$ is related to the magnitude of the momentum $k$ carried by either pion in their center-of-mass frame.  Each of these quantities can be determined from the finite-volume $\pi\pi$ energy, $E_{\pi\pi}$ obtained from the lattice calculation:
\begin{equation}
\begin{aligned}
s = E_{\pi\pi}^2 - \vec P^2_{tot}, \quad
\gamma = \frac{E_{\pi\pi}}{\sqrt{s}}, \quad 
k^2 = \frac{s}{4} - m_\pi^2, \quad 
q = k \frac{L}{2\pi}.
\label{eq:CM-formulae}
\end{aligned}
\end{equation}

The vector of integers, $\Vec{d}$, is related to the total momentum $\vec P_{tot}$ by
\begin{equation}
\begin{aligned}
\Vec{d}=\frac{L}{2\pi}\Vec{P}_{tot}
\end{aligned}
\end{equation}
and $Z_{00}^{\Vec{d},\gamma}(\hat{s},q^2)$ is the generalized L\"uscher's zeta function, which is defined as 
\begin{equation}
\begin{aligned}
Z_{00}^{\Vec{d},\gamma}(\hat{s},q^2) = \frac{1}{\sqrt{4\pi}}\sum_{\vec r \, \in \mathcal{N}^{\vec d, \gamma}}(\vec{r}\,^2-q^2)^{-\hat{s}}\,,
\label{eq:zeta}
\end{aligned}
\end{equation}
where the conventional argument $s$ of the zeta function is replaced here by $\hat{s}$ to remove the possible confusion with the square of the center-of-mass energy and the set $\mathcal{N}^{\vec d, \gamma}$ is defined as
\begin{equation}
\begin{aligned}
\mathcal{N}^{\vec d, \gamma}=\Bigl\{\vec r\,\Bigr|\vec r=\hat{\gamma}^{-1}\Bigl(\Vec{n}+\Vec{d}/2+\Vec{\ell}/2\Bigr),\Vec{n}\in \mathbb{Z}^3\Bigr\}.
\end{aligned}
\end{equation}
Here the vector $\vec \ell$ represents the effect of the boundary conditions. If the particle satisfies periodic boundary conditions in the $i^{th}$ direction then $\ell_i=0$, while with anti-periodic boundary conditions we have $\ell_i=1$. The quantity $\hat\gamma^{-1}$ is a linear transformation on 3-vectors defined as
\begin{equation}
\begin{aligned}
\hat{\gamma}^{-1}(\vec n) = \frac{1}{\gamma}  \vec n_\parallel + \vec n_\perp\,,
\end{aligned}
\end{equation}
using the notation of Ref.~\cite{Yamazaki:2004qb}.
\par

As defined in Eq.~\eqref{eq:zeta} the generalized zeta function diverges at $\hat{s}=1$ and needs to be expressed differently to be evaluated at $\hat{s}=1$. We use a simple generalization of a formula given in Ref.~\cite{Yamazaki:2004qb} to do this. Combining all of these formulae we can obtain the $\pi\pi$ scattering phase shift at the energy $\sqrt{s}$ from the finite-volume energy eigenvalue $E_{\pi\pi}$ determined from our lattice calculation. Note that in obtaining Eq~\eqref{equ:luscher_phase_shift} we are implicitly neglecting the contributions to the scattering of partial waves with $l \geq 1$. In the stationary frame, assuming the $\pi\pi$ operators are constructed in the trivial representation of the cubic group, cubic symmetry prevents states with $1 \leq l < 4$ from contributing~\cite{Luscher:1990ux}. The interaction strength in the $l\geq 4$ channels is known to be small and these interactions can be safely neglected~\cite{Luscher:1990ux}. 

However in the moving frame the relativistic length contraction naturally breaks the cubic symmetry down to a smaller group, the trivial representation of which also allows for contributions from $d$-wave ($l=2$) interactions. Previous calculations~\cite{Wang:2019nes} have shown that the phase shifts in the $l=2$ channel are small around the kaon mass, which will be further suppressed in the moving frame calculation where $\sqrt{s}$ is smaller than $m_K$ so we can therefore continue to assume $s$-wave dominance. As described above, the G-parity boundary conditions also break the cubic symmetry but the effects can be suppressed with a careful choice of $\pi\pi$ operator. Any systematic errors arising from this source are discussed further in Section~\ref{sec:systematic_error_cubic}.

\subsection{Calculation technique}

As shown above, the $\pi\pi$ scattering phase shift is related to the energy of a finite-volume $\pi\pi$ state, or more specifically to the ``pion momentum'' $k$ carried by either pion outside the range of the strong force.  However, on a discretized lattice with anti-periodic boundary conditions ({\it i.e.}~a case where the single-pion ground-state has non-zero momentum), the determination of $k$ from the measured $\pi\pi$ energy must be performed carefully.  If the $\pi\pi$ interaction is relatively weak then $k$, which is a measure of that interaction, will be close to its free field value and we must take precautions that the potentially small difference between $k$ and its free field value is nevertheless large when compared with the discretization errors associated with the spatial momenta of the pions in our calculation.  However, as can be seen in Eq.~\eqref{eq:CM-formulae}, $k$ is determined from differences of larger quantities and care must be taken to insure that the quantities being subtracted have, to the degree possible, common finite lattice spacing errors so that these errors will largely cancel in the difference.  Specifically the quantities being subtracted should be chosen so that their difference will vanish in the limit that the $\pi\pi$ interactions vanish, even when computed at finite lattice spacing. \par

This cancellation of finite lattice spacing errors can be accomplished by working with two related quantities determined from our calculation:  $\Delta E=E_{\pi\pi}-2E_\pi$, which measures the $\pi\pi$ interaction strength, and $E_\pi$, the lowest energy of a moving pion. Using $\Delta E$ for example, the effects of the finite lattice spacing upon the pion dispersion relation that enter both $E_{\pi\pi}$ and $2E_{\pi}$ will largely cancel, leaving only the subtler effects of the discretization upon the two-pion interaction itself.  Even for the case of non-zero total momentum $\vec P_{tot}$, we will exploit our choice of anti-periodic boundary conditions in all three directions and use for $E_\pi$ the ground-state, single-pion energy.  Each of the three non-zero total momenta that we study can be formed from two pions carrying the minimum allowed momenta $\vec p = (\pm 1, \pm 1,\pm 1)\pi/L$ so that $2E_\pi$ will be the minimum energy of two interacting pions in the limit in which that interaction vanishes. \par

Thus, the quantities $\Delta E$ and $E_\pi$ will be computed on the lattice, and systematic errors estimated to account for the residual effects of the finite lattice spacing. The results are finite-volume predictions for $\Delta E$ and $E_\pi$ in the continuum limit, albeit with an unphysical pion mass, and in Section~\ref{sec:systematic_error} we will estimate and propagate the systematic discretization errors on these quantities. (While it would be better to determine $\Delta E$ and $E_\pi$ by performing calculations at multiple lattice spacings and taking the continuum limit, this is at present beyond our available resources.)  Adopting this strategy to account for the discretization effects, we can then apply the generalization of L\"uscher's finite-volume quantization condition without ambiguity using the continuum dispersion relation
\begin{eqnarray} 
k^2 &=& (\Delta E + 2E_\pi)^2/4 - E_\pi^2 + 3(\frac{\pi}{L})^2-\frac{1}{4}\vec P_{tot}^2 \nonumber \\
      &=& \frac{\Delta E^2 }{4} + \Delta E E_\pi  + 3(\frac{\pi}{L})^2-\frac{1}{4}\vec P_{tot}^2\,,
\label{equ:def_pion_mom_k}
\end{eqnarray} 
where we have continued to assume anti-periodic boundary conditions in three directions and note that the last two terms on the right-hand side of Eq.~\eqref{equ:def_pion_mom_k} are exactly known.  \par

The second line in Eq.~\eqref{equ:def_pion_mom_k} demonstrates the purpose of this rearrangement: our result for $k$ is proportional to the small quantity $\Delta E$ (a measure of the $\pi\pi$ interaction strength) plus other kinematic quantities that are determined without finite lattice spacing error.  This guarantees that at finite lattice spacing the phase shift determined in this way from the quantization condition will vanish when $\Delta E\to 0$ so that the fractional finite lattice spacing errors expected in $\Delta E$ can be directly propagated to determine the corresponding error in $\delta(s)$.

We thereby obtain a value for the phase shift at an unphysical pion mass, from which a prediction for the physical phase shift can be obtained by assigning suitable systematic errors for discretization effects and the unphysical pion mass as will be discussed in Section~\ref{sec:systematic_error}. \par

\subsection{Phase shift results (statistical error only)}
\label{sec:phase_shift_result_stat_err_only}
In this section we tabulate our results for the $\pi\pi$ scattering phase shifts including their statistical errors, computed according to the method described above.  For each result we must specify the energy at which the phase shift takes the quoted value and we choose to assign the appropriate $\sqrt{s}$ in such a way as to minimize the error introduced by the unphysical pion mass, 143 MeV, at which our calculation is performed, a value 6\% larger than the 135 MeV which we adopt as the physical pion mass in this paper. For values of $\sqrt{s}$ on the order of the kaon mass, the error associated with this unphysical 8 MeV shift in the pion mass is small and is estimated using chiral perturbation theory in Section~\ref{sec:systematic_error}.  However, since the $I=0$ and 2 phase shifts vanish when $\sqrt{s} = 2m_\pi$, this unphysical pion mass error can become large as $\sqrt{s}$ approaches $2m_{\pi,\mathrm{unphy}} > 2m_{\pi,\mathrm{phy}}$.  This effect can be easily eliminated if we view our computed phase shifts as functions of the pion momentum $k$ rather than $\sqrt{s}$, since a calculation of the phase shifts will give results which vanish at $k=0$ independent of the pion mass. 

Thus, for each computed value of the phase shift we use the measured lattice $\pi\pi$ energy and lattice pion mass to obtain the relative momentum $k$, and then when presenting our results for the phase shifts assign an energy determined by combining this momentum in the continuum limit with the physical pion mass by applying the dispersion relation 
\begin{equation}
s = 4(k^2 + m_{\pi,\textrm{phy}}^2)\,.
\label{equ:def_sqrt_s}
\end{equation} 
The effect of the unphysical pion mass on the actual strength of the interaction ({\it i.e.}~upon the phase shift itself) is small and is treated as a systematic error that we estimate using chiral perturbation theory, as discussed in Sec.~\ref{sec:systematic_error}.

In Table~\ref{table:phase_shift_stat_only} we list the phase shifts calculated from 4(3) different momenta of the center of mass for the $I=2(0)$ channel and calculate the corresponding $\sqrt{s}$ using Eq.~\eqref{equ:def_sqrt_s}.   Here we only include the statistical error. The full error budget will be discussed in Section~\ref{sec:systematic_error}. We do not provide a result for the phase shift of the $I=0$ channel in the case where $P_{tot} = (2,2,2)\frac{\pi}{L}$.  At this lowest value of $\sqrt{s}$ the attractive interaction between two pions results in a center-of-mass $\pi\pi$ energy that lies below $2m_\pi$.

Instead, we calculate the scattering length for both isospin channels with the moving frame where $P_{tot} = (2,2,2)\frac{\pi}{L}$. For the $I=2$ channel, we start with the following expansion of the phase shift as a function of the relative momentum $k$:
\begin{equation}
    k\textrm{cot}(\delta_0) = -\frac{1}{a_0} + \frac{1}{2}r_{\textrm{eff}}k^2 + O(k^3)
\end{equation}
where $a_0$ is the scattering length and $r_{\textrm{eff}}$ is the effective range. Since $k$ is very small where $P_{tot} = (2,2,2)\frac{\pi}{L}$, we can keep only the leading term:
\begin{equation}
    a_0 = -\frac{\delta_0}{k}
\end{equation}

The result is listed in Table~\ref{table:phase_shift_stat_only}. For the $I=0$ channel, we generalize Eq. (1.3) in Ref.~\cite{Luscher:1986pf} to a moving frame, and keep only the leading term while neglecting all terms with higher power of $a_0/L$. 

This approximation is reasonable since the contribution from these higher-order terms is of O(10) smaller than the statistical error and the systematic error discussed later. The result is also shown in Table~\ref{table:phase_shift_stat_only}. We point out that the scattering length for $I=0$ channel given in that Table is inconsistent with the experimental value at $3\sigma$ level if only the statistical error is considered. This inconsistency mostly comes from the excited state systematic error, as will be discussed in Sec.~\ref{sec:systematic_error}.

\begin{table}[tbp]
 \begin{tabular}{ | c | c | c | c | c | c | c | c |}
 \hline
 $P_{tot}$ & I & $aE_{\pi\pi}$ & k(MeV) & $\sqrt{s}$(MeV) & $\delta$ & $m_{\pi}a_0$ \\
 \hline
 $(0,0,0)\frac{\pi}{L}$ & 2 & 0.4153(4) &  248.4(3) &  565.4(5) & -$11.0(2)^\circ$ &\\
 \hline
 $(2,0,0)\frac{\pi}{L}$ & 2 & 0.4045(3) &  197.9(2) &  479.1(3) & -$7.96(23)^\circ$ &\\
 \hline
 $(2,2,0)\frac{\pi}{L}$ & 2 & 0.4001(3) &  138.4(3) &  386.7(4) & -$4.48(40)^\circ$ &\\
 \hline
 $(2,2,2)\frac{\pi}{L}$ & 2 & 0.3984(3) &  14.4(2.1) &  271.5(4) & -$0.32(20)^\circ$ & -0.055(15)\\
 \hline\hline
 $(0,0,0)\frac{\pi}{L}$ & 0 & 0.3479(11) &  193.0(9) &  471.0(1.5) &  $32.3(1.0)^\circ$ & \\
 \hline
 $(2,0,0)\frac{\pi}{L}$ & 0 & 0.3774(23) &  170.6(2.4) &  435.1(3.8) &  $24.0(3.4)^\circ$ &\\
 \hline
 $(2,2,0)\frac{\pi}{L}$ & 0 & 0.3895(17) &  123.2(2.6) &  365.6(3.4) &  $18.0(4.5)^\circ$ &\\
 \hline
 $(2,2,2)\frac{\pi}{L}$ & 0 & 0.3972(4) &  &  & & 0.072(38)\\
 \hline
 \end{tabular}
 \caption{The phase shifts with statistical errors only for 4(3) different total momenta for the $I=2(0)$ channel and the corresponding $\sqrt{s}$, together with the $I=2(0)$ scattering length calculated from moving frame calculation with total momentum $(2,2,2)\frac{\pi}{L}$. Here the statistical error of each phase shift is obtained not by simply propagating the statistical error of $E_{\pi\pi}$, but a more elaborate method discussed in Sec.~\ref{sec:error_budget} which removes the uncertainty of the energy at which we quote the phase shift.}
 \label{table:phase_shift_stat_only}
\end{table}

\subsection{Lellouch-L\"uscher factor}
\label{sec:phase_shift:LL}
In our companion calculation of the $I=0$ $K\to\pi\pi$ matrix elements~\cite{Abbott:2020hxn}, an important ingredient is the Lellouch-L\"uscher factor~\cite{Lellouch:2000pv}, which removes both the difference in normalization between states defined in finite and infinite volume and the leading power-law finite-volume corrections to the finite-volume matrix element. This factor is defined as:
\begin{equation}
\begin{aligned}
F^2 = \frac{4\pi m_K E^2_{\pi\pi}}{k^3}(k\frac{\partial\delta_I}{\partial k}+q\frac{\partial\phi^{\vec d, \gamma}}{\partial q})\,,
\end{aligned}
\end{equation}
where $\delta_I$ is the isospin $I$, $s$-wave $\pi\pi$ phase shift an $\phi^{\vec d, \gamma}$ is defined in Eq.~\eqref{equ:luscher_phase_shift}. This formula should be evaluated at $E_{\pi\pi} = m_K$. 

The moving frame calculation enables us to determine the phase shifts at various energies,
which allows us to perform an {\it ab initio} measurement of $\frac{\partial\delta_0}{\partial k}$ using a finite-difference approximation. We now focus on the $I=0$ case since our calculation has been tuned to give $s$ close to $m_K^2$ for the $I=0$, $\pi\pi$ ground state.  We approximate the factor $F$ using two different methods. In the first we subtract the values of $\delta_0$ at $P_{tot} = (0,0,0)\frac{\pi}{L}$ and $P_{tot} = (2,0,0)\frac{\pi}{L}$, which gives
\begin{equation}
    \frac{\partial\delta_0}{\partial k} = 0.372(153)
\end{equation}
and in the second method we replace the second total momentum with $P_{tot} = (2,2,0)\frac{\pi}{L}$, and obtain
\begin{equation}
    \frac{\partial\delta_0}{\partial k} = 0.205(63)\,.
\end{equation}
Both results are consistent with 
\begin{equation}
    \frac{\partial\delta_0}{\partial k} = 0.276(1)\,,
\end{equation}
which is calculated from the dispersive analysis~\cite{Colangelo:2001df,Abbott:2020hxn}. Note we have not attempted to account for systematic effects arising from the finite-difference approximation or other effects here. Nevertheless we find good agreement between our lattice results and the dispersive prediction, albeit with large statistical errors. These results are also presented in Ref.~\cite{Abbott:2020hxn} where the dispersive result was used for the final analysis. Note that these values differ slightly (within errors) due to different choices of fit range and the finite-difference approximation being applied there to the phase shift is a function of energy rather than a function of $k$.\par


\section{Systematic error analysis}
\label{sec:systematic_error}
There are several sources of systematic error which affect our results: the breaking of cubic symmetry by our G-parity boundary conditions, the non-zero lattice spacing of our single gauge ensemble, the unphysical value of our pion mass and contamination of our multi-operator, multi-state fits due to the presence of additional excited states, not included in our fit. In this section, we describe our procedure for estimating the size of these errors.  The full error budget for the phase shifts we obtain is given at the end of this section and a comparison is made with the dispersive predictions~\cite{Colangelo:2001df}.

\subsection{Cubic symmetry breaking}
\label{sec:systematic_error_cubic}
One distinguishing feature of our calculation is our choice of boundary conditions: we use G-parity instead of the standard periodic boundary conditions commonly used in other $\pi\pi$ scattering calculations. As discussed in Sec.~\ref{sec:lattice_detail} and in Ref.~\cite{Christ:2019sah}, G-parity boundary terms in the quark action break the usual cubic symmetry of our lattice action and cubic volume.  We will distinguish two possible effects of this breaking of cubic symmetry by the boundary conditions: the effects on the finite-volume eigenstates of the transfer matrix and the limitations on the symmetry properties of interpolating operators constructed from the quark fields. \par

Since the physical states in our finite volume are pions which obey cubically anti-periodic boundary conditions, we expect that the effects of this quark-level cubic asymmetry will be suppressed exponentially in the linear size of our spatial volume.  Local phenomena will not be affected by these boundary terms but only phenomena which span the entire volume.  This consideration should apply to the size of the corrections to the standard $\pi\pi$ finite-volume quantization condition, reducing these G-parity cubic symmetry breaking effects to the size of other finite-volume corrections.   \par

Of greater concern is our inability to confidently use cubic symmetry when interpreting the rotational quantum numbers of the states produced by our interpolating operators.  The G-parity breaking of cubic symmetry limits the selection of quark momenta that can be introduced when constructing interpolating operators resulting in operators which contain a mixture of representations of the cubic group.  The only solution to this problem which we have found is an empirical one: we must carefully construct pion interpolating operators to reduce the mixing of different cubic symmetry representations below the level that we are able to observe. \par

As described in Ref.~\cite{Christ:2019sah}, a numerical investigation on single-pion correlation functions has been performed on a smaller lattice, which suggests that if we construct these pion interpolating operators with a single choice of quark momentum assignment chosen from the set of allowed quark momenta (for example choice 1 of Appendix~\ref{sec:appendix_quark_momentum} of the present paper), then we observe a clear cubic symmetry breaking effect in the overall normalization of the corresponding two-point functions. In that appendix we also introduce a second choice with the same total momentum but with different assignments of quark momentum. We observe that if we construct our pion interpolating operators by averaging the two momentum choices, then the resulting cubic symmetry breaking becomes sub-statistical. Since this cubic symmetry breaking is purely due to the boundary condition, it will be further suppressed by the larger volume used in the current study, and is therefore negligible in this work. While the normalization of the two single pion operators carrying momenta which are related by cubic symmetry show small differences, the pion energies are always the same providing evidence for the assertion in the preceding paragraph that the spectrum of the transfer matrix shows only exponentially small cubic asymmetry. \par

We can also calculate the size of the cubic symmetry breaking in our $\pi\pi$ interpolating operators directly by studying the overlap between interpolating operators belonging to different representations of the cubic group.  We will focus on the stationary frame since in the moving frame calculation the $s$ and $d$-waves are coupled to each other even if we have exact cubic symmetry. If we have exact cubic symmetry, we can project all three groups of $\pi\pi$ interpolating operators (they are $\pi\pi(111,111)$, $\pi\pi(311,311)$ and $\sigma$) onto the $A_1$ and $T_2$ representations, which primarily map onto the $l=0$ and $l=2$ representations of the continuum rotation group, respectively, and for each group of operators these two representations will be orthogonal. However, if the symmetry group is reduced to the $D_{3d}$ group, the $T_2$ representation of $O_h$ group will no longer be irreducible and will contain the $A_1$ representation of $D_3d$ group, making a nonzero overlap between operators constructed according to the $A_1$ and $T_2$ representations of the $O_h$ group possible. The size of this overlap can then serve as a measure of the cubic symmetry breaking.  \par

We start by considering the two projections of the  $\pi\pi(111,111)$ operators and define their overlap as the average
\begin{equation}
    C_{a,a}^{I,T_2,A_1}(t) =  \frac{1}{T}\sum_{t_{\textrm{src}}}\langle O_{\pi\pi(111,111)}^{I,T_2}(t+\Delta+t_{\textrm{src}})^\dagger O_{\pi\pi(111,111)}^{I,A_1}(t_{\textrm{src}})\rangle\,,
\label{eq:cubic-breaking}
\end{equation} 
where $\Delta = 4$, as introduced earlier in Eq.~\eqref{eq:delta-conventions}.  Here and below we follow a convention similar to that introduced in Section~\ref{sec:pipi_energy} in which the labels $a$, $b$ and $c$ correspond to $\pi\pi(111,111)$, $\pi\pi(311,311)$ and $\sigma$ respectively.  If the cubic symmetry breaking effects are negligible, then $C_{a,a}^{I,T_2,A_1}(t)$ will be consistent with 0.  

Since we are interested in the size of these cubic symmetry breaking effects relative to the correlation functions from which we obtain our results, we will present the normalized correlator
\begin{equation}
    R_{a,a}^{I,T_2,A_1}(t) =  \frac{C_{a,a}^{I,T_2,A_1}(t)}{\sqrt{C_{a,a}^{I,T_2}(t)  C_{a,a}^{I,A_1}(t)}}\,.
\label{eq:normalized_overlap_ratio}
\end{equation} 
where $C_{a,a}^{I,T_2}(t)$ and $C_{a,a}^{I,A_1}(t)$ are defined by:
\begin{equation}
    C_{a,a}^{I,\mathcal{R}}(t) =  \frac{1}{T}\sum_{t_{\textrm{src}}}\langle O_{\pi\pi(111,111)}^{I,\mathcal{R}}(t+\Delta+ t_{\textrm{src}})^\dagger O_{\pi\pi(111,111)}^{I,\mathcal{R}}(t_{\textrm{src}})\rangle\,,
\end{equation} 
where $\mathcal{R} =A_1$ or $T_2$.  The ratio $R_{a,a}^{I,T_2,A_1}(t)$ in Eq.~\eqref{eq:normalized_overlap_ratio} provides an estimate of the fractional contamination in the correlation functions which we study that results from cubic symmetry breaking.  While we cannot be sure of the quantum numbers of the dominant state which propagates in the mixed correlator given in Eq.~\eqref{eq:cubic-breaking}, the ratio given in Eq.~\eqref{eq:normalized_overlap_ratio} divides by the time dependence implied by the arithmetic mean of what we expect to be the lowest masses in the $A_1$ and $T_2$ channels.

\begin{figure}
\centering
\begin{minipage}{0.45\textwidth}
  \centering
  \includegraphics[width=1.0\linewidth]{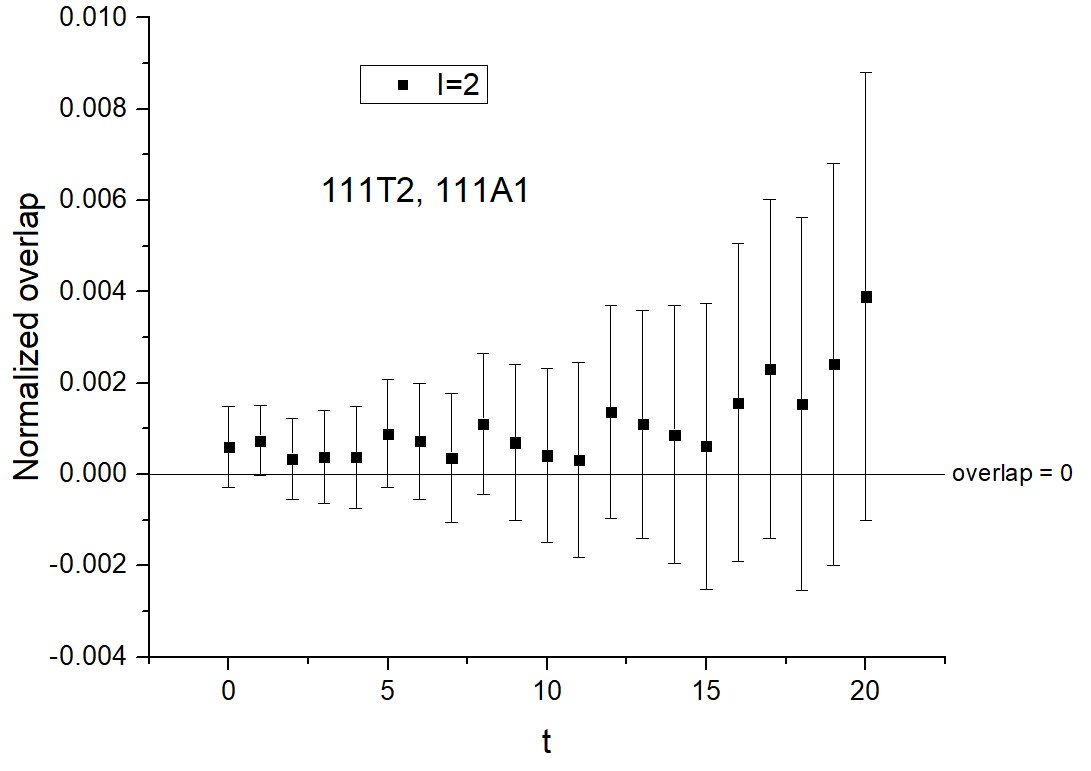}
  \label{fig:test1}
\end{minipage}%
\begin{minipage}{0.45\textwidth}
  \centering
  \includegraphics[width=1.0\linewidth]{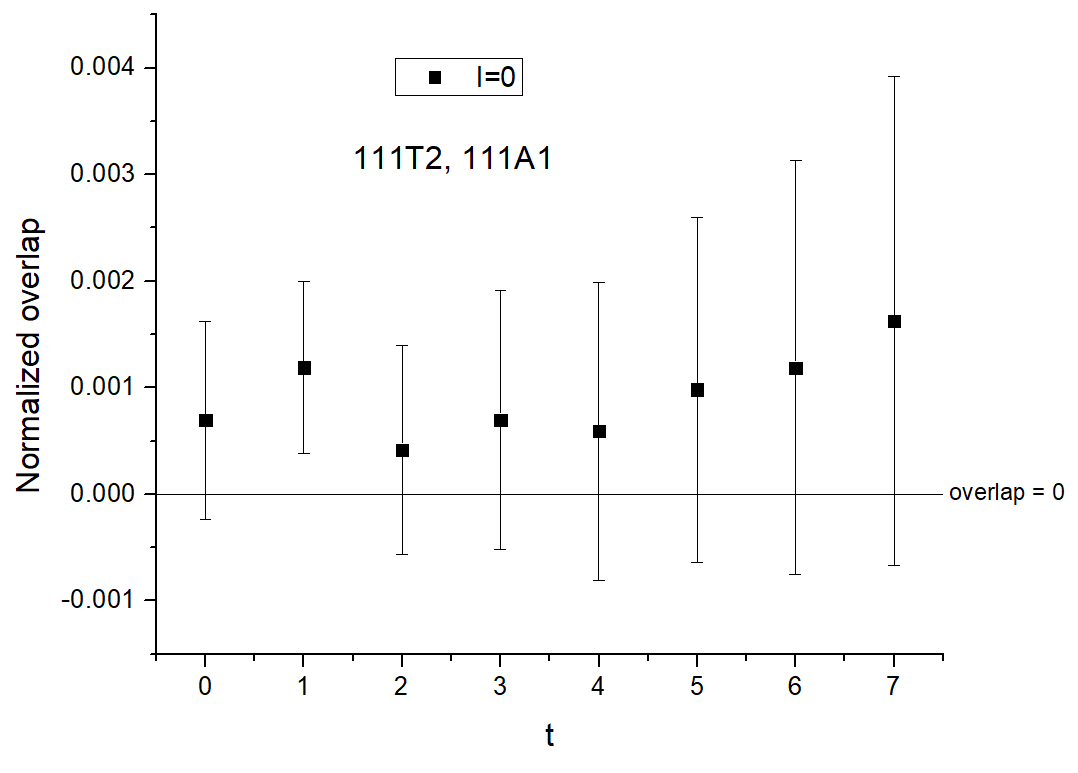}
  \label{fig:test2}
\end{minipage}
\caption{The overlap amplitudes between the $\pi\pi^{A_1}(111,111)$ and $\pi\pi^{T_2}(111,111)$ operators in the isospin $I=2$ (left) and $I=0$ (right) channels. The overlaps amplitudes are consistent with zero at all time separations which implies negligible cubic symmetry breaking for the $\pi\pi(111,111)$ interpolating operators.}
\label{fig:avg_overlap_111_111}
\end{figure}

\begin{figure}[tb]
\centering
\includegraphics[width=0.48\textwidth]{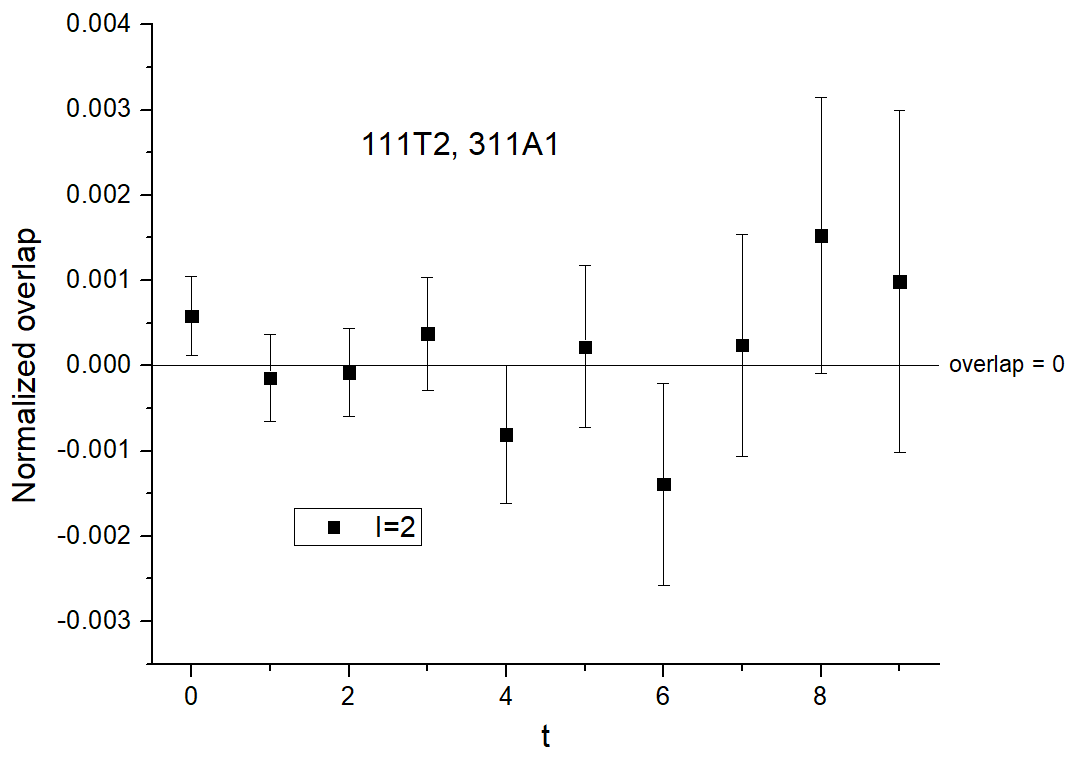}
\includegraphics[width=0.48\textwidth]{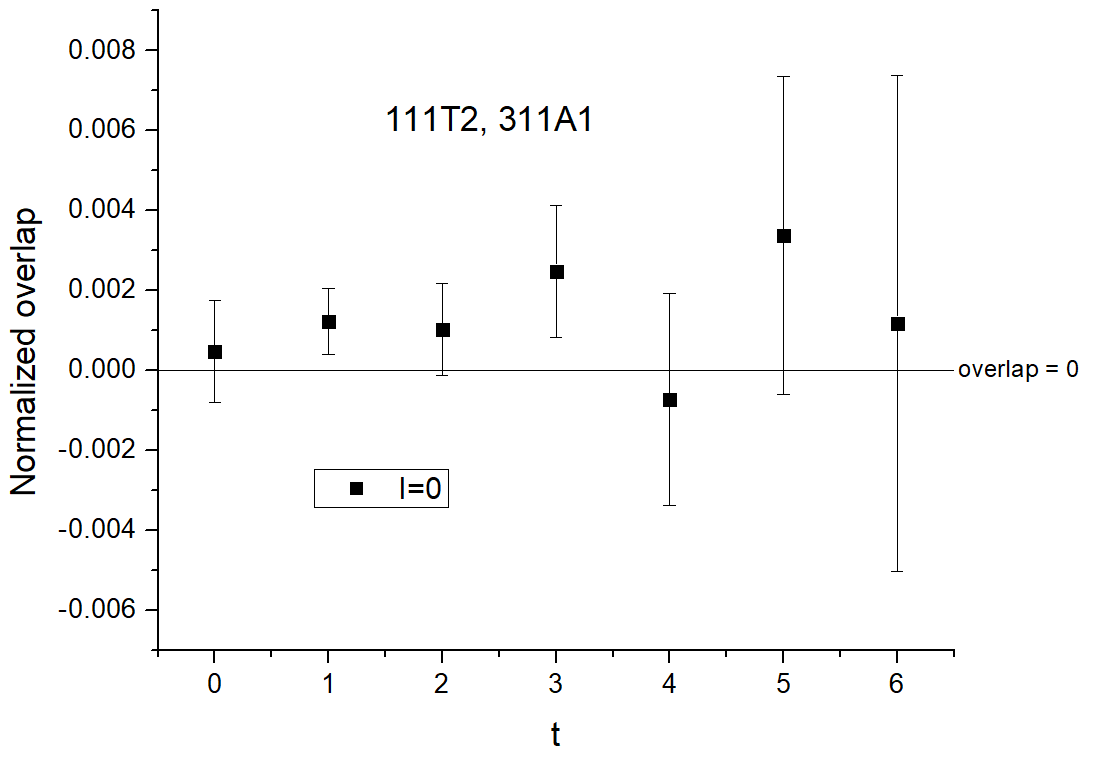}\\
\includegraphics[width=0.48\textwidth]{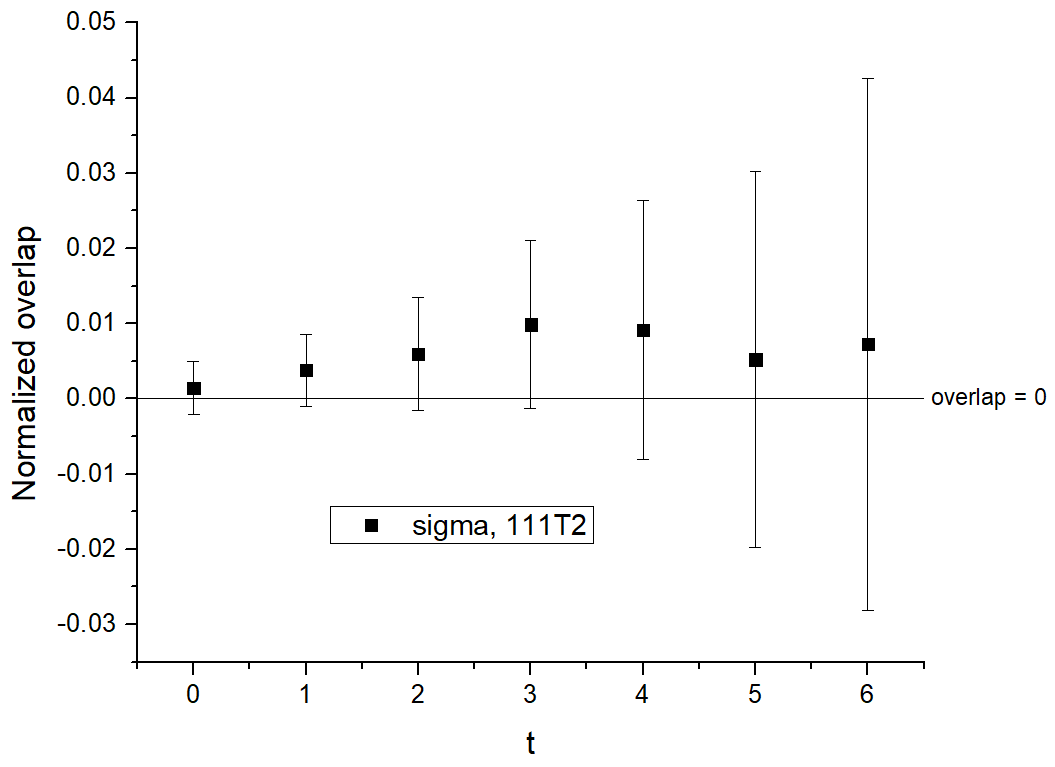}
\caption{Upper: the overlap amplitudes between the normalized $\pi\pi^{T_2}(111,111)$ operator and the normalized $\pi\pi^{A_1}(311,311)$ operator in the isospin $I=2$ (left) and $I=0$ (right) channels. Lower: the overlap amplitude between the normalized $\sigma$ operator and the normalized $\pi\pi^{T_2}(111,111)$ operator. All overlap amplitudes are consistent with zero with errors more than two orders of magnitude smaller than one which implies a negligible cubic symmetry breaking for these $\pi\pi(311,311)$ and $\sigma$ interpolating operators. \label{fig:avg_overlap_A1_T2}}
\end{figure}

The results are shown in Fig.~\ref{fig:avg_overlap_111_111}, where the left panel shows the normalized overlap amplitude for the $I=0$ channel, and the right panel shows that for the $I=2$ channel. Here and in the later graphs shown in Figure~\ref{fig:avg_overlap_A1_T2} we choose the time ranges to best present our results.  We exclude large times because the statistical errors become very large and would require a highly compressed scale to display.  However, in each case sufficiently large times are shown that the signal from the states which we study should be an important contributor to the correlation function, so the small size of $R_{aa}^{I,T_2,A_1}(t)$ for those later times implies at most a fractional percent contamination of our results from cubic symmetry breaking.  Of special interest is the size of $ R_{a,a}^{I,T_2,A_1}(t)$ for $t=0$ and 1 where the statistical errors are very small and cubic symmetry breaking is not visible at the tenth of a percent scale.

Having verified the approximate cubic symmetry of the $\pi\pi(111,111)$ operator, we next calculate the overlap amplitude between the $\pi\pi(111,111)$ operator in the $T_2$ representation and the other $(311,311)$ and $\sigma$ operators in the $A_1$ representation by evaluating:  
\begin{equation}
    C_{b/c,a}^{I,A_1,T_2}(t) = \sum_{t_{\textrm{src}}}\langle O_{\pi\pi(311,311)/\sigma}^{I,A_1}(t+\Delta+t_{\textrm{src}})^\dagger O_{\pi\pi(111,111)}^{I,T_2}(t_{\textrm{src}}) \rangle.
\end{equation}
The results are most easily interpreted if we again examine the normalized ratio
\begin{equation}
    R_{b/c,a}^{I,A_1,T_2}(t) =  \frac{C_{b/c,a}^{I,A_1,T_2}(t)}{\sqrt{C_{b/c,b/c}^{I,A_1}(t)  C_{a,a}^{I,T_2}(t)}}\,.
\end{equation} 

The results are shown in Fig.~\ref{fig:avg_overlap_A1_T2}, where the upper panel shows the overlap between the $\pi\pi^{A_1}(311,311)$ and $\pi\pi^{T_2}(111,111)$ interpolating operators in the $I=0$ and $I=2$ channels, and the lower panel shows the overlap between the $I=0$, $\sigma$ and $\pi\pi^{T_2}(111,111)$ interpolating operators. Similar to Fig.~\ref{fig:avg_overlap_111_111}, all three overlap amplitudes are consistent with 0 at the fractional percent level, which suggests both the $\pi\pi(311,311)$ and the $\sigma$ operators obey approximate cubic symmetry. \par

Based on the above results, we conclude that if we construct the pion interpolating operators by averaging the two sets of quark momentum assignments as described in the appendix, we can achieve accurate cubic symmetry at the meson level, despite the symmetry breaking at the quark level.  We therefore do not assign any systematic error arising from cubic symmetry breaking. \par



\subsection{Finite lattice spacing}

Fundamental to the connection between the scattering phase shifts and the two-particle finite-volume energies is the recognition that it is the interaction between the particles, described by a non-zero scattering phase shift, that causes the two-particle energy in finite volume to be shifted away from the simple spectrum of non-interacting particles in a box.  When adopting formulae to determine the scattering phase shifts from the two-particle finite-volume energies in Section~\ref{sec:phase_shift} we were careful to preserve this connection for non-zero lattice spacing.  \par

Specifically Eq.~\eqref{equ:def_pion_mom_k} determines the relative center of mass momentum $k$ between the two pions in the finite-volume $\pi\pi$ ground state that enters L\"uscher's quantization condition as a function of the energy difference $\Delta E$ between the finite volume $\pi\pi$ energy and that of two non-interacting pions.  This difference would vanish in the absence of interactions, even at non-zero lattice spacing.  We then assign a relative systematic error to this measured energy difference that is of the same size as is found for other similar quantities computed on this ensemble for which a continuum limit has been evaluated.  Thus, we use 
\begin{equation}
\frac{\mathrm{Error(\Delta E)}}{\Delta E} = c a^2,
\label{eq:DeltaE-a2}
\end{equation}
where $c$ is chosen from the finite lattice spacing errors reported in Ref.~\cite{Blum:2014tka}.  In detail we use the average ChPTFV value of the magnitudes of $c_f^{ID}$, $c_{f^{(K)}}^{ID}$, $c_{w_0,a}^{ID}$ and $c^{ID}_{\sqrt{t_0},a}$, given in Table XVII in that paper which are the four coefficients which describe the $a^2$ finite lattice spacing errors for these four different physical quantities computed with the same lattice action and gauge coupling as used here. This gives a relative error of 1.6\% for $\Delta E$ which we round up to 2\%. \par

Note this relative error is usually a small quantity, and therefore a small absolute error when compared with $E_{\pi}$ and $E_{\pi\pi}$ since we are calculating the $\pi\pi$ scattering phase shift at relatively low energies (near the kaon mass). The error determined from Eq.~\eqref{eq:DeltaE-a2} is then propagated in the standard way to obtain the $O(a^2)$ error for the scattering phase shift which we use for each of our four values of total momentum. \par

\subsection{Finite volume}

Finite volume affects the energy of $\pi\pi$ states in two ways. The first effect results in the quantized finite-volume energies, is described by the L\"uscher quantization condition and can be viewed for large $L$ as a power law effect. The second effect falls exponentially with the system size and is caused by the interaction radius being a finite fraction of the system size or, equivalently, the effect of off-shell singularities when the Poisson summation formula is used to estimate finite-volume effects.  This second effect is usually much smaller than the first and is the source of the systematic error considered here. This exponentially suppressed correction for the $I=2$ channel for periodic boundary conditions can be formulated as~\cite{Bedaque:2006yi}:
\begin{equation}
    (k\cot\delta(s))_L = (k\textrm{cot}\delta(s))_\infty + \Delta_{FV}\,,
\end{equation}
where 
\begin{equation}
    \Delta_{FV} = -\frac{m_\pi}{\sqrt{2\pi}}\sum_{\Vec{n},|\Vec{n}|\neq 0}\frac{e^{-|\Vec{n}|m_\pi L}}{\sqrt{|\Vec{n}|m_\pi L}}\left[1-\frac{227}{24}\frac{1}{|\Vec{n}|m_{\pi} L}+...\right]
\end{equation}
for the case of near-zero relative momentum.  According to Fig. 2 from Ref~\cite{Bedaque:2006yi}, this correction introduces an approximate 1\% relative error in the scattering length for a volume with periodic boundary conditions but the same size and physical parameters as the volume with G-parity boundary conditions studied here.  \par

For our G-parity boundary condition lattice, since the pion satisfies anti-periodic boundary conditions, the formula for $\Delta_{FV}$ has to be modified as follows:
\begin{equation}
    \Delta_{FV} = -\frac{m_\pi}{\sqrt{2\pi}}\sum_{\Vec{n},|\Vec{n}|\neq 0}\frac{(-1)^{n_x+n_y+n_z}e^{-|\Vec{n}|m_\pi L}}{\sqrt{|\Vec{n}|m_\pi L}}\left[1-\frac{227}{24}\frac{1}{|\Vec{n}|m_{\pi} L}+...\right]\,.
\end{equation}
\par
This leads to a relative error of approximately 0.6\%.  We round this number up to 1\% and adopt it as an estimate of the finite volume effects for our more general case which includes the $I=0$ channel, non-zero $\pi\pi$ relative momentum and non-zero total momentum.

\subsection{Unphysical kinematics}
\label{sec:unphykin}
The pion mass which we measured on this ensemble is 142.3 MeV, which is 5\% larger than our choice for the physical pion mass (135 MeV) at which we wish to determine the scattering phase shifts.  We deal with this pion-mass mismatch in two steps. In the first step we shift the $\pi\pi$ energy at which we quote the phase shift, as has been discussed in Sec.~\ref{sec:phase_shift}, by expressing the phase shift as a function of the two-pion relative momentum $k$ in the center-of-mass system and then identifying this value of $k$ with a $\pi\pi$ energy using the physical pion mass.  We view this correction, which will be large for energies near the $\pi\pi$ threshold, as the most important effect of this pion-mass mismatch.  In the second step we account for the remaining effects of this pion mass mismatch as a systematic error in our result for the phase shift.\par

\begin{figure}
\centering
\begin{minipage}{0.45\textwidth}
  \centering
  \includegraphics[width=1.0\linewidth]{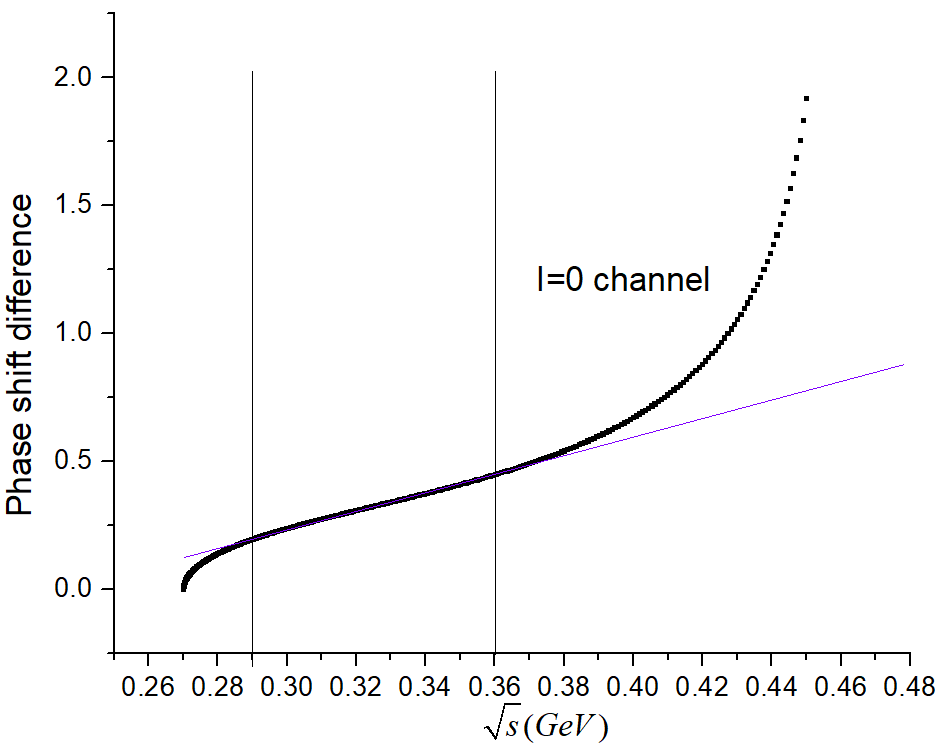}
\end{minipage}%
\begin{minipage}{0.45\textwidth}
  \centering
  \includegraphics[width=1.0\linewidth]{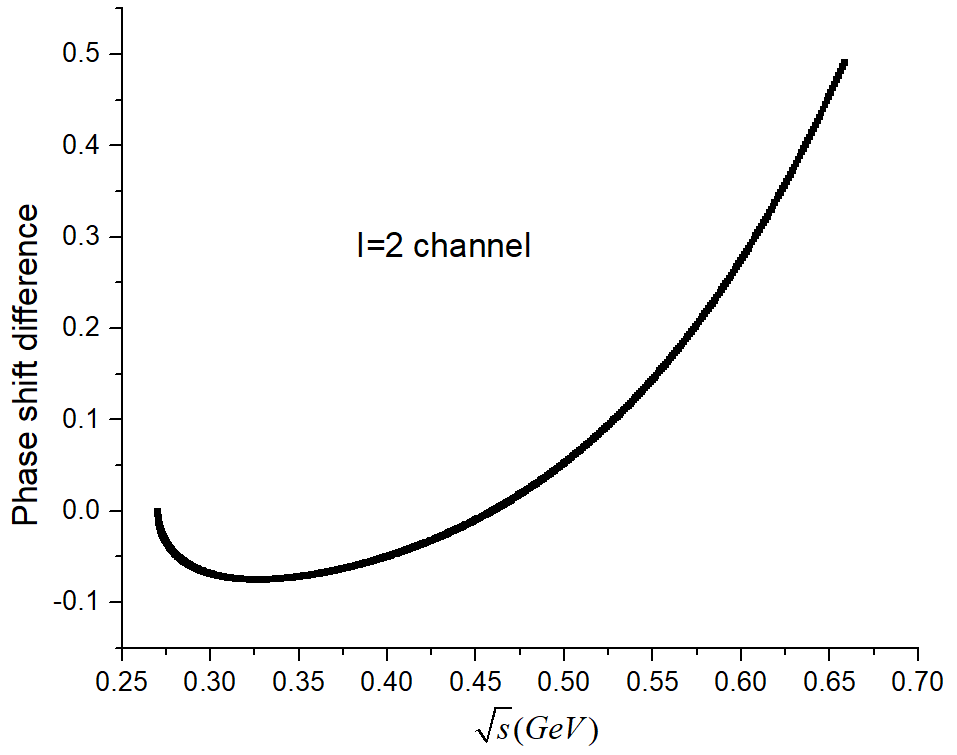}
\end{minipage}
\caption{A ChPT calculation of the difference between the scattering phase shift evaluated at the physical pion mass and the pion mass calculated for our ensemble, $\delta_I(m_\pi = 142~\mathrm{MeV}) - \delta_I(m_\pi = 135~\mathrm{MeV})$, as a function of $\sqrt{s}$ defined in Eq.~\eqref{equ:def_sqrt_s} and shown in degrees. For the $I=0$ channel, at large $\sqrt{s}$, the ChPT calculation begins to break down while at lower $\sqrt{s}$, the relation is approximately linear, which is consistent with the dispersive prediction. The straight line is a linear fit to the ChPT result in the region 290\,MeV $\leq \sqrt{s} \leq$ 360\,MeV.}
\label{fig:unphysical_error}
\end{figure}

We estimate the remaining pion-mass-mismatch error by using ChPT to calculate the difference between the scattering phase shift evaluated at these two different pion masses but at the same value for $k$. The NLO ChPT prediction for the scattering amplitude of both the I=0 and I=2 channels for small relative momenta are listed in Appendix~\ref{sec:appendix_chpt_phase_shift}, and the predicted phase shift difference as a function of $\sqrt{s}$ is plotted in Fig.~\ref{fig:unphysical_error}. \par

\begin{table}
 \begin{tabular}{ | c | c | c | c | c | c | c |}
 \hline
 $P_{tot}$ & I & $\sqrt{s}$ & $\delta\phi_{\textrm{unphy}}$ & Method \\
 \hline
 $(0,0,0)\frac{\pi}{L}$ & 0 & 471.0(1.5) & $0.833^\circ$ & Linear extrapolation \\ 
 \hline
 $(2,0,0)\frac{\pi}{L}$ & 0 &  435.1(3.8) & $0.708^\circ$ & Linear extrapolation \\ 
 \hline
 $(2,2,0)\frac{\pi}{L}$ & 0 &  365.6(3.4) & $0.474^\circ$ & ChPT \\ 
 \hline\hline
 $(0,0,0)\frac{\pi}{L}$ & 2 &  565.4(5) & $0.181^\circ$ & ChPT \\ 
 \hline
 $(2,0,0)\frac{\pi}{L}$ & 2 & 479.1(3)  & $0.025^\circ$ & ChPT \\ 
 \hline
 $(2,2,0)\frac{\pi}{L}$ & 2 & 386.7(4) & $0.057^\circ$ & ChPT \\ 
 \hline
 $(2,2,2)\frac{\pi}{L}$ & 2 &  271.5(4) & $0.020^\circ$ & ChPT \\ 
 \hline
 \end{tabular}
 \caption{The assigned values for the systematic error resulting from our unphysical pion mass and the methods used to determine them.}
 \label{table:unphysical_error}
\end{table}

There is a remaining uncertainty in this approach that must be resolved. The ChPT calculation is only valid for small $k$, a condition not valid for our stationary calculation, which results in the rapid rise of the phase shift difference in the $I=0$ channel when $\sqrt{s} > 380$ MeV. Similar behavior is seen also in the $I=2$ channel, although the breakdown appears to occur more slowly as a function of $\sqrt{s}$, suggesting the ChPT result for this channel can be considered sufficiently reliable for energies in our range of interest. We modify our systematic error determination for $\sqrt{s} \approx m_K$ to resolve this issue. Notice that the dispersive prediction, whose range of validity is expected to extend above that of ChPT, shows a relation between the phase shift and $\sqrt{s}$ that is close to linear for a broad range of $\pi\pi$ energy, up to and including the kaon mass~\cite{Colangelo:2001df}.  This suggests that the relation between the $\sqrt{s}$-dependence of the difference between the two phase shifts with different pion masses will also be linear. 
Thus, we can use ChPT to determine the phase shift difference at relatively small $\pi\pi$ energy and then linearly extrapolate to higher energies.  Here we perform a linear fit to the ChPT prediction in the range 290 MeV$ \le \sqrt{s} \le 360$ MeV. For the $I=2$ channel the ChPT result remains linear over the range of our data so we simply use the ChPT value to determine this I=2 systematic error. Our assignments of these unphysical-pion-mass errors are listed in Table.\ref{table:unphysical_error}. \par

\subsection{Higher partial wave correction}
In Sec.~\ref{sec:Luscher_formula_derivation} where we derive L\"uscher's quantization condition, we assume that the contributions to the $s$-wave phase shift from higher partial waves are negligible. For the stationary frame calculation where the leading order correction comes from the partial wave with $l=4$, this contribution should be highly suppressed because of the large angular momentum. For moving frame calculations, the leading order correction comes from D-wave where the neglected phase shift may be larger. For completeness, we should estimate the contribution to the systematic error from neglecting this D-wave phase shift.\footnote{We thank the referee for suggesting that we include an estimate of this source of systematic error}

According to Eq.~105 from Ref.~\cite{Rummukainen:1995vs}, under the condition where the D-wave phase shift is small, the correction to the $s$-wave phase shift after including the D-wave phase shift is:
\begin{equation}
    \Delta_0(p) = -\frac{m_{20}^2}{m_{00}^2+1}\delta_2(p)
\end{equation}
where 
\begin{equation}
    m_{00} = \frac{1}{\pi^{3/2}\gamma q}Z_{00}^{\vec{d},\gamma}(1,q^2)
\end{equation}
and
\begin{equation}
    m_{20} = -\frac{1}{\pi^{3/2}\gamma q^3}Z_{20}^{\vec{d},\gamma}(1,q^2)
\label{eq:m20}
\end{equation}
Here we generalize the numerical recipes for evaluating L\"uscher's zeta function given in Ref.~\cite{Yamazaki:2004qb} to the general cases where $l \neq 0$ to evaluate $Z_{20}^{\vec{d},\gamma}(1,q^2)$ in Eq.~\eqref{eq:m20}. The D-wave phase shift, $\delta_2(p)$, can be evaluated by assuming that the energy in the center of mass frame is small in our moving frame calculation so that we can approximate the D-wave phase shift by using the D-wave scattering length:
\begin{equation}
    \delta_2^I(p) = \left(\frac{p^2}{m_\pi^2}\right)^2\frac{p}{\sqrt{p^2+m_\pi^2}}a_2^I
\end{equation}
where $a_2^I$ is the D-wave scattering length for isospin-I. The values we used are listed in Tab.~\ref{tab:d-wave_scattering_length}, which were obtained from Ref.~\cite{donoghue2014dynamics}, Table VI-4.

We can then calculate the $s$-wave phase shift correction, $\Delta_0(p)$, and the results are listed in Tab.~\ref{tab:s-wave_phase_shift_correction_from_d-wave}.

\begin{table}[h!]
\centering
 \begin{tabular}{ | c | c |}
 \hline
 $m_\pi a_2^{I=0}$ & $17\times 10^{-4}$ \\
 \hline
 $m_\pi a_2^{I=2}$ & $1.3\times 10^{-4}$ \\
 \hline
 \end{tabular}
 \caption{The D-wave scattering length for isospin 0 and 2 channel}
 \label{tab:d-wave_scattering_length}
\end{table}
\begin{table}
\centering
 \begin{tabular}{ | c | c | c | c | c |}
 \hline
 $P_{tot}$ & I & $\sqrt{s}$ & $\delta_0$ & $\Delta_0(p)$ \\
 \hline
 $(2,0,0)\frac{\pi}{L}$ & 0 &  435.1 & $24.0^\circ$ & $-0.05^\circ$ \\ 
 \hline
 $(2,2,0)\frac{\pi}{L}$ & 0 &  365.6 & $18.0^\circ$ & $-0.27^\circ$ \\ 
 \hline\hline
 $(2,0,0)\frac{\pi}{L}$ & 2 & 479.1  & $-7.96^\circ$ & $-0.01^\circ$ \\ 
 \hline
 $(2,2,0)\frac{\pi}{L}$ & 2 & 386.7 & $-4.48^\circ$ & $-0.02^\circ$ \\ 
 \hline
 $(2,2,2)\frac{\pi}{L}$ & 2 &  271.5 & $-0.32^\circ$ & $0^\circ$ \\ 
 \hline
 \end{tabular}
 \caption{The $s$-wave phase shift correction from including a non-zero D-wave phase shift. The D-wave phase shift is approximated using D-wave scattering length. Note the D-wave phase shift does not contribute for the (2,2,2) case because of symmetry. }
 \label{tab:s-wave_phase_shift_correction_from_d-wave}
\end{table}
\subsection{Excited state contamination}

In Sec.~\ref{sec:pipi_energy} we tried different fit ranges to find a balance between minimizing the excited state contamination error and the statistical error.  For the preferred fit ranges shown in Tables~\ref{tab:I=2} and \ref{tab:I=0}, there is no obvious evidence that the neglected excited states give a significant contribution to the fitting result, as can be seen in the energy plots in Fig.~\ref{Fig:stationary} and Fig.~\ref{Fig:moving}.  As discussed in Ref.~\cite{Wang:2019nes}, the presence of an apparent plateau in the fitted energy as a function of the lower limit of the fitting range does not imply that we can neglect excited state contamination since the noisy data may make it difficult to find the ``true'' plateau before the effects of an omitted excited state are obscured by the noise. An important example is our previous result for the ground state $I=0$ $\pi\pi$ energy obtained from a portion of the current ensemble~\cite{Bai:2015nea} where we significantly underestimated this error. \par

A potentially more robust approach to estimating the error from omitted excited states than examining the dependence of the fitted results on $t_{\textrm{min}}$ is to explicitly include an extra excited state in the fit and to determine the size of the systematic shift in the resulting ground-state energy, an approach we call performing an extra-state fit.  Of course, in most cases such an extra-state fit will require some additional assumptions since if this fit could have been easily performed we would have included this extra state in our preferred fit.  However, while such extra assumptions may have been inappropriate for the determination of the preferred central value, their introduction may be a reasonable approach to estimate an excited-state systematic error. \par

The difficulty associated with an extra state fit can be illustrated by the single-operator analysis used in Ref.~\cite{Bai:2015nea} to determine the energy of the stationary $I=0$, ground state.  Table~\ref{tab:1op2state} lists the results from extra-state fits to the single-operator data presented in that earlier paper.  In this case our preferred single-state fit gave a ground state energy $E_0 = 0.3606(74)$ using the fitting range $6-25$.  We can see that for $t_{\textrm{min}} < 6$ adding the extra state does not give a ground state energy that can be resolved from the result of the  preferred single-state fit.  In addition the overlap factors between the operator and the first excited state are consistent with 0.  We cannot perform the extra-state fit within the same fit range as the one we chose for the preferred single-state fit ($6-25$) since the fitting procedure does not converge. We conclude that such (unconstrained) extra-state fits can be misleading for data containing multiple nearby states with limited statistics and/or a rapid loss of signal to noise as a function of time where the available fitting range is insufficiently long to adequately distinguish the energy separations among the multiple nearby states. \par

\begin{table}
 \begin{tabular}{ | c | c | c | c |}
 \hline
 $t_{\textrm{min}}$ & 3 & 4 & 5 \\
 \hline
$A_0$ & 0.1589(107) & 0.1555(133) & 0.1538(48)\\
\hline
$A_1$ & 0.05974(7064) & 0.02679(2305) & $7822(6959)\times 10^{3}$\\
\hline
$E_0$ & 0.3666(124) & 0.3643(133) & 0.3606(75) \\
\hline
$E_1$ & 1.207(725) & 0.8429(4071) & 4.835(173)\\
\hline
$B$ & $-753(2366)\times 10^{-7}$ & $-784(2362)\times 10^{-7}$ & $-793(2223)\times 10^{-7}$\\
\hline
 \end{tabular}
 \caption{Single-operator two-state fit results with fit ranges $t_{\textrm{min}} - 25$ for the stationary, $I=0$ case.  The amplitudes $A_0$ and $A_1$ are the overlap factors between the operator and the ground and first excited state, $E_0$ and $E_1$ are their energies while $B$ is the around-the-world constant introduced in an analogous fashion as in Eqs.~\eqref{eq:fit-form-single-op} and \eqref{eq:fit-form}.  The fit procedure fails when $t_{\textrm{min}} \geq 6$. The single-state fit with fit range $6-25$ gives $E_0 = 0.3606(74)$ which is consistent with the ground state energy in all three extra-state fits.}
  \label{tab:1op2state}
\end{table}

One strategy to resolve this problem is to fix one or more of the parameters in the extra-state fit. The first parameter we might fix is the energy of the extra state. 

A prediction of this finite-volume energy can be obtained by finding the intersects between L\"uscher's formula and a phenomenological model, or a fit to our lattice results, for the scattering phase shift as a function of energy. For simplicity we chose to compute this estimate using the dispersive predictions of Ref.~\cite{Colangelo:2001df}. Unfortunately we found that fixing this extra-state energy alone typically does not solve the problem: sometimes the fitting procedure continues to fail while in those cases where the fit converges, it gives a ground-state energy whose statistical error is several times larger than the statistical error on the result from the preferred fit, which suggests that such an approach may over-estimate the excited state error, conflating it with the statistical error.   \par

Thus, further assumptions are needed to obtain a statistically meaningful estimate of this excited-state contamination error.  Because of the very different pattern of operator-eigenstate overlaps we will adopt two different strategies, one for the $I=0$ stationary $\pi\pi$ state and the second for the remaining seven cases: the three $I=0$ calculations with non-zero total momenta and all four $I=2$ calculations. The method applied for the latter we will refer to as ``method A", and that applied in the special $I=0$ stationary case as ``method B". \par

For the three moving-frame $I=0$ and all the $I=2$ calculations, the operator-eigenstate overlap matrix is close to diagonal and the two-point function of a given operator with itself is well described by one exponential coming from a single energy eigenstate.  We call the operator which couples primarily to the ground state the ``ground-state operator", and the operator whose two-point function is dominated by the $n^{\textrm{th}}$ excited state the ``$n^{\textrm{th}}$ excited operator''.  We can then make the reasonable assumption, that because of the small value of the overlap factors between the ground-state operator and those excited states that we include in the preferred fit, the extra-state contribution to these small overlap factors can be neglected. That means that we can focus on the Green's function constructed from the ground-state operator only where the most important effect of an omitted extra state is to change the two parameters (overlap factor and energy) associated with the ground-state. \par

The argument above suggests that we can perform a fit to the two-point function constructed from the ground-state operator in which the number of states included is one more than the number in the preferred fit, and that we can fix the overlap factors between the ground-state operator and the excited states in that fit to those already determined by the preferred fit.  Since these overlap factors are small, we can also fix the energies of those excited states to their values from the preferred fit. To summarize, if our preferred fit involved $n$ operators and $m$ states, we will fit the single ground-state operator two-point function to an expression which includes $m+1$ states: the $m$ states which appear in our preferred fit and the new extra state.  The choice of whether or not to include an around-the-world constant is the same as the choice made for the preferred fit.  In this new fit, we fix the extra-state energy to that given by the dispersive prediction and the parameters that are associated with the first $m-1$ excited state energies and their overlap amplitudes with the ground state to those values determined by the preferred fit. \par

It should be noticed that since the preferred fit and this extra-state fit are performed on data constructed using the same resampling method, these parameters should be fixed sample-by-sample.  Thus, for each bootstrap/jackknife sample, the fixed parameters we used will vary: they are the ones obtained from exactly the same sample used when we perform the preferred fit. That leaves us with three (or four) free parameters to be determined in this extra-state fit: the overlap between the ground-state operator and the ground state, the ground-state energy, the overlap between the ground-state operator and the $m^{\textrm{th}}$ excited state (the extra state) and possibly an around-the-world constant. The results are shown in Table~\ref{tab:I=2_exc_err} and Table~\ref{tab:I=0_exc_err}.  \par

Unfortunately, as described this method gives us a statistical error on the ground-state energy that is still much larger than that given by the preferred fit.  This may be the understandable consequence of trying to obtain information about two states from a single operator Green's function, a case where the number of states exceeds the number of operators. Some improvement must be made. One possibility is to fix not only the energy of the extra state, but also its overlap factor with the ground-state operator. In contrast to the  energy of the excited state, we do not have external information that could give us a reasonable theoretical value for this overlap amplitude.  However, we can argue that the result from the above fit gives us a reasonable estimate for the size of this overlap factor.  \par

Thus, in the notation introduced in Eq.~\eqref{eq:fit-form} we view  $[A_{am}-\delta_{A_{am}}, A_{am}+\delta_{A_{am}}]$ as a reasonable interval for the overlap amplitude between the operator $O_a$ (using $a$ to label the ground-state  operator) and the $m^{\textrm{th}}$ excited state (our extra state). Here $\delta_{A_{am}}$ is the statistical error on the quantity $A_{am}$ found in the fit described above.  We then perform three fits, where the only difference between them and the one above is that we fix the overlap factor between the ground-state operator and the $m^{\textrm{th}}$ excited state using $A_{am}, A_{am}-\delta_{A_{am}}$ and $A_{am}+\delta_{A_{am}}$ (Notice that for each bootstrap/jackknife sample, this overlap factor will \textbf{NOT} vary). We then calculate the difference between these three ground-state energies and the ground-state energy obtained from the preferred fit.  The largest difference gives our estimate of the excited-state error. The results are shown in Table~\ref{tab:I=2_exc_err} for the $I=2$ channel and Table~\ref{tab:I=0_exc_err} for the $I=0$ channel. As mentioned above, we refer to this method of estimating the excited-state systematic error, based on the near-diagonal character of the matrix of operator-eigenstate overlap amplitudes, as  ``method $A$''. \par

While this method gives a statistically precise result for the excited-state contamination error for the three moving-frame $I=0$ calculations and all four $I=2$ calculations, the diagonal pattern of the operator-eigenstate overlap matrix upon which it is based is not found for the stationary $I=0$ channel, where the overlap factors between the various normalized operators and eigenstates have similar sizes.  Consequently, we cannot apply the method $A$ above in this case. We therefore adopt a different approach, referred to as ``method $B$''. While we are unable to perform a convergent extra-state fit (3-operator-3-state fit) with the fit range that we chose for the preferred fit (6-15), we can perform that fit using a fit range with a smaller $t_{\textrm{min}}$, provided the extra-state energy is fixed.  For convenience, we again give the extra-state energy the value predicted by the dispersive result for the $I=0$ phase shift.  This fit gives us values for the overlap amplitudes for the extra state and each of the three operators being studied.  We then perform a 3-operator-3-state fit with the same fit range as the preferred fit (6-15), while fixing all the information about the extra excited state to that obtained from the fit with the decreased value of $t_{\textrm{min}}$. Note that in this case we did not observe a significant increase in the error on the ground-state energy after applying this procedure, hence it was not necessary to perform a second step holding these overlap factors fixed to the extrema of their error bars as was the case for method A above. 

To summarize, the eight parameters that are allowed to vary in this final fit are the six overlap amplitudes between the three operators and the ground and first excited states, and the energies of these two states. We then calculate the energy difference (for present purposes labeled as $\Delta E$) between the ground-state energy obtained from this extra-state fit and the preferred fit and the statistical error on this difference ($\delta \Delta E$).  We then use $\delta E_{\mathrm{exc}} = \Delta E + \delta \Delta E$ as our estimate of the systematic error on the ground-state energy resulting from excited-state contamination.  The results are listed in the right-most column of Table~\ref{tab:I=0_exc_err}. This small estimate for the excited state error for the $I=0$ stationary case is supported by the determinant test result in Sec~\ref{sec:pipi_energy} where the normalized determinant of the three dimensional matrix of the Green's function is consistent with zero, suggesting that the $3^{\textrm{rd}}$ state is difficult to resolve with a fit range where $t_{\textrm{min}}$ is larger than 5.\par

\begin{table}[ht]
\begin{tabular}{ | c | c | c | c | c |}
  \hline
   I=2 channel & $P_{tot} = (2,2,2)\frac{\pi}{L}$ & $P_{tot} = (2,2,0)\frac{\pi}{L}$ & $P_{tot} = (2,0,0)\frac{\pi}{L}$ & $P_{tot} = (0,0,0)\frac{\pi}{L}$\\
  \hline
  Fit range & 10-25 & 12-25 & 11-25 & 10-25 \\
  \hline
  Fit strategy & 1op-4state & 1op-4state & 1op-4state & 1op-3state\\
  \hline
  $A_{a0}$ &  0.3935(10) &  0.2768(8) &  0.1934(3) &  0.4206(13)   \\
\hline
  $A_{a1}$ &  \pmb{0.004684(565)} &  \pmb{0.007011(548)} &  \pmb{0.009301(455)} &   \pmb{0.01240(1021)}  \\
\hline
  $A_{a2}$ &  \pmb{.001209(1890)} &  \pmb{0.005350(1812)} &  \pmb{0.005249(1482)} &  \pmb{0.1641(1176)}  \\
\hline
  $A_{a3}$ & \pmb{-0.07301(3318)} &  \pmb{0.03847(3424)} & \pmb{-0.00001(3132)} &   -  \\
\hline
  $E_0$ &  0.3981(4) &  0.4000(4) &  0.4045(3) & 0.4152(6)   \\
\hline
  $E_1$ &  \pmb{0.5453(7)} &  \pmb{0.5480(10)} &  \pmb{0.5514(9)} &   \pmb{0.7128(170)}  \\
\hline
  $E_2$ &  \pmb{0.6902(28)} &  \pmb{0.6874(40)} &  \pmb{0.6916(48)} &  \pmb{0.9169(0.0000)}   \\
\hline
  $E_3$ &  \pmb{0.6923(0.0000)} & \pmb{ 0.6934(0.0000)} &  \pmb{0.8047(0.0000)} &   -  \\
\hline
  $B_{aa}$ &  $8118(71)\times 10^{-9}$ &  $4036(35)\times 10^{-9}$ &  $1981(16)\times 10^{-9}$ &  $9380(152)\times 10^{-9}$   \\
\hline
  $p$-value & 0.078 & 0.157  & 0.268 &  0.683 \\
  \hline
  $\delta E_{\textrm{exc}}$ & 0.0007 & 0.0004 & 0.0002 & 0.0006\\
  \hline
 \end{tabular}
  \caption{Results from the fits used to determine the excited state error for the ground-state energies in the $I=2$ channel. We use method $A$ described in the text to estimate the error for all four total momenta.  As described in the text, these results are obtained in two stages.  First all of the quantities in bold font except for the extra-state amplitudes ($A_{a3}$ in columns 2-4 and $A_{a2}$ in column 5) are held fixed for the first stage fits.  The results from the first stage fits for these four amplitudes are shown in this table.  For the second stage fits these four amplitudes are also held fixed at their central values and at their central values plus and minus the statistical error shown.  The largest difference between the resulting ground-state energy and the ground-state energy obtained in the preferred fit is shown in the final row and is our estimate of the excited state error.}
  \label{tab:I=2_exc_err}
\end{table}

\begin{table}
\begin{tabular}{ | c | c | c | c | c | }
  \hline
   I=0 channel & $P_{tot} = (2,2,2)\frac{\pi}{L}$ & $P_{tot} = (2,2,0)\frac{\pi}{L}$ & $P_{tot} = (2,0,0)\frac{\pi}{L}$ & $P_{tot} = (0,0,0)\frac{\pi}{L}$\\
  \hline
  Fit range & 6-12 & 8-15 & 7-15 & 6-15\\
  \hline
  Fit strategy & 1op-4state & 1op-4state & 1op-4state & 3op-3state\\
  \hline
   $A_{a0}$ &  0.3861(25) &  0.2633(36) &  0.1760(37) &  0.3638(50)   \\
\hline
  $A_{a1}$ & \pmb{-0.02647(391)} & \pmb{-0.04909(1123)} & \pmb{-0.05431(776)} &   -0.1958(120)  \\
\hline
  $A_{a2}$ & \pmb{-0.01354(312)} & \pmb{-0.03005(552)} & \pmb{-0.02450(274)} &  \pmb{0.01537(2140)}   \\
\hline
  $A_{a3}$ & -0.06485(3778) & -0.03121(4065) & -0.04877(3298) &   -  \\
\hline
  $A_{b0}$ & - & - & - &   0.003244(368)  \\
\hline
  $A_{b1}$ & - & - & - &   0.03248(709)  \\
\hline
  $A_{b2}$ & - & - & - &  \pmb{0.05802(618)}  \\
\hline
  $A_{c0}$ & - & - & - &   $-4396(72)\times 10^{-7}$  \\
\hline
  $A_{c1}$ & - & - & - &    $-3644(264)\times 10^{-7}$ \\
\hline
  $A_{c2}$ & - & - & - &  \pmb{$9394(5883)\times 10^{-8}$}  \\
\hline
  $E_0$ &  0.3972(12) &  0.3898(22) &  0.3769(39) &   0.3474(13)  \\
\hline
  $E_1$ &  \pmb{0.5264(37)} &  \pmb{0.5148(91)} &  \pmb{0.5032(75)} &  0.5484(208)   \\
\hline
  $E_2$ &  \pmb{0.6881(93)} &  \pmb{0.6788(252)} &  \pmb{0.6514(183)} &   \pmb{0.6623(0.0000)}  \\
\hline
  $E_3$ &  \pmb{0.6649(0.0000)} &  \pmb{0.6610(0.0000)} &  \pmb{0.6851(0.0000)} &   -  \\
\hline
  $p$-value & 0.212 & 0.545 & 0.953  &   0.082 \\
  \hline
  $\delta E_{\textrm{exc}}$ &  0.0017 & 0.0019  & 0.0052 &  0.0010\\
  \hline
 \end{tabular}
  \caption{Results used to determine the excited state error for the ground-state energies in the $I=0$ channel.  We use method $A$ for the three moving-frame results given in columns 2-4, as described for the $I=2$ states in the text, so the explanation in the caption to Table~\ref{tab:I=2_exc_err} applies. Method $B$ is used for the stationary frame calculation whose results are shown in column 5. In the first stage of this procedure, only the extra-state energy $E_2$ is held fixed and the other three overlap amplitudes ($A_{a2}$, $A_{b2}$ and $A_{c2}$), all four shown in bold font, are determined to have the values shown, using the fit range 4-15. In the second stage, these four quanties are fixed to the values shown in the table and the preferred 3-operator-2-state fit carried out with the data effectively shifted by the fixed extra-state contribution. The difference between the ground-state energies determined from this extra-state fit and the preferred fit gives the excited-state error listed in the final row, as described in the text.}
  \label{tab:I=0_exc_err}
\end{table}

\subsection{Error budget}
\label{sec:error_budget}
We will now combine all of the errors detailed in the earlier sections to provide values for the seven $\pi\pi$ phase shifts that we have computed at specific energies and their corresponding errors.  We divide these errors into two categories. The first are the measurement errors associated with our lattice calculation of the finite-volume $\pi\pi$ energies, which includes the statistical error and the excited state contamination error.  As is discussed below, this category of error requires special attention since when we use L\"uscher's finite volume formalism to determine the scattering phase shift, errors in this category lead to correlated errors in both the value for the phase shift and the energy at which the phase shift is determined.  \par

We refer to the second category of errors as tuning errors. These include the finite volume error, finite lattice spacing error, the unphysical kinematics error, and the higher partial wave correction error.  We assign these errors directly to the seven phase shifts. They represent discrepancies between our physical results as presented and those we actually obtain.  For example, we describe our results as phase shifts in the continuum limit, but in reality the number we obtain contains finite lattice spacing errors.  Similarly we intend our phase shift results to be for the case where $m_\pi=135$ MeV.  In reality, our calculation was done for a different pion mass and we have made a small correction to the energy at which the phase shift is quoted to compensate for the shifted $\pi\pi$ threshold arising from our incorrect mass and, as was discussed in Section~\ref{sec:unphykin}, the remaining errors were estimated using ChPT. \par

Our first category of errors, the measurement errors, requires special treatment if we are to account for the tight correlation between the way the errors on the measured finite-volume energy shift affect both the implied value of the phase shift and the energy at which that phase shift is quoted. The issue is that we do not know the exact value of the finite volume $\pi\pi$ energy $E_{\rm FV}$, only that it lies with a certain confidence within the range $\bar E_{\rm FV} \pm \Delta E_{\rm FV}$, where $\bar E_{\rm FV}$ is the central value of the energy and $\Delta E_{\rm FV}$ is the error. As a result, naively propagating the error through the evaluation of the L\"uscher function would produce a result for which uncertainties exists on both the phase shift and the energy at which it is evaluated, and for which the errors are completely correlated (through the L\"uscher function).  This unsatisfactory situation can be remedied by transforming the result such that the phase shift is quoted at a fixed value of the energy and the uncertainty exists only on the phase shift. To achieve this we note that the allowed finite-volume energies $E_{\rm FV}$ mark the intersections of the curve $\delta(E)$, which describes the energy dependence of the phase shift, and the L\"uscher curve $W(E)$ relating the energy to the phase shift at a fixed lattice size:

\begin{equation}
\delta(E_{\mathrm{FV}}) = W(E_{\mathrm{FV}}).
\label{eq:FV-quant}
\end{equation}

We imagine that the true intersect occurs at some energy $E$ that is close to our measured central value $\overline{\raisebox{3.3 mm}{}\raisebox{3.3 mm}{}E}_{\rm FV}$ and perform a first-order Taylor expansion in $E$ of both sides of Eq.~\eqref{eq:FV-quant} about this central value
\begin{equation}
\delta(\overline{\raisebox{3.3 mm}{}E}_{\rm FV}) + \frac{ d\delta(E) }{dE} \Big{|}_{ \overline{\raisebox{2.3 mm}{}E}_{\rm FV} } (E - \overline{\raisebox{3.3 mm}{}E}_{FV})  \approx W(\overline{\raisebox{3.3 mm}{}E}_{\rm FV}) + \frac{ dW(E) }{dE} \Big{|}_{\overline{\raisebox{2.3 mm}{}E}_{\rm FV} } (E - \overline{\raisebox{3.3 mm}{}E}_{FV})\,.
\end{equation}
Rearranging the above we obtain an expression for the phase shift evaluated at our measured finite-volume energy $\overline{\raisebox{3.3 mm}{}E}_{\rm FV}$
\begin{equation}
\delta(\overline{\raisebox{3.3 mm}{}E}_{\mathrm{FV}}) = W(\overline{\raisebox{3.3 mm}{}E}_{\mathrm{FV}}) + \left\{\frac{dW(E)}{dE} 
                 -\frac{d\delta(E)}{dE}\right\}_{\overline{\raisebox{2.3 mm}{}E}_{\mathrm{FV}}} 
                  \left(E-\overline{\raisebox{3.3 mm}{}E}_{\mathrm{FV}}\right).
\label{eq:FV-expand}
\end{equation}

\begin{figure}
\centering
\includegraphics[width=1.0\linewidth]{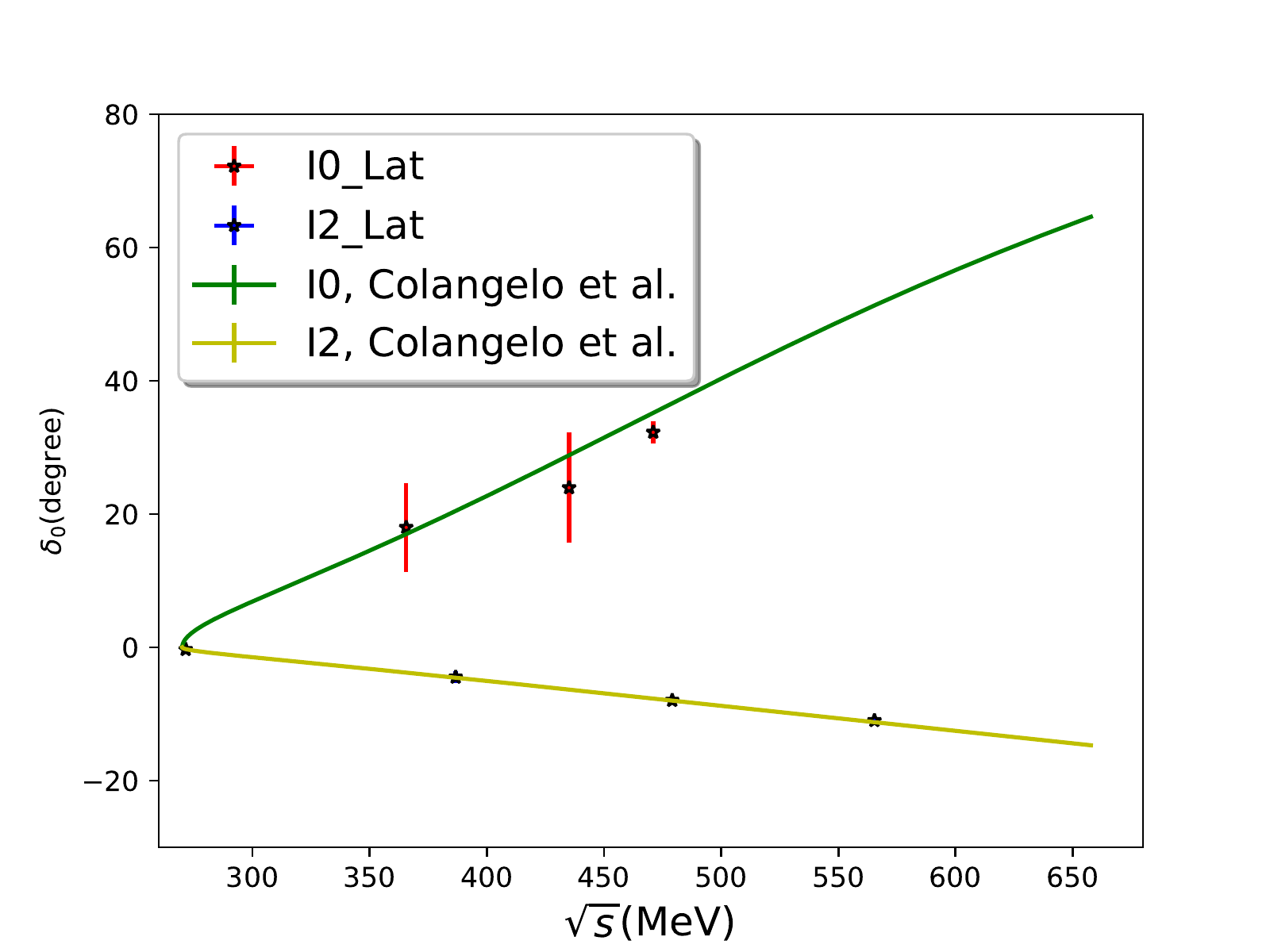}
\caption{A graph of our results for the seven phase shifts for $I=0$ and $I=2$ as a function of energy.  Shown also on the plot are the corresponding dispersive results~\cite{Colangelo:2001df}.  Note: the errors are not shown for the dispersive results.}
\label{fig:final}
\end{figure}

\begin{figure}
\centering
\includegraphics[width=1.0\linewidth]{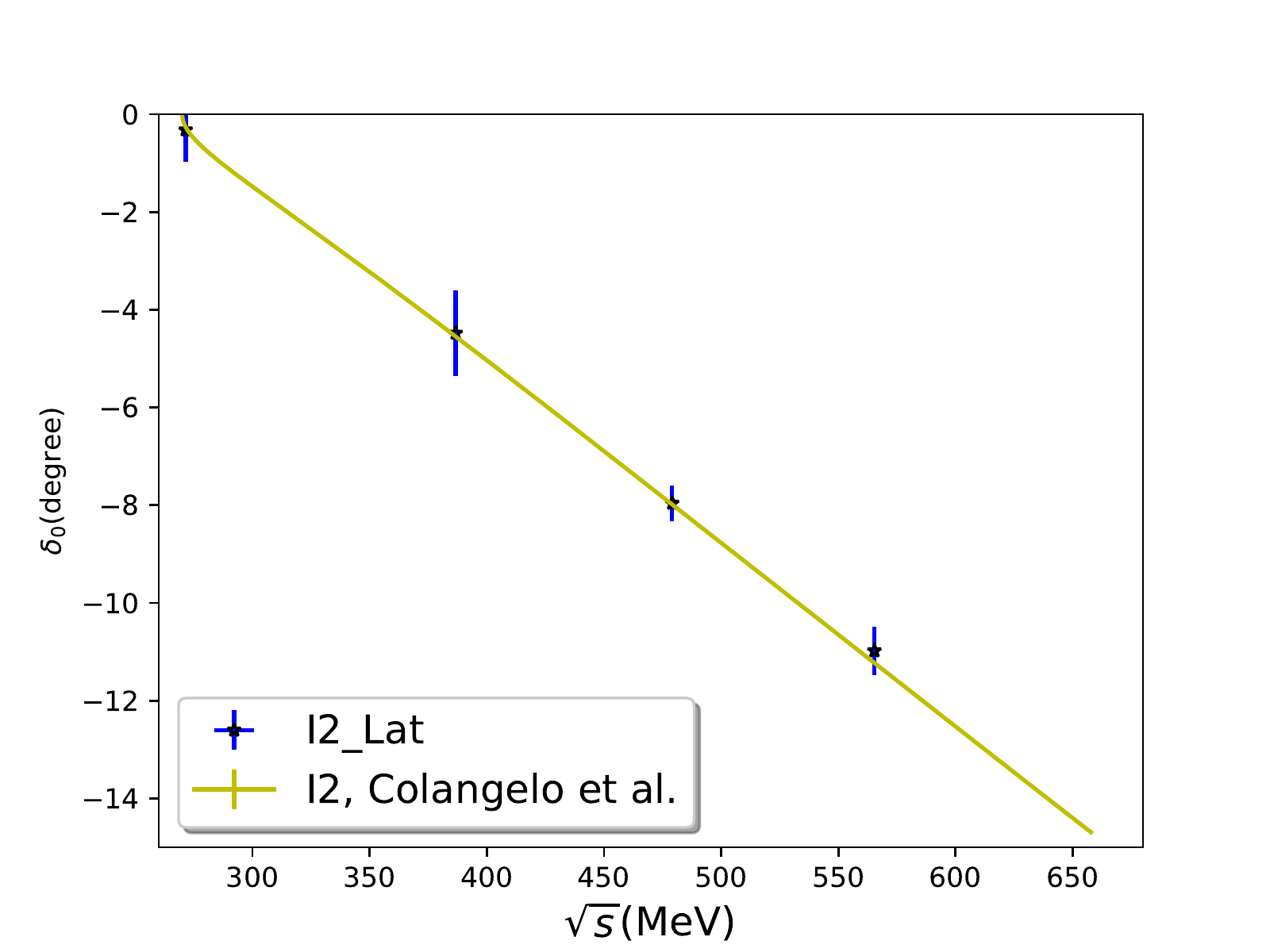}
\caption{The results for the $I=2$ phase shifts together with the corresponding dispersive results~\cite{Colangelo:2001df} that are shown in Fig.~\ref{fig:final}, but here with an expanded scale.}
\label{fig:final_I2}
\end{figure}

This approach requires an estimate of the derivative of the phase shift $\delta(E)$ with respect to its energy.  While this could be obtained from a fit to our data, we find it simplest to use the result from the dispersive analysis~\cite{Colangelo:2001df} which agrees well with our data as can be seen from the comparisons shown in Figures~\ref{fig:final} and \ref{fig:final_I2}.  \par

Thus, the statistical and excited state contamination errors on the measured finite-volume energies are each converted to errors on the phase shift at the fixed energy $E=\overline{\raisebox{3.3 mm}{}E}_{\mathrm{FV}}$ using the following relations: \par

\begin{equation}
\Delta\delta_{\mathrm{stat/exc}} = \left|\frac{dW(E)}{dE} 
                 -\frac{d\delta(E)}{dE}\right|_{\overline{\raisebox{2.3 mm}{}E}_{\mathrm{FV}}} \Delta E_{\mathrm{stat/exc}}\,,
\label{eq:FV-meas-error}
\end{equation}
where $\Delta E_{\mathrm{stat}}$ and $\Delta E_{\mathrm{exc}}$ are the statistical and excited state contamination errors that are assigned to the measured finite-volume energy.  The three tuning errors $\Delta\delta_{\mathrm{dis}}$, $\Delta\delta_{\mathrm{FV}}$, $\Delta\delta_{\mathrm{unphy}}$ coming from non-zero lattice spacing, finite volume and unphysical pion mass are assigned directly to the phase shifts.  All of these errors are listed in Table \ref{tab:sys_err}, where the total systematic error is the combination in quadrature of the systematic errors shown in columns 5 through 9. 

Another topic that we should discuss is the systematic error on the scattering length of both $I=0$ and $I=2$ channels. Similar to the discussion above, the systematic error is dominated by the excited state error, so we neglect all other sources of systematic error. The results are summarized in Tab.~\ref{tab:scattering_length_result}. It can be shown that, despite the fact that the scattering length of $I=0$ channel is inconsistent with the experimental value if we only consider the statistical error, as we show in Sec.~\ref{sec:phase_shift_result_stat_err_only}, the final systematic error for $I=0$ channel is three times larger than the central value, and by including this error, the lattice results are consistent with the experimental value. Our large excited state error in moving frame calculation, and the resulting large relative error on the energy difference between $E_{\pi\pi}$ and $2m_{\pi}$, leads to such a large error. 

\begin{table}
\begin{tabular}{ | c | c | c | c || c | c | c | c | c |}
  \hline
   $P_{tot}$ & I & $\sqrt{s}$(MeV) & $\delta$ & $\Delta\delta_{\textrm{dis}}$ & $\Delta\delta_{\textrm{FV}}$ & $\Delta\delta_{\textrm{unphy}}$ & $\Delta\delta_{\textrm{exc}}$ & $\Delta\delta_{\ell=2}$ \\
  \hline
 $(0,0,0)\frac{\pi}{L}$ & 0 & 471.0 & 32.3(1.0)(1.4) & 0.64  & 0.32 & 0.83 & 0.90 & 0.0\\ 
 \hline
 $(2,0,0)\frac{\pi}{L}$ & 0 &  435.1 & 24.0(3.4)(7.6) & 0.46  & 0.23 & 0.71 & 7.6 & 0.05\\ 
 \hline
 $(2,2,0)\frac{\pi}{L}$ & 0 &  365.6 & 18.0(4.5)(4.9) & 0.36  & 0.18 & 0.47 & 4.9 & 0.27\\ 
 \hline\hline
 $(0,0,0)\frac{\pi}{L}$ & 2 &  565.4 & -10.98(22)(44) & 0.20  & 0.10 & 0.18 & 0.34 & 0\\ 
 \hline
 $(2,0,0)\frac{\pi}{L}$ & 2 & 479.1  & -7.96(23)(29) &  0.16  & 0.08 & 0.03 & 0.23 & 0.01\\ 
 \hline
 $(2,2,0)\frac{\pi}{L}$ & 2 & 386.7 & -4.48(40)(77) & 0.09  & 0.04 & 0.06 & 0.76 & 0.02\\ 
 \hline
 $(2,2,2)\frac{\pi}{L}$ & 2 &  271.5 & -0.32(20)(63) & 0.01  & 0.00 & 0.02 & 0.63 & 0.0\\ 
 \hline
 \end{tabular}
 \caption{The final error budget for each of the seven $\pi\pi$ scattering phase shifts determined in this paper. Here each of the energies ($\sqrt{s}$) is the center-of-mass energy at which the phase shift $\delta$ has been determined, adjusted to correct for the unphysical pion mass according to the procedure described in Section~\ref{sec:phase_shift}. The right-most five columns are explained in the text.  All of the angles appearing in this table are expressed in degrees.  Two errors are shown with the phase shift results in the fourth column.  The first is statistical and is given in Table~\ref{table:phase_shift_stat_only}.  The second is systematic and is the average in quadrature of the individual errors shown in columns 5-8.}
\label{tab:sys_err}
\end{table}

\begin{table}
\begin{tabular}{ | c | c | c |}
  \hline
   I & $m_\pi a_0$(lat) & $m_\pi a_0$(exp) \\
  \hline
 0 & 0.072(38)(210) & 0.220(5)\\ 
 \hline
 2 & -0.055(15)(68) & -0.0444(10)\\ 
 \hline
 \end{tabular}
 \caption{The final results for the $\pi\pi$ scattering length determined in this paper. Here $a_0$(exp) is the experimental value of the scattering length quoted from Ref.~\cite{Colangelo_2000}}
\label{tab:scattering_length_result}
\end{table}

\section{Conclusions}
\label{sec:conclusions}
In this paper we have presented in detail a lattice calculation of the $\pi\pi$ scattering phase shift for both the $I=0$ and $I=2$ channels. Our final results are presented in Table~\ref{tab:sys_err} and illustrated in Figures~\ref{fig:final} and \ref{fig:final_I2}.  This calculation is performed using a physical pion mass with G-parity boundary conditions in all three spatial directions.  This results in a single-pion, finite-volume ground state with non-zero momentum, which is crucial for the closely related calculation presented in Ref.~\cite{Abbott:2020hxn} for the $I=0$ $K\to\pi\pi$ decay amplitude and $\epsilon'$. \par

Compared with our 2015 calculation~\cite{Bai:2015nea} of the same quantities, the current calculation is based on a 3.4 times larger number of configurations and incorporates two additional $\pi\pi$ interpolating operators, one of which is a four-quark operator constructed from two pion interpolating operators which each carry larger-than-minimum momenta, while the other is a scalar two-quark operator (the sigma operator).  With these improvements we obtain an $I=0$ $\pi\pi$ scattering phase shift at 471 MeV of $32.3^\circ(1.0)(1.4)$\footnote{Note this number is slightly different from the number given in Ref.~\cite{Abbott:2020hxn}. This is because we have now included an estimate of the error due to the unphysical pion mass, resulting in a correction to the $\pi\pi$ energy and we have refined our excited state error estimation.}. Comparing this result with the one presented five years ago~\cite{Bai:2015nea}, we have the following improvements: i) The statistical error is improved by a factor of 5. ii) We are able to provide a more reliable and detailed systematic error analysis.   iii) We have been able to resolve the $3\sigma$ discrepancy between our earlier result for this phase shift and that predicted by a dispersive analysis~\cite{Colangelo:2001df} so that our current results agree well with the dispersive prediction (cf. Fig.~\ref{fig:final}). The discrepancy is now understood to have resulted from excited $\pi\pi$ state contamination, which was underestimated in Ref.~\cite{Bai:2015nea} and is now under much greater control. \par

More specifically, we have employed a concrete procedure for estimating the error resulting from a nearby excited state that was not included in our fit. As discussed in Sec.~\ref{sec:systematic_error}, we introduce one additional state into our fit but with an energy fixed to that given by the dispersive calculation~\cite{Colangelo:2001df} and with overlap factors to our operators carefully estimated, so as to avoid introducing instability in the fits or inflating the statistical error. The resulting shift in the ground-state energy then provides a meaningful indicator of the size of the corresponding systematic error. \par

In addition to computing the $I=0$ phase shift for two pions with zero total momentum, we also perform a moving-frame calculation with three different total momenta. The observation~\cite{Kelly:2019wfj} that three types of lattice symmetry can be used to significantly reduce the number of contractions was exploited to reduce the contraction time by a factor of seven.  The resulting values of the $\pi\pi$ phase shifts at lower energies not only allow us to perform a comparison with dispersive and chiral perturbation theory predictions but also give us an independent evaluation of the Lellouch-L\"uscher correction needed to obtain the $K\to\pi\pi$ decay amplitude from a finite-volume lattice QCD calculation.  Because of the critical role played by the sigma interpolating operator in the stationary frame calculation, we will include a sigma operator with non-zero total momentum in future work.  This operator might be expected to strongly couple to more states in the fit, in contrast to the $\pi\pi(311,311)$ operator, and may significantly reduce the errors as it did for the stationary case.  An additional future goal is to extrapolate the result to the continuum limit, since in this work, the calculation is performed on a single $32^3$ lattice. We are now preparing such a scaling calculation using two lattice volumes with spatial extents of $48^3$ and $64^3$. \par

In this paper, we have used a combination of bootstrap and jackknife methods~\cite{Kelly:2019yxg} together with correlated fits to determine the ground-state energies, the operator-state overlap amplitudes and the analysis of the excited-state error. This method allows us to estimate the goodness-of-fit, which provides guidance as to which model and fitting range we should choose (see Sec.~\ref{sec:pipi_energy}).  We also compared our multi-state fitting with the GEVP method and found that our fitting procedure gave consistent statistical errors for the $I=0$ case than did our implementation of the GEVP method.  We did not attempt to estimate the systematic errors resulting from the GEVP approach.  The GEVP method may well excel when more states and operators are included. \par

\section*{Acknowledgments}

We would like to thank our RBC and UKQCD collaborators for their ideas and
assistance and to specifically acknowledge the contributions of Chulwoo Jung 
and Taku Izubuchi.  The generation of the gauge configurations used in this work 
was primarily performed using the IBM BlueGene/Q (BG/Q) installation at BNL 
(supported by the RIKEN BNL Research Center and BNL),  the Mira computer 
at the ALCF (as part of the Incite program), Japan's KEKSC 1540 computer 
and the STFC DiRAC machine at the University of Edinburgh 
(STFC grants ST/R00238X/1, ST/S002537/1, ST/R001006/1), with additional generation
performed using the NCSA Blue Waters machine at the University of
Illinois. The majority of the measurements and analysis, including the NPR
calculations, were performed using the Cori supercomputer at NERSC, with
contributions also from the Hokusai machine at ACCC RIKEN and the BG/Q
machines at BNL.  

The software used in this calculation was developed from a combination of 
CPS~\cite{Jung:2014ata} \\~\href{https://github.com/RBC-UKQCD/CPS}{https://github.com/RBC-UKQCD/CPS} and Grid~\cite{Boyle:2016lbp}~\href{https://github.com/paboyle/Grid}{https://github.com/paboyle/Grid}.

T.B. was supported by U.S. DOE grant \#DE-SC0010339, N.H.C., R.D.M. and T.W. by U.S. DOE grant \#DE-SC0011941.   C.L. and A.S. were supported in part by the US DOE contract \#DE-SC0012704. D.H. was supported by a SCIDAC grant "Computing  the Properties of Matter with Leadership Computing Resources.  A.S.M. was supported in part by US DOE Contract DESC0012704(BNL) and by a DOE Office of Science Early Career Award.  DJM was supported in part by the U.S. DOE grants \#DE-SC0010339 and \#DE-SC0011090.  C.T.S. was partially supported by an Emeritus Fellowship from the Leverhulme Trust and by STFC (UK) grants\,ST/P000711/1 and ST/T000775/1. 

\appendix

\section{Quark level momentum distribution}
\label{sec:appendix_quark_momentum}
\input{appendix_quark_momentum.tex}

\section{Chpt prediction for phase shift}
\label{sec:appendix_chpt_phase_shift}
\input{appendix_chpt_phase_shift.tex}

\section{Contractions for \texorpdfstring{$\pi\pi$}{} and \texorpdfstring{$\sigma$}{} operators}
\label{sec:appendix_contraction}
\input{appendix_contraction}

\FloatBarrier
\bibliography{main}

\end{document}

%% file: user_def.tex
\newcommand{\ba}{\begin{eqnarray}}
\newcommand{\ea}{\end{eqnarray}}
\newcommand{\bas}{\begin{eqnarray*}}
\newcommand{\eas}{\end{eqnarray*}}
\newcommand{\be}{\begin{equation}}
\newcommand{\ee}{\end{equation}}
\newcommand{\bes}{\begin{equation*}}
\newcommand{\ees}{\end{equation*}}
\newcommand{\bi}{\begin{itemize}}
\newcommand{\ei}{\end{itemize}}

\newcommand{\bcentre}{\begin{center}}
\newcommand{\ecentre}{\end{center}}

\font\tenmsb=msbm10 scaled\magstep1
\font\sevenmsb=msbm7 scaled\magstep1
\font\fivemsb=msbm5 scaled\magstep1
\newfam\msbfam
\textfont\msbfam=\tenmsb
\scriptfont\msbfam=\sevenmsb
\scriptscriptfont\msbfam=\fivemsb

\def\ln{\mathop{\rm ln}}		    

\def\su3{SU(3)}

\def\tD{\mbox{D}\kern-0.65em\raise0.15ex\hbox{/}\kern0.15em} 
\def\sD{\mbox{\scriptsize D}\kern-0.5em\raise0.15ex\hbox{\scriptsize/}}
\def\ssD{\mbox{\tiny D}\kern-0.42em\raise0.15ex\hbox{\tiny/}}

\def\dslash{\hbox{\(\partial\)}\kern-0.5em\raise0.15ex\hbox{/}} 

\def\su3{SU(3)}

\def\nicefrac#1#2{\leavevmode\kern.1em\raise.5ex\hbox{\the\scriptfont0 #1}\kern-
.1em/\kern-.15em\lower.25ex\hbox{\the\scriptfont0 #2}}

\newcommand{\bea}{\begin{eqnarray}}
\newcommand{\eea}{\end{eqnarray}}

\newcommand{\colvectwo}[2]{\left(\begin{array}{c}#1\\#2\end{array}\right)}

%% file: appendix_quark_momentum.tex
\subsection{Pion operator}

In Sec.~\ref{sec:overview:ops}, Eqs.~\eqref{eq:pion-1}-\eqref{eq:pion-3} we detail the interpolating operators for the charged and neutral pions. The pion momentum, $\vec P = \vec p + \vec q$ is the sum of the momenta $\vec p$ and $\vec q$ assigned to the quark and antiquark, respectively. Given a pion momentum, there are multiple ways of distribute momentum between quark and anti-quark component. 

As shown in Ref.~\cite{Christ:2019sah} the allowed quark momenta (for G-parity BCs in 3 directions) are \begin{equation} 
\pm\frac{\pi}{2L}(1,1,1) + \frac{2\pi}{L}\vec n\,, 
\end{equation} 
where $\vec n$ is a vector of integers. While any combination of $\vec p$ and $\vec q$ satisfying this condition result in valid pion interpolating operators, we observed in Ref.~\cite{Christ:2019sah} that the cubic symmetry breaking manifest in the operator amplitudes between pion states of total momentum related by cubic rotations is dramatically suppressed by averaging over pairs of bilinear operators with the same total momentum but with different assignments of quark momenta. The specific criteria for selecting those momenta are discussed in more detail in that paper; here in Table~\ref{tab:quark_mom_choice_pion} we list only the two choices for each of the 32 total momenta. The momentum distribution is listed below for all 32 pions we use (in units of $\pi/2L$):

\begin{table}
\centering
\scalebox{0.8}{
\begin{tabular}{|c|c|c|}
\hline
Pion momentum & quark momentum, choice 1&  quark momentum, choice 2\\
\hline
( 2, 2, 2) & ( 1, 1, 1) + ( 1, 1, 1) & (-1,-1,-1) + ( 3, 3, 3)\\
\hline
(-2,-2,-2) & (-1,-1,-1) + (-1,-1,-1) & ( 1, 1, 1) + (-3,-3,-3)\\
\hline
( 2, 2,-2) & ( 1, 1, 1) + ( 1, 1,-3) & (-1,-1,-1) + ( 3, 3,-1)\\
\hline
( 2,-2, 2) & ( 1, 1, 1) + ( 1,-3, 1) & (-1,-1,-1) + ( 3,-1, 3)\\
\hline
(-2, 2, 2) & ( 1, 1, 1) + (-3, 1, 1) & (-1,-1,-1) + (-1, 3, 3)\\
\hline
(-2,-2, 2) & (-1,-1,-1) + (-1,-1, 3) & ( 1, 1, 1) + (-3,-3, 1)\\
\hline
(-2, 2,-2) & (-1,-1,-1) + (-1, 3,-1) & ( 1, 1, 1) + (-3, 1,-3)\\
\hline
( 2,-2,-2) & (-1,-1,-1) + ( 3,-1,-1) & ( 1, 1, 1) + ( 1,-3,-3)\\
\hline
( 2, 2, 6) & ( 1, 1, 1) + ( 1, 1, 5) & (-1,-1,-1) + ( 3, 3, 7)\\
\hline
( 2, 6, 2) & ( 1, 1, 1) + ( 1, 5, 1) & (-1,-1,-1) + ( 3, 7, 3)\\
\hline
( 6, 2, 2) & ( 1, 1, 1) + ( 5, 1, 1) & (-1,-1,-1) + ( 7, 3, 3)\\
\hline
(-2,-2,-6) & (-1,-1,-1) + (-1,-1,-5) & ( 1, 1, 1) + (-3,-3,-7)\\
\hline
(-2,-6,-2) & (-1,-1,-1) + (-1,-5,-1) & ( 1, 1, 1) + (-3,-7,-3)\\
\hline
(-6,-2,-2) & (-1,-1,-1) + (-5,-1,-1) & ( 1, 1, 1) + (-7,-3,-3)\\
\hline
( 2, 2,-6) & ( 1, 1, 1) + ( 1, 1,-7) & (-1,-1,-1) + ( 3, 3,-5)\\
\hline
( 2,-6, 2) & ( 1, 1, 1) + ( 1,-7, 1) & (-1,-1,-1) + ( 3,-5, 3)\\
\hline
(-6, 2, 2) & ( 1, 1, 1) + (-7, 1, 1) & (-1,-1,-1) + (-5, 3, 3)\\
\hline
(-2,-2, 6) & (-1,-1,-1) + (-1,-1, 7) & ( 1, 1, 1) + (-3,-3, 5)\\
\hline
(-2, 6,-2) & (-1,-1,-1) + (-1, 7,-1) & ( 1, 1, 1) + (-3, 5,-3)\\
\hline
( 6,-2,-2) & (-1,-1,-1) + ( 7,-1,-1) & ( 1, 1, 1) + ( 5,-3,-3)\\
\hline
(-2, 2, 6) & ( 1, 1, 1) + (-3, 1, 5) & (-1,-1,-1) + (-1, 3, 7)\\
\hline
( 2, 6,-2) & ( 1, 1, 1) + ( 1, 5,-3) & (-1,-1,-1) + ( 3, 7,-1)\\
\hline
( 6,-2, 2) & ( 1, 1, 1) + ( 5,-3, 1) & (-1,-1,-1) + ( 7,-1, 3)\\
\hline
( 2,-2,-6) & (-1,-1,-1) + ( 3,-1,-5) & ( 1, 1, 1) + ( 1,-3,-7)\\
\hline
(-2,-6, 2) & (-1,-1,-1) + (-1,-5, 3) & ( 1, 1, 1) + (-3,-7, 1)\\
\hline
(-6, 2,-2) & (-1,-1,-1) + (-5, 3,-1) & ( 1, 1, 1) + (-7, 1,-3)\\
\hline
( 2,-2, 6) & ( 1, 1, 1) + ( 1,-3, 5) & (-1,-1,-1) + ( 3,-1, 7)\\
\hline
(-2, 6, 2) & ( 1, 1, 1) + (-3, 5, 1) & (-1,-1,-1) + (-1, 7, 3)\\
\hline
( 6, 2,-2) & ( 1, 1, 1) + ( 5, 1,-3) & (-1,-1,-1) + ( 7, 3,-1)\\
\hline
(-2, 2,-6) & (-1,-1,-1) + (-1, 3,-5) & ( 1, 1, 1) + (-3, 1,-7)\\
\hline
( 2,-6,-2) & (-1,-1,-1) + ( 3,-5,-1) & ( 1, 1, 1) + ( 1,-7,-3)\\
\hline
(-6,-2, 2) & (-1,-1,-1) + (-5,-1, 3) & ( 1, 1, 1) + (-7,-3, 1)\\
\hline
\end{tabular}
}
\caption{The quark/anti-quark momenta choices for all 32 pion total momenta. For each total momenta, two momentum choices are given to suppress the cubic symmetry given. All momenta are given in units of $\pi/2L$.}
\label{tab:quark_mom_choice_pion}
\end{table}

Recall that in addition to the above, we also symmetrize the momentum between the quark and antiquark by averaging the assignments $(\vec p,\vec q)$ and $(\vec q, \vec p)$. Thus in practice our pion interpolating operators comprise an average over a total of four quark field bilinears.

\subsection{\texorpdfstring{$\sigma$}{s} operator}
In this work we use the $\sigma$ operator only in the case of zero total momentum, and as a result the momenta assigned to the quark and anti-quark fields must be equal and opposite. We construct an operator that is symmetric under cubic rotations by averaging over 8 orientations of the quark momentum. The list of momenta assigned to the quark operator are given in Table.~\ref{tab:quark_mom_choice_sigma}. 
\begin{table}
\begin{center}
\begin{tabular}{|c|c|c|}
\hline
Index & quark momentum\\
\hline
1& ( 1, 1, 1)\\
  \hline
2& (-1,-1,-1)\\
  \hline
3& (-3, 1, 1)\\
  \hline
4& ( 3,-1,-1)\\
  \hline
5& ( 1,-3, 1)\\
  \hline
6& (-1, 3,-1)\\
  \hline
7& ( 1, 1,-3)\\
  \hline
8& (-1,-1, 3)\\                                                                                                                        
\hline
\end{tabular}
\caption{The 8 orientations of quark momentum we average to get the $\sigma$ operator with zero total momentum. The anti-quark momentum in each case is the reverse of the quark momentum. All momenta are given in units of $\pi/2L$.}
\label{tab:quark_mom_choice_sigma}
\end{center}
\end{table}

%% file: appendix_chpt_phase_shift.tex
In this appendix we present the partial wave amplitude $t_{l=0}^{I}$ results from the next-to-leading-order (NLO) ChPT. These amplitudes are connected with the scattering phase shift by
\begin{equation}
    t_{l=0}^{I} = \left(\frac{s-4}{s}\right)^{1/2}e^{i\delta^I_{l=0}(s)}\textrm{sin}[\delta^I_{l=0}(s)]
\end{equation}
\begin{equation}
\begin{aligned} 
t_{0}^{0}\left(m_{\pi}, f_{\pi}, k\right)=\frac{7 m_{\pi}^{2}}{16 \pi f_{\pi}^2}\left(1+\frac{8}{7} \frac{k^{2}}{m_{\pi}^{2}}\right)+\frac{m_{\pi}^{4}}{8 \pi f_{\pi}^{4}}\left[\left(5 b_{1}+12 b_{2}+48 b_{3}+32 b_{4}+\frac{83}{24 \pi^{2}}\right)\right.\\
\left.+\left(8 b_{2}+96 b_{3}+96 b_{4}+\frac{269}{36 \pi^{2}}\right) \frac{k^{2}}{m_{\pi}^{2}}+\left(\frac{176}{3} b_{3}+\frac{272}{3} b_{4}+\frac{311}{54 \pi^{2}}\right) \frac{k^{4}}{m_{\pi}^{4}}\right] \\
+\frac{m_{\pi}^{4}}{256 \pi^{3} f_{\pi}^{4}} \frac{1}{\sqrt{1+\frac{m_{\pi}^{2}}{k^{2}}}}\left(49+112 \frac{k^{2}}{m_{\pi}^{2}}+64 \frac{k^{4}}{m_{\pi}^{4}}\right) \log \left(\frac{1-\sqrt{1+\frac{m_{\pi}^{2}}{k^{2}}}}{1+\sqrt{1+\frac{m_{\pi}^{2}}{k^{2}}}}\right) \\
+\frac{m_{\pi}^{4}}{1152 \pi^{3} f_{\pi}^{4}} \sqrt{1+\frac{m_{\pi}^{2}}{k^{2}}}\left(27+64 \frac{k^{2}}{m_{\pi}^{2}}+112 \frac{k^{4}}{m_{\pi}^{4}}\right) \log \left(\frac{\sqrt{1+\frac{m_{\pi}^{2}}{k^{2}}}-1}{\sqrt{1+\frac{m_{\pi}^{2}}{k^{2}}}+1}\right) \\
-\frac{m_{\pi}^{4}}{128 \pi^{3} f_{\pi}^{4}}\left(1-\frac{1}{24} \frac{m_{\pi}^{2}}{k^{2}}\right)\left\{-4[\log (2)]^{2}+\left[\log \left(\frac{4 k^{2}}{m_{\pi}^{2}}\right)\right]^{2}-\left[\log \left(\frac{\sqrt{1+\frac{m_{\pi}^{2}}{k^{2}}}-1}{\sqrt{1+\frac{m_{\pi}^{2}}{k^{2}}}+1}\right)\right]^{2}\right. \\
+4 \log \left(\frac{4 k^{2}}{m_{\pi}^{2}}\right) \log \left(1+\sqrt{1+\frac{m_{\pi}^{2}}{k^{2}}}\right)+4\left[\log \left(1+\sqrt{1+\frac{m_{\pi}^{2}}{k^{2}}}\right)\right]^{2} \\
\left.+2 \log \left(\frac{4 k^{2}}{m_{\pi}^{2}}\right) \log \left(\frac{\sqrt{1+\frac{m_{\pi}^{2}}{k^{2}}}-1}{\sqrt{1+\frac{m_{\pi}^{2}}{k^{2}}}+1}\right)+4 \log \left(1+\sqrt{1+\frac{m_{\pi}^{2}}{k^{2}}}\right) \log \left(\frac{\sqrt{1+\frac{m_{\pi}^{2}}{k^{2}}}-1}{\sqrt{1+\frac{m_{\pi}^{2}}{k^{2}}}+1}\right)\right\}
\end{aligned}
\end{equation}
\begin{equation}
\begin{aligned}
t_{0}^{2}\left(m_{\pi}, f_{\pi}, k\right)=- \frac{m_{\pi}^{2}}{8 \pi f_{\pi}^{2}}\left(1+2 \frac{k^{2}}{m_{\pi}^{2}}\right)+\frac{m_{\pi}^{4}}{4 \pi f_{\pi}^{4}}\left[\left(b_{1}+16 b_{4}+\frac{43}{96 \pi^{2}}\right)\right.\\ 
\left.-\left(2 b_{2}-48 b_{4}-\frac{199}{144 \pi^{2}}\right) \frac{k^{2}}{m_{\pi}^{2}}+\left(\frac{16}{3} b_{3}+\frac{112}{3} b_{4}+\frac{265}{216 \pi^{2}}\right) \frac{k^{4}}{m_{\pi}^{4}}\right] \\ +\frac{m_{\pi}^{4}}{64 \pi^{3} f_{\pi}^{4}} \frac{1}{\sqrt{1+\frac{m_{\pi}^{2}}{k^{2}}}}\left(1+4 \frac{k^{2}}{m_{\pi}^{2}}+4 \frac{k^{4}}{m_{\pi}^{4}}\right) \log \left(\frac{1-\sqrt{1+\frac{m_{\pi}^{2}}{k^{2}}}}{1+\sqrt{1+\frac{m_{\pi}^{2}}{k^{2}}}}\right) \\
+\frac{m_{\pi}^{4}}{1152 \pi^{3} f_{\pi}^{4}} \sqrt{1+\frac{m_{\pi}^{2}}{k^{2}}}\left(27+112 \frac{k^{2}}{m_{\pi}^{2}}+88 \frac{k^{4}}{m_{\pi}^{4}}\right) \log \left(\frac{\sqrt{1+\frac{m_{\pi}^{2}}{k^{2}}}-1}{\sqrt{1+\frac{m_{\pi}^{2}}{k^{2}}}+1}\right) \\
+\frac{m_{\pi}^{4}}{256 \pi^{3} f_{\pi}^{4}}\left(1+\frac{13}{12} \frac{m_{\pi}^{2}}{k^{2}}\right)\left\{-4[\log (2)]^{2}+\left[\log \left(\frac{4 k^{2}}{m_{\pi}^{2}}\right)\right]^{2}-\left[\log \left(\frac{\sqrt{1+\frac{m_{\pi}^{2}}{k^{2}}}-1}{\sqrt{1+\frac{m_{\pi}^{2}}{k^{2}}+1}}\right)\right]^{2}\right. \\
+4 \log \left(\frac{4 k^{2}}{m_{\pi}^{2}}\right) \log \left(1+\sqrt{1+\frac{m_{\pi}^{2}}{k^{2}}}\right)+4\left[\log\left (1+\sqrt{1+\frac{m_{\pi}^{2}}{k^{2}}}\right)\right]^{2} \\
\left.+2 \log \left(\frac{4 k^{2}}{m_{\pi}^{2}}\right) \log \left(\frac{\sqrt{1+\frac{m_{\pi}^{2}}{k^{2}}}-1}{\sqrt{1+\frac{m_{\pi}^{2}}{k^{2}}}+1}\right)+4 \log \left(1+\sqrt{1+\frac{m_{\pi}^{2}}{k^{2}}}\right) \log \left(\frac{\sqrt{1+\frac{m_{\pi}^{2}}{k^{2}}}-1}{\sqrt{1+\frac{m_{\pi}^{2}}{k^{2}}}+1}\right)\right\} \,.
\end{aligned}
\end{equation}

The parameters $b_1-b_4$ are linear combinations of low energy constants defined in Ref~\cite{Bijnens:1995yn}, and we took their values from Ref~\cite{Colangelo:2001df}. In this work, these expressions are used to estimate the unphysical pion mass error. The two pion masses we use are the lattice pion mass $m^{\textrm{lat}}_{\pi} = 142.3$ MeV and physical pion mass $m^{\textrm{phy}}_{\pi} = 135$ MeV.

%% file: appendix_contraction.tex
In this appendix we list the contraction formula for each diagram introduced in Sec.~\ref{sec:overview}. The first four diagrams are associated with the product of two $\pi\pi$ interpolating operators, where the four time slices are the time coordinates of the four single-pion interpolating operators, which are $t_{\textrm{src}}-4, t_{\textrm{src}}, t_{\textrm{snk}}$ and $t_{\textrm{snk}}+4$, respectively. The final four expressions correspond to the cases where at least one of the source or sink operators is a $\sigma$ operator. The quantity $P_{t_a, t_b}$ is the G-parity quark propagator from $t_a$ to $t_b$ while the flavor-spin matrix $S_1$ is defined as $S_1 = \sigma_3\gamma_5$.  These eight amplitudes are obtained from the following contractions:
\begin{equation}
    C = \frac{1}{2}Tr\left \{ P_{t_1, t_3} S_1 P_{t_3, t_2} S_1 P_{t_2,t_4} S_1 P_{t_4, t_1} S_1 \right \}
\end{equation}
\begin{equation}
    D = \frac{1}{2} \left [ \left ( \frac{1}{2}Tr\left \{ P_{t_1, t_3} S_1 P_{t_3, t_1} S_1 \right \}  \right   ) \cdot \left ( \frac{1}{2}Tr\left \{ P_{t_2, t_4} S_1 P_{t_4, t_2} S_1 \right \}  \right   ) + (t_3 \leftrightarrow t_4) \right ]
\end{equation}
\begin{equation}
    R = \frac{1}{2} \left [ \frac{1}{2}Tr\left \{ P_{t_1, t_2} S_1 P_{t_2, t_3} S_1 P_{t_3,t_4} S_1 P_{t_4, t_1} S_1 \right \} + (t_3 \leftrightarrow t_4)  \right ]
\end{equation}
\begin{equation}
    V =  \left ( \frac{1}{2}Tr\left \{ P_{t_1, t_2} S_1 P_{t_2, t_1} S_1 \right \}  \right   ) \cdot \left ( \frac{1}{2}Tr\left \{ P_{t_3, t_4} S_1 P_{t_4, t_3} S_1 \right \}  \right   )
\end{equation}
\begin{equation}
    C_{\sigma\sigma} = Tr\left \{ P_{t_1, t_2} P_{t_2, t_1}\right \} 
\end{equation}
\begin{equation}
    V_{\sigma\sigma} = \left (Tr\left \{ P_{t_1, t_1}\right \} \right )\cdot \left ( Tr\left \{ P_{t_2, t_2}\right \} \right )
\end{equation}
\begin{equation}
    C_{\sigma\pi\pi} = i \cdot Tr\left \{ P_{t_1, t_0} P_{t_0, t_2} S_1 P_{t_2, t_1} S_1\right \} 
\end{equation}
\begin{equation}
    V_{\sigma\pi\pi} = i \cdot Tr\left \{ P_{t_0, t_0} \right \} \cdot Tr\left \{ P_{t_1, t_2} S_1 P_{t_2, t_1} S_1\right \}  \,.
\end{equation}